%% file: main.tex
\documentclass[12pt]{iopart}

\usepackage[dvipsnames]{xcolor}

\usepackage{fancyhdr}
\usepackage[UKenglish]{babel}
\usepackage [autostyle, english = british]{csquotes}
\usepackage{hyperref}
\usepackage{capt-of}
\usepackage{makeidx}
\usepackage{setspace}
\usepackage{placeins}
\usepackage{algorithm}
\usepackage{algorithmic}

\expandafter\let\csname equation*\endcsname\relax 
\expandafter\let\csname endequation*\endcsname\relax
\usepackage{amsmath}
\usepackage{siunitx}
\usepackage{graphicx}
\usepackage{float}
\usepackage{subcaption}
\usepackage[labelfont=bf]{caption}
\usepackage{amsfonts}

\newcommand{\rev}[1]{{ #1}}

\usepackage{xr}
\makeatletter
\newcommand*{\addFileDependency}[1]{
  \typeout{(#1)}
  \@addtofilelist{#1}
  \IfFileExists{#1}{}{\typeout{No file #1.}}
}
\makeatother

\newcommand*{\myexternaldocument}[1]{%
    \externaldocument{#1}%
    \addFileDependency{#1.tex}%
    \addFileDependency{#1.aux}%
}
\myexternaldocument{SM}

\begin{document}

\title[Coupled differentiation and division of stem cells]{Coupled differentiation and division of embryonic stem cells inferred from clonal snapshots}

\author{Liam J. Ruske}
\address{Rudolf Peierls Centre for Theoretical Physics, University of Oxford, Parks Road, Oxford, OX1 3PU, UK}
\ead{Liam.Ruske@physics.ox.ac.uk}

\author{Jochen Kursawe}
\address{School of Mathematics and Statistics, University of St Andrews, North Haugh, St Andrews, UK}
\ead{jochen.kursawe@st-andrews.ac.uk}

\author{Anestis Tsakiridis}
\address{Centre for Stem Cell Biology, Department of Biomedical Science, The University of Sheffield, Sheffield, United Kingdom}
\ead{a.tsakiridis@sheffield.ac.uk}

\author{Valerie Wilson}
\address{MRC Centre for Regenerative Medicine, The University of Edinburgh,
Edinburgh BioQuarter, 5 Little France Drive, Edinburgh, EH164UU, UK}
\ead{V.Wilson@ed.ac.uk}

\author{Alexander G. Fletcher}
\address{School of Mathematics and Statistics, University of Sheffield, Hicks Building, Hounsfield Road, Sheffield, S10 1RB, UK}
\address{Bateson Centre, University of Sheffield, Sheffield S10 2TN, UK}
\ead{a.g.fletcher@sheffield.ac.uk}

\author{Richard A. Blythe}
\address{SUPA, School of Physics and Astronomy,
University of Edinburgh, James Clerk Maxwell Building, Peter Guthrie Tait Road, Edinburgh, EH9 3FD, UK}
\ead{R.A.Blythe@ed.ac.uk}

\author{Linus J. Schumacher}
\address{MRC Centre for Regenerative Medicine, The University of Edinburgh,
Edinburgh BioQuarter, 5 Little France Drive, Edinburgh, EH164UU, UK}
\ead{Linus.Schumacher@ed.ac.uk}

\vspace{10pt}
\begin{indented}
\item[]April 2020
\end{indented}


\begin{abstract} 
The deluge of single-cell data obtained by sequencing, imaging and epigenetic markers has led to an increasingly detailed description of cell state. 
However, it remains challenging to identify how cells transition between different states, in part because data are typically limited to snapshots in time. 
A prerequisite for inferring cell state transitions from such snapshots is to distinguish whether transitions are coupled to cell divisions. 
To address this, we present two minimal branching process models of cell division and differentiation in a well-mixed population. 
These models describe dynamics where differentiation and division are coupled or uncoupled. 
For each model, we derive analytic expressions for each subpopulation's mean and variance and for the likelihood, allowing exact Bayesian parameter inference and model selection in the idealised case of fully observed trajectories of differentiation and division events.
In the case of snapshots, we present a sample path algorithm and use this to predict optimal temporal spacing of measurements for experimental design. 
We then apply this methodology to an \textit{in vitro} dataset assaying the clonal growth of epiblast stem cells in culture conditions promoting self-renewal or differentiation. 
Here, the larger number of cell states necessitates approximate Bayesian computation. 
For both culture conditions, our inference supports the model where cell state transitions are coupled to division. 
For culture conditions promoting differentiation, our analysis indicates a possible shift in dynamics, with these processes becoming more coupled over time.
\end{abstract}

%
%
%
%
%

\clearpage

\section{Introduction}

Changes in gene expression underlie many aspects of cellular behaviour in tissue development, homeostasis, and regeneration. 
The concept of discrete cell states is intended to capture the distinct patterns of gene expression that are observed within tissues over time. 
The deluge of single cell data is leading to an increasingly detailed description of cell state \cite{Casey2020,Morrisdev169748}. 
There are various ways to interrogate a cell's state in different contexts in vivo and in vitro. 
Modern technologies such as scRNAseq produce vast amounts of data but are costly, laborious to analyse, and relatively noisy. 
Older techniques, such as immunofluorescent stainings, where cell state can be defined by the co-expression of a small number of genes/proteins, are cheaper and simpler and thus still remain heavily used.

While there are numerous techniques for classifying cell states based on single cell data, and ordering them along pseudotime trajectories \cite{Saelens2018,Tritschlerdev170506}, such classification only offers very limited insight into the \emph{dynamics} of cell state transitions \cite{Tritschlerdev170506,Weinreb2017}. The ability to quantify the dynamics of cell state transitions promises greater insights into cell heterogeneity and how differentiation varies with culture conditions, and could help steer targeted differentiation of stem cells in vitro. Currently, the data remain limited to snapshots in most cases, with only one or few time-point measurements available, and a loss of cell identity across measurements due to their destructive nature. Fluorescent reporters of gene expression for live-imaging circumvent this, but their availability is relatively limited and they allow  observation of at most a few genes.

Correct tissue development and regeneration requires balanced cell proliferation and differentiation at the tissue level to ensure the right cell type in the right place at the right amount. However, which dynamics at the cellular level give rise to this population behaviour is in many cases still unclear. In particular, deciding how or whether cell state transitions are coupled to, or occur independently of, cell division is a pre-requisite for quantifying cell state transition rates. 
Determining evidence for such coupling from snapshot data is an incompletely solved problem \cite{greulich2016dynamic}.

If we only have access to snapshots of clonal colony growth, how can one distinguish whether cell division and cell state transitions are coupled? The answer to this question will affect how well one can infer cell state transition rates, as assuming a ``wrong'' underlying model will impair the quality of the inference. Conversely, can we inform experimental design to maximise the information gained from experimental data?
To address this, we combine biophysical theory of stochastic population dynamics with Bayesian parameter inference and model comparison. As an example we choose the well-established system of epiblast stem cells (EpiSCs) \cite{tsakiridis2014distinct}. EpiSCs are self-renewing, pluripotent cell lines that are an \textit{in vitro} equivalent of the embryonic epiblast tissue. In vivo, these cells begin to differentiate (undergo lineage commitment) during gastrulation.

We start with minimal branching process models of cell division and cell state transition (e.g. reversible differentiation), and derive the population dynamics and exact likelihoods for inference based on complete trajectory data (observing every cell division/state transition). We then present a sample path algorithm for inference based on snapshot data. We verify our methods using simulated data and show how the information gained from experimental data can be improved by adaptively spacing the snapshots in time. Next, we extend our models and inference pipeline to EpiSC clonal assay data. In this case, the sample path algorithm becomes too computationally intensive, due to the larger cell state space. Using an approximate Bayesian computation sequential Monte-Carlo scheme, we find that inference favours coupling of cell division and transitions. We compare inferred transition rates in two experimental conditions, promoting self-renewal and differentiation, and find a bias towards increased expression of two transcription factors (T(Bra) and FoxA2) and decreased expression of another (Sox2) under conditions promoting differentiation, in line with published results of a previous experimental study that cultured EpiSCs in vitro \cite{tsakiridis2014distinct}.

\section{Methods}
\input{methods.tex}

\input{results.tex}

\section{Discussion}
\input{discussion.tex}

\section*{References}

\bibliographystyle{unsrt}
\bibliography{mainbib}

\clearpage

\end{document}


\title{Supplementary Material}

\vspace{10pt}

\appendix
\section{Mapping parameters from Model U to Model C}

In the following we show that for any given parameter set $(\lambda_{A}, \lambda_{B}, k_{AB}, k_{BA})$ in Model U, there exists a set of parameters $(\lambda_{AA}, \lambda_{AB}, \lambda_{BB})$ in Model C for which the asymptotic behaviour ($t\rightarrow \infty$) of the means $\langle N_{A} \rangle$, $\langle N_{B} \rangle$ is identical in both models. Following equation~(\ref{mABY}), the asymptotic behaviour (($\beta t \gg 1$)) of mean populations in Model U is given by
\begin{eqnarray}
\langle N_{A} \rangle (t) &= \frac{1}{2} \left[\left(1+\frac{\gamma}{2 \beta}\right)N_{A0} + \frac{k_{BA}}{\beta} N_{B0} \right]  e^{(\alpha+\beta) t}, \\
\langle N_{B} \rangle (t) &= \frac{1}{2} \left[\left(1-\frac{\gamma}{2 \beta}\right)N_{B0} + \frac{k_{AB}}{\beta} N_{A0} \right] e^{(\alpha+\beta) t}, \label{appx1}
\end{eqnarray}
where $\alpha$, $\beta$, $\gamma$ are constants depending on Model U parameters and given by equation~(\ref{mABY}). 
By comparing equations~(\ref{appx1}) to the mean populations in Model C, given by
\begin{eqnarray}
	\langle N_{A} \rangle (t) &= N_{A}(0) e^{(\lambda_{AA}-\lambda_{BB})t}, \\
	\langle N_{B} \rangle (t) &= N_{B}(0) + N_{A}(0) \frac{2 \lambda_{BB} + \lambda_{AB}}{\lambda_{AA}-\lambda_{BB}} \left( e^{(\lambda_{AA}-\lambda_{BB})t} - 1 \right),
	\label{mABX_appendixA1}
\end{eqnarray}
it can be shown that mean population in Model C can be mapped to the ones of Model U by setting
\begin{eqnarray}
	\lambda_{AA} &= \alpha + \beta + \frac{f(\bi{\lambda})-m}{2}, \\
	\lambda_{BB} &= \frac{f(\bi{\lambda})-m}{2}, \\
	\lambda_{AB} &= m,
	\label{mABX_appendixA2}
\end{eqnarray}
where $m$ is a free parameter which can be chosen within the interval $0<m<f(\bi{\lambda})$ and $f(\bi{\lambda})$ is a function depending on Model U parameters:
\begin{eqnarray}
	f(\bi{\lambda}) &= \frac{\alpha+\beta}{2} \left[ \frac{k_{AB}}{2\beta} + \frac{\beta-\gamma/2}{ k_{BA}} \left(1-\frac{\gamma}{2 \beta}\right) \right].
\end{eqnarray}
\clearpage

\section{Properties of the  probability generating function}
Given a two-variable probability distribution $p_{i,j}(t)$ we define the corresponding \emph{probability generating function} as

\begin{equation}
G(z_{A},z_{B},t) := \sum_{i,j=0}^{\infty} p_{i,j}(t) (z_{A})^{i} (z_{B})^{j}.
\label{Genf}
\end{equation}

This allows any desired moment of $N_A(t)$ and $N_B(t)$ to be obtained by differentiation:

\begin{equation}
\fl \langle (N_{A})^{k} (N_{B})^{l} \rangle (t) = \sum_{i,j=0}^{\infty} i^{k} j^{l} p_{i,j}(t) %
	= \left(z_{A} \frac{\partial}{\partial z_{A}} \right)^{k}  \left(z_{B} \frac{\partial}{\partial z_{B}} \right)^{l} G(z_{A},z_{B},t) \Bigr|_{\boldsymbol{z}=1}.
	\label{Genprop2}
\end{equation}

To obtain the governing equation for the probability generating function, we multiply both sides of the master equation by $z_A^{N_A} z_B^{N_B}$ and sum over all $N_A\ge0$ and $N_B\ge0$. The initial condition $p_{i,j}(0) = \delta_{i,N_{A0}} \delta_{j,N_{B0}}$ translates to an initial condition on $G(z_A,z_B,t)$ of $G(z_A,z_B,0)=z_A^{N_{A0}} z_B^{N_{B0}}$. 

Similarly, we can calculate the PMF from the generating function by
\begin{eqnarray}
	p_{k,l}(t) &= Pr\left[N_{A}(t)=k,N_{B}(t)=l\right] \nonumber\\
	&= \sum_{i,j=0}^{\infty} p_{i,j}(t) \delta_{i,k} \delta_{j,l} = \frac{1}{k! \cdot l!}  G^{(k,l)}(0,0,t) ,
	\label{Genprop1}
\end{eqnarray}
where $G^{(k,l)}:= (\partial_{z_{A}})^{k} (\partial_{z_{B}})^{l} G(z_{A},z_{B},t)$ denotes the partial derivative of the PGF~\cite{johnson2005univariate}.
\rev{Later when deriving the approximate expressions of the mean populations $\langle \mathbf{\Phi} \rangle (t)$ for the extended Model C, we used the following expression for the transition matrix $\mathbf{M}$:  
\newcommand\scalemath[2]{\scalebox{#1}{\mbox{\ensuremath{\displaystyle #2}}}}
\begin{equation}
\hspace*{-2cm}
\scalemath{0.6}{
\begin{pmatrix}
    \theta_{0}-\theta_{T+}-\theta_{F+}-\theta_{S+} & 2 \theta_{S-} & 2 \theta_{F-} & 0 & 2 \theta_{T-} & 0 & 0 & 0\\
    2 \theta_{F+} & \theta_{0}-\theta_{T+}-\theta_{F-}-\theta_{S+} & 0 & 2 \theta_{S-} & 0 & 2 \theta_{T-} & 0 & 0\\
    2 \theta_{S+} & 0 & \theta_{0}-\theta_{T+}-\theta_{F+}-\theta_{S-} & 2 \theta_{F-} & 0 & 0 & 2 \theta_{T-} & 0\\
    0 & 2 \theta_{S+} & 2 \theta_{F+} & \theta_{0}-\theta_{T+}-\theta_{F-}-\theta_{S-} & 0 & 0 & 0 & 2 \theta_{T-}\\
    2 \theta_{T+} & 0 & 0 & 0 & \theta_{0}-\theta_{T-}-\theta_{F+}-\theta_{S+} & 2 \theta_{F-} & 2 \theta_{S-} & 0\\
    0 & 2 \theta_{T+} & 0 & 0 & 2 \theta_{F+} & \theta_{0}-\theta_{T-}-\theta_{F-}-\theta_{S+}  & 0 & 2 \theta_{S-}\\
    0 & 0 & 2 \theta_{T+} & 0 & 2 \theta_{S+} & 0 & \theta_{0}-\theta_{T-}-\theta_{F+}-\theta_{S-} & 2 \theta_{F-}\\
    0 & 0 & 0 & 2 \theta_{T+} & 0 & 2 \theta_{S+} & 2 \theta_{F+} & \theta_{0}-\theta_{T-}-\theta_{F-}-\theta_{S-}\\
\end{pmatrix}
}
\label{M_C}
\end{equation}
Similarly, the transition matrix $\mathbf{M}$ for the extended Model U is defined as:  
\begin{equation}
\hspace*{-2cm}
\scalemath{0.6}{
\begin{pmatrix}
    \theta_{0}-\theta_{T+}-\theta_{F+}-\theta_{S+} & \theta_{S-} & \theta_{F-} & 0 &  \theta_{T-} & 0 & 0 & 0\\
     \theta_{F+} & \theta_{0}-\theta_{T+}-\theta_{F-}-\theta_{S+} & 0 &  \theta_{S-} & 0 &  \theta_{T-} & 0 & 0\\
     \theta_{S+} & 0 & \theta_{0}-\theta_{T+}-\theta_{F+}-\theta_{S-} &  \theta_{F-} & 0 & 0 &  \theta_{T-} & 0\\
    0 &  \theta_{S+} &  \theta_{F+} & \theta_{0}-\theta_{T+}-\theta_{F-}-\theta_{S-} & 0 & 0 & 0 &  \theta_{T-}\\
     \theta_{T+} & 0 & 0 & 0 & \theta_{0}-\theta_{T-}-\theta_{F+}-\theta_{S+} &  \theta_{F-} &  \theta_{S-} & 0\\
    0 &  \theta_{T+} & 0 & 0 &  \theta_{F+} & \theta_{0}-\theta_{T-}-\theta_{F-}-\theta_{S+}  & 0 &  \theta_{S-}\\
    0 & 0 &  \theta_{T+} & 0 &  \theta_{S+} & 0 & \theta_{0}-\theta_{T-}-\theta_{F+}-\theta_{S-} &  \theta_{F-}\\
    0 & 0 & 0 &  \theta_{T+} & 0 &  \theta_{S+} &  \theta_{F+} & \theta_{0}-\theta_{T-}-\theta_{F-}-\theta_{S-}\\
\end{pmatrix}
}
\label{M_U}
\end{equation}
}
\clearpage

\section{Scaling of the COV} 

\subsection{Combined data of trajectory ensemble}

In the following, we will consider data from several trajectories (ensemble) and determine how the COV scales with an ensemble size $n$. This might be useful for experimental design: for example consider a system from which data was measured and used for parameter inference. The scaling behavior of the COV allows one to estimate how many additional measurements are required to double the accuracy of the parameter inference. 
The posterior distribution $p(\theta_{\nu}|\xi_{1},\ldots,\xi_{n})$ given complete-data of several trajectories ${\xi_{1},\ldots,\xi_{n}}$ is:
\begin{eqnarray}
	&p(\theta_{\nu}|\xi_{1},\ldots,\xi_{n}) \propto L(\theta_{\nu}|\xi_{1}) \cdot \ldots \cdot L(\theta_{\nu}|\xi_{n})\cdot q(\theta_{k}) \nonumber \\
	\Rightarrow &\theta_{\nu}|\xi_{1},\ldots,\xi_{n} \sim Ga\left(a_{\nu}+\sum_{i=1}^{n} r_{\nu,i}, b_{\nu} + \sum_{i=1}^{n} \int_{0}^{T} N_{\nu,i}(t) dt\right),
	\label{comp_lhood_ens}
\end{eqnarray}
where $r_{\nu,i}$ and $N_{\nu,i}$ are reaction numbers and populations in trajectory $i$. Assuming that all trajectories ${\xi_{1},\ldots,\xi_{n}}$ are sampled from the same system, $r_{\nu,i}$ should be on average the same for all trajectories: $\langle r_{\nu,i} \rangle = \langle r_{\nu} \rangle$. The COV is therefore:
\begin{eqnarray}
	\langle c_{v}\rangle(\theta_{\nu}|\xi_{1},\ldots,\xi_{n}) &=& \left\langle \frac{1}{\sqrt{a_{\nu}+\sum_{i=1}^{n} r_{\nu,i}}} \right\rangle \nonumber\\
	&\approx& \left[a_{\nu}+\sum_{i=1}^{n} \langle r_{\nu,i}\rangle\right]^{-1/2} \approx \left[n\cdot \langle r_{\nu}\rangle\right]^{-1/2}.
	\label{cv_N}
\end{eqnarray}
Since we do not know the full distribution of $r_{\nu}$ as a function of reaction constants $\boldsymbol{{\theta}}$ and $T$, we approximate the mean of the function in line 1 by the function of the mean in line 2 (mean-field approximation). This approximation works well as long as the function approximately scales linear with $\sum_{i} r_{\nu,i}$ in the range in which $p(\sum_{i} r_{\nu,i})\gg0$. If we assume $\sum_{i} r_{\nu,i}$ is sufficiently large and the spread of $p(\sum_{i} r_{\nu,i})\gg0$ is not too large, this mean-field approximation should work sufficiently well. We also used $n \langle r_{\nu} \rangle \gg a_{\nu}$ in line 3, since we are interested in the asymptotic behaviour $n\rightarrow\infty$. The scaling of $\langle c_{v}\rangle$ with ensemble size $n$ is shown in Figure~\ref{Cv_X}b for rate constant $\theta_{1}$ of Model C. The proposed $n^{-1/2}$ scaling coincides with the ensemble mean of $c_{v}$ surprisingly well, despite the mean-field approximation.

\subsection{Varying observation period T} \label{inhom_P_app}

We are now interested how the COV $c_{v}$ scales with observation period $T$. For a single trajectory, $\langle c_{v} \rangle \approx \langle r_{k}\rangle^{-1/2}$ (see \ref{cv_N}), where we assumed $\langle r_{\nu}\rangle \gg a_{\nu}$ since we are interested in the asymptotic behaviour $T\rightarrow\infty$. We therefore need to derive an expression for the mean number of observed reactions $\langle r_{\nu}\rangle$. We have defined reaction events as a Poisson process with transition rate $h_{\nu}$. The transition rates $h_{\nu}\propto N_{\nu}$ depend on the current population $N_{\nu}$, which change with every occurred reaction. To make some progress, we may approximate this process by replacing the current population $N_{\nu}$ with its mean value $\langle N_{\nu}\rangle (t)$. The resulting process is a Poisson process with time dependent transition rate $h^{*}_{\nu} = \theta_{\nu} \cdot \langle N_{\nu}\rangle (t)$, also called \textit{Inhomogenous Poisson Process}. In this case, the total number of reactions $r_{k}$ is Poisson distributed~\cite{ross1996stochastic}: 
\begin{equation}
	p(r_{\nu}) \approx \frac{(\Lambda)^{r_{\nu}}}{{r_{\nu}!}} e^{-\Lambda} \quad \text{with}   \Lambda := \int_{0}^{T} h^{*}_{\nu}(t) dt = \theta_{\nu} \int_{0}^{T} \langle N_{\nu}\rangle(t) dt,
	\label{nreact_approx}
\end{equation} 
with mean $\langle r_{\nu} \rangle=\Lambda$. The mean COV for an ensemble of $n$ trajectories is therefore:
\begin{eqnarray}
	\langle c_{v} \rangle(\theta_{\nu}|\xi_{1},\ldots,\xi_{n}) \approx \frac{1}{\sqrt{n \theta_{\nu} \int_{0}^{T} \langle N_{\nu} \rangle (t) dt}} ,
	\label{cv_T}
\end{eqnarray}
where the mean COV of $\theta_{\nu}$ depends explicitly on the rate constant itself (additionally, $\langle N_{\nu} \rangle$ may also depend on $\theta_{\nu}$). 
This result can be useful for experimental design again: a rough estimate for the magnitude of $\theta_{k}$ might be obtained from either prior knowledge or from a posteriors of quick trail runs. 
Equation~(\ref{cv_T}) then allows us to estimate how long a full experiment should run in order to get parameter posteriors of desired precision. 

\subsection{Parameter estimation in Models C and U}

\begin{figure}
	\centering
	\subcaptionbox{}{\includegraphics[width=0.45\textwidth]{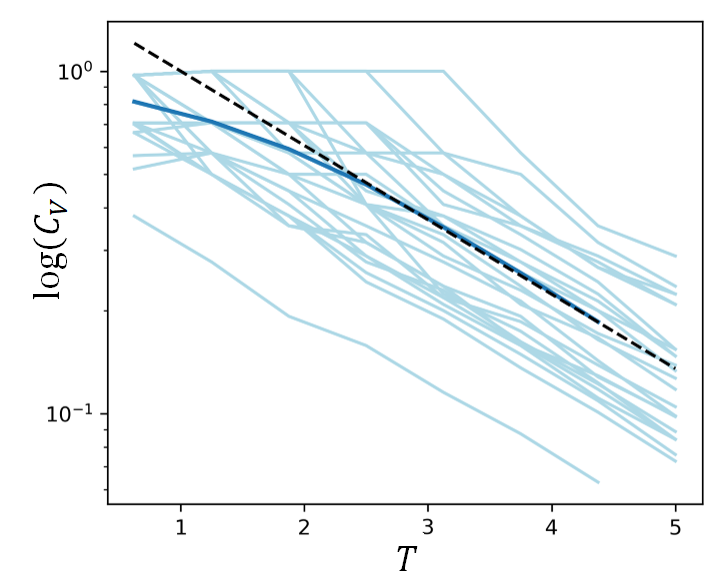}}
	\subcaptionbox{}{\includegraphics[width=0.45\textwidth]{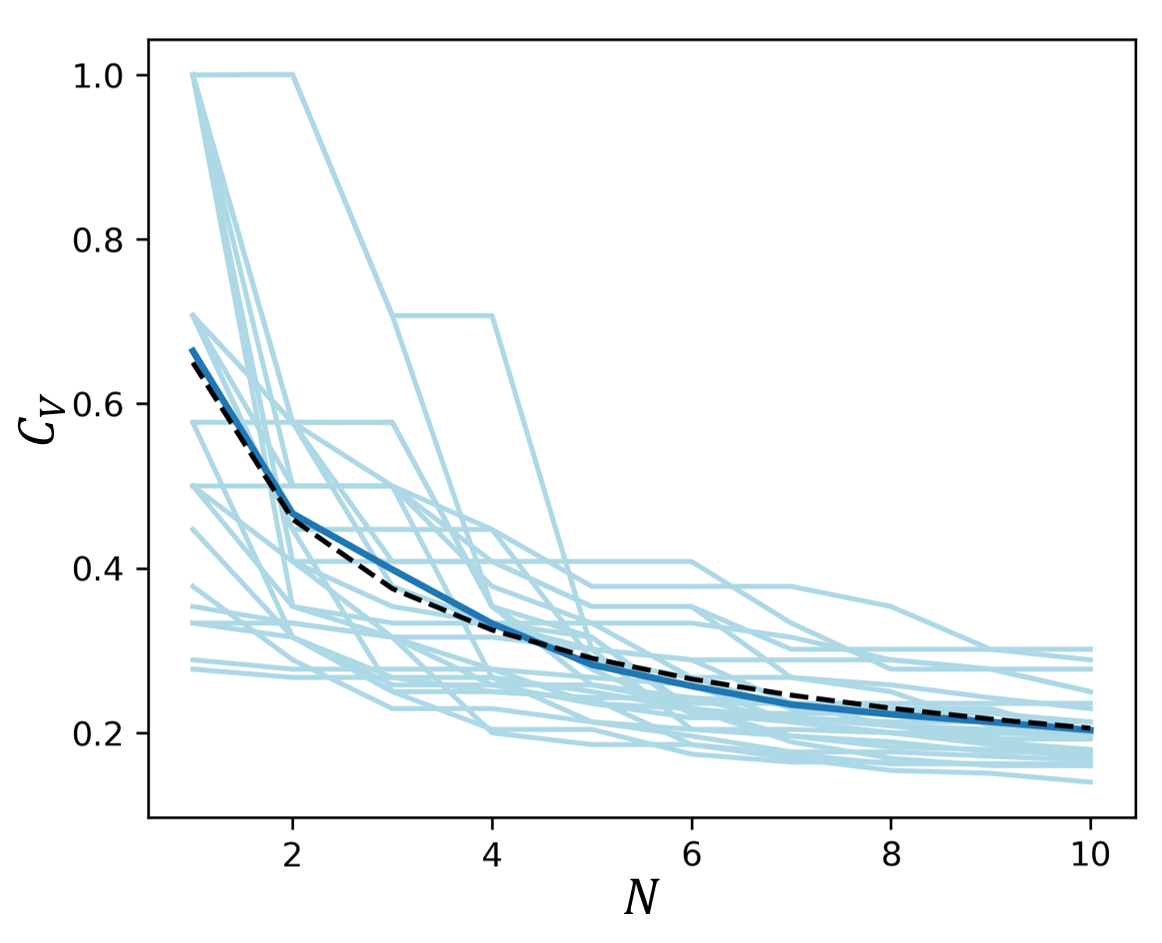}}
	\caption{Dependence of coefficient of variation of rate constant $\theta_{1}$ on observation time $T$ and ensemble size $N$. Trajectories were generated from Model C using parameters $(\theta_{1},\theta_{2},\theta_{3})=(1,1,0)$. (a) Dependence of $c_{v}$ on observation time $T$. Solid light-blue and dark blue lines show a set of trajectories and their mean, $\langle c_{v} \rangle$, respectively. The dashed black line shows the asymptotic expression given in equation~(\ref{X_cv_T_appendix}). 
	(b) Dependence of $c_{v}$ on ensemble size $N$. Solid light-blue and dark blue lines show a set of trajectories and their mean, $\langle c_{v} \rangle$, respectively. The dashed black line shows an $N^{-1/2}$ scaling.} 
	\label{Cv_X}
\end{figure}

As derived in equation~(\ref{mABX}), the mean number of cells in state A in Model C,
\begin{equation}
\langle N_{A} \rangle (t) = N_{A0} e^{t \Delta \lambda},
\label{mABX_2}
\end{equation}
either grows or decays over time, depending on the sign of $\Delta \lambda$. Combining equation~(\ref{mABX_2}) and (\ref{cv_T}), the asymptotic behaviour for $T \Delta\lambda \gg1$ is given by:
\begin{eqnarray}
		\langle c_{v} \rangle(\theta_{k}|\xi_{1},\ldots,\xi_{n}) \propto
		\left\{
		\begin{array}{ll}
		\frac{\exp\left(- T \Delta \lambda /2\right)}{\sqrt{N_{A0} n}} \quad & \Delta \lambda >0, \\[10pt]
		\frac{1}{\sqrt{N_{A0} n}} \quad & \Delta \lambda <0.
		\end{array} 
		\right.
		\label{X_cv_T_appendix}
\end{eqnarray}
The posterior precision therefore does not increase with observation time $T$ if $\Delta \lambda <0$. This makes sense, since A-type cells will always die out for $\lambda_{BB}>\lambda_{AA}$ (see equation~(\ref{pext1})). 
The scaling of $\langle c_{v}\rangle$ with observation time $T$ is shown in Figure~\ref{Cv_X}a for rate constant $\theta_{1}$ of Model C. 
The proposed exponential scaling $e^{-\Delta \lambda \cdot T/2}$ coincides well with the ensemble mean of $c_{v}$ for large observation times $T$. 
For small times $T$, both the mean-field approximation~(\ref{cv_N}) and $T \Delta\lambda \gg1$ break down.

We derived the following asymptotic behaviour of the means $\langle N_{A}\rangle$ for Model U:
\begin{equation}
\langle N_{A} \rangle (t), \langle N_{B} \rangle (t)  \propto e^{(\alpha+\beta) t} ,
\label{mABY_3}
\end{equation}
where $\alpha$ and $\beta$ are functions of rate constants $\boldsymbol{{\theta}}$ (see \ref{mABY2}). Since $\alpha+\beta\geq0$, the mean populations always grow exponentially. Combining equations~(\ref{mABY_3}) and (\ref{cv_T}), the asymptotic behaviour for $(\alpha+\beta) T\gg1$ is given by
\begin{equation}
\langle c_{v} \rangle(\theta_{k}|\xi_{1},\ldots,\xi_{n}) \propto
\frac{\exp\left(-(\alpha+\beta)T/2 \right)}{n}. 
\label{Y_cv_T_appendix}
\end{equation}

\clearpage

\section{Sampling of trajectories with fixed endpoints}

For sampling trajectories with fixed end points, we approximate the true stochastic process of Models C and U by an inhomogeneous Poisson process with time-dependent transition rates $h_{\nu}^{*}(t)$.

\subsection{Proposing valid reaction numbers}

This section closely follows the derivation presented in~\cite{wilkinson2011stochastic}. Let $\bi{r}$ be the vector whose entries $r_{\nu}$ are the number of $\nu$-type reactions occurring between two time points $t_{0}$ and $t_{1}$ with population snapshots $\bi{N}_{0}$ and $\bi{N}_{1}$, receptively. The stoichiometric matrix $\mathcal{S}_{l,k}$ denotes the change of population $l$ caused by reaction $k$. The reaction vector $\bi{r}$ therefore needs to fulfil the following equation:
\begin{equation}
\bi{N}_{1} = \bi{N}_{0} + \mathcal{S} \bi{r} \quad \Rightarrow \quad \mathcal{S} \bi{r} = \bi{N}_{1} - \bi{N}_{0}.
\end{equation}
We assume that the stoichiometric matrix $\mathcal{S}$ is of full rank, since otherwise we can simplify the model by reducing the number of model dimensions. Then, we can partition the matrix $\mathcal{S}=(\tilde{\mathcal{S}},\mathcal{S}^{'})$ where $\tilde{\mathcal{S}}$ is an invertible matrix. In the same way, we partition the reaction vector $\bi{r} = (\bi{\tilde{r}},\bi{r^{'}})$, which yields:
\begin{eqnarray}
\mathcal{S} \bi{r} &=& (\tilde{\mathcal{S}},\mathcal{S}^{'})  
\left( {\begin{array}{*{20}c}
\bi{\tilde{r}} \\ \bi{r^{'}}  
 \end{array} } \right) = \tilde{\mathcal{S}}  \bi{\tilde{r}} + \mathcal{S}^{'}  \bi{r^{'}}  =\bi{N}_{1} - \bi{N}_{0} \nonumber\\
\Rightarrow \qquad \bi{\tilde{r}} &=& \tilde{\mathcal{S}}^{-1}  \left[ \bi{N}_{1} - \bi{N}_{0} - \mathcal{S}^{'}  \bi{r^{'}} \right].
\label{reaction_vector}
\end{eqnarray}
A good strategy for proposing valid reaction numbers $\bi{r}$ is therefore the following: Starting from a valid solution $\bi{r}$, we first randomly perturb reaction numbers corresponding to $\bi{r}^{'}$ in an arbitrary way, and then calculate the residual elements $\bi{\tilde{r}}$ accordingly to (\ref{reaction_vector}). Since solutions of equation~(\ref{reaction_vector}) are not necessarily always positive, we have to make sure that all elements of $\bi{r}$ are non-negative, otherwise we reject the proposal immediately.

\subsection{Sampling reaction times} \label{Appendix_reactiontimes}

Given a reaction vector $\bi{r}$ with $r_{k}$ being the numbers of type-$k$ reactions, we sample the corresponding $r_{k}$ reaction times $0\leq \tau_{j} \leq \Delta t$ from an an inhomogeneous Poisson process with transition rate $h_{k}^{*}(t)$~\cite{Pasupathy_generatingnonhomogeneous}.

Since populations $\bi{N}(0)$, $\bi{N}(\Delta t)$ are known at the end points of the trajectory, so are the transition rates $h_{k}^{*}= \theta_{k}\cdot N_{k}$. The question is now: how should the transition rates $h_{k}^{*}(t)$ vary between these values to accurately approximate the true process? Because we know that mean populations $\langle N_{k} \rangle (t)$ are either constant, exponentially growing or decaying, a natural choice for $h_{k}^{*}(t)$, $0\leq t \leq \Delta t$ would be an exponential function:
\begin{equation}
	h_{k}^{*}(t) = \theta_{k}\cdot N_{k}(0)  \left(\frac{N_{k}(T)}{N_{k}(0)}\right)^{t/\Delta t}  .
	\label{exp_hazard_appendix}
\end{equation} 

\subsection{Setting up the acceptance ratio}

The Poisson approximation step is taken into account by selecting an appropriate Metropolis-Hastings acceptance probability, ensuring that the MCMC output is actually drawn from the correct posterior 
(see Metropolis-Hastings-Green Algorithm in~\cite{brooks2011handbook}). To construct the acceptance probability, we require the complete-data likelihood of trajectories $\xi$ in the inhomogeneous Poisson model $L_{A}(\xi|\boldsymbol{\theta})$. 
For arbitrary transition rates $h_{k}^{*}(t)$, the complete-data likelihood of $n$ reaction events in range $[0,\Delta t]$ is given by
\begin{equation}
L_{P}(\xi|\boldsymbol{\theta}) = \left[\prod_{i=1}^{n} h_{\nu_{i}}^{*}(t_{i}) \right]  e^{-\Lambda_{0}(\Delta t)} \quad \text{with} \quad \Lambda_{0}(\Delta t) := \int_{0}^{\Delta t} h^{*}_{0}(t)  dt  ,
\end{equation}
where $h^{*}_{0}=\sum_{\nu} h^{*}_{\nu}$ is the combined transition rate and $\nu_{i}$ is the type of the $i$-th reaction. Plugging in the exponential transition rate (\ref{exp_hazard_appendix}) yields:
\begin{eqnarray}
&L_{P}(\xi|\boldsymbol{\theta}) = \left[\prod_{i=1}^{n} \theta_{\nu_{i}} N_{\nu_{i}}(0)  \left(\frac{N_{\nu_{i}}(T)}{N_{\nu_{i}}(0)}\right)^{t_{i}/\Delta t} \right]  e^{-\Lambda_{0}(\Delta t)} \nonumber \\
&\text{with} \quad  \Lambda_{0}(\Delta t) = \Delta t \sum_{k} \theta_{k}  \frac{\Delta N_{k}}{\Delta \log(N_{k})}  ,
\end{eqnarray}
where $\Delta N_{k} = N_{k}(\Delta t)-N_{k}(0)$ and $\Delta \log(N_{k})=\log(N_{k}(\Delta t))-\log(N_{k}(0))$. We are now ready to construct the Metropolis-Hastings step for updating trajectories $\xi$ with fixed endpoints on interval $[0,\Delta t]$. First we propose a new set of valid reaction numbers $\bi{r^{\star}}$ from a proposal distribution $f(\bi{r^{\star}}|\bi{r})$. We chose to update $\bi{r^{\star}}=\bi{r}+\zeta$ by adding to each element a random integer $\zeta$ drawn from a symmetric Skellam distribution~\cite{johnson1969discrete} with variance $2  \omega$. The tuning parameter $\omega$ was chosen such that the proposals $\bi{r^{\star}}$ are accepted on average with a probability of $\approx 30 \%$. Then, conditional on $\bi{r^{\star}}$, we will sample a trajectory (reaction times $\tau^{\star}_{j}$) from the approximate Poisson process with exponential transition rates (\ref{exp_hazard_appendix}). The trajectory $\xi^{\star}:=\{\bi{r^{\star}},\tau^{\star}_{i}\}$ is then accepted with probability $\text{min}(1,A)$ with
\begin{eqnarray}
	A = \left[\frac{L_{\rm CD}(\xi^{\star}|\boldsymbol{{\theta}})}{L_{P}(\xi^{\star}|\boldsymbol{\theta})}  \frac{q(\bi{r^{\star}})}{f(\bi{r^{\star}}|\bi{r})}\right] \left/ \left[\frac{L_{\rm CD}(\xi|\boldsymbol{{\theta}})}{L_{P}(\xi|\boldsymbol{\theta})}  \frac{q(\bi{r})}{f(\bi{r}|\bi{r^{\star}})}\right]\right.  ,
	\label{acc-prob}
\end{eqnarray}
where $L_{\rm CD}$ is the complete-data likelihood of the true model (see (\ref{LCD})) and $q(\bi{r})=\prod_{\nu} q_{\nu}(r_{\nu})$ is the probability that $\bi{r}$ reactions were observed in the inhomogeneous Poisson model. 
The terms $q_{\nu}$ are the PMF of the Poisson distribution:
\rev{
\begin{equation}
	q_{\nu}(r_{\nu}) = \frac{\Lambda_{\nu}(\Delta t)^{r_{\nu}}}{r_{\nu}!}  e^{-\Lambda_{\nu}(\Delta t)} \quad \text{with} \quad \Lambda_{\nu}(\Delta t) = \Delta t \cdot \theta_{\nu}  \frac{\Delta N_{\nu}}{\Delta \log(N_{\nu})}  .
\end{equation}
}
Note that the the acceptance probability (\ref{acc-prob}) depends on the current parameter vector $\boldsymbol{\theta}$, which is drawn at every iteration of the MCMC. 
The overall structure of this algorithm was originally motivated by~\cite{wilkinson2011stochastic} and modified by us to fitS Model c AND u. 
It is an application of the general Metropolis-Hastings-Green Algorithm as described in~\cite{brooks2011handbook}. 

\clearpage

\section{Bayes factor calculation using population snapshots}
\subsection{The harmonic mean identity} \label{sect_HM}

For a given model $M$, some data $y$ and a parameter $\theta$, we have from Bayes' theorem that
\begin{equation}
\frac{p(\boldsymbol{\theta}|M)}{p(\boldsymbol{\xi}|M)} = \frac{p(\boldsymbol{\theta}|\boldsymbol{\xi},M)}{p(\boldsymbol{\xi}|\boldsymbol{\theta},M)} \;.
\end{equation}
Integrating both sides over $\boldsymbol{\theta}$ gives us the identity
\begin{equation}
	\frac{1}{p(\boldsymbol{\xi}|M)} =  \mathbb{E}\left[\left.\frac{1}{p(\boldsymbol{\xi}|\boldsymbol{\theta},M)} \right|\boldsymbol{\xi},M\right].
\end{equation}
This is useful because the average is taken over the posterior distribution of the parameters $\boldsymbol{\theta}$, conditioned on a particular set of observations $\boldsymbol{\xi}$. This is exactly what the MCMC algorithm generates, and hence we can estimate the marginal likelihood $p\left(\{\boldsymbol{N}(t_{i})\}|M\right)$ as
\begin{equation}
	\hat{p}(\{\boldsymbol{N}(t_{i})\}|M) = \left[\frac{1}{B} \sum_{m=1}^{B} \frac{1}{L_{SS}(\{\boldsymbol{N}(t_{i})\}|\boldsymbol{\theta}^{m})}\right]^{-1} \;,
	\label{HM_est}
\end{equation}
where $\boldsymbol{\theta}^{m}$ is the $m^{\rm th}$ sample of the parameter vector obtained from the MCMC algorithm. Unfortunately, the snapshot likelihood $L_{SS}(\{\boldsymbol{N}(t_{i})\}|\boldsymbol{\theta}^{m})$ needs to be evaluated by a Monte Carlo (MC) estimate for each parameter vector $\boldsymbol{\theta}^{m}$ separately by:
\begin{equation}
	\hat{L}_{SS}(\{\boldsymbol{N}(t_{i})\}|\boldsymbol{\theta}^{m}) = \frac{1}{B^{'}} \sum_{k=1}^{B^{'}} \delta(\{ \boldsymbol{N}(t_{i}) - \boldsymbol{N}^{sim}_{k}(t_{i}) \})  ,
\end{equation}
where $\boldsymbol{N}^{sim}_{k}(t_{i})$ is a simulated trajectory from the respective model with parameters $\boldsymbol{\theta}^{m}$. The delta function $\delta(\cdot)$ equals one if simulated and observed populations coincide at snapshot times. This resulting in a computationally intense MC within MCMC algorithm, which is however still less computationally intensive than the other presented methods (thermodynamic integration, product space search). 

\subsection{Thermodynamic integration} 

\textit{Thermodynamic integration} is another method to estimate marginal likelihoods of models. It is based on the fact that the logarithm of the marginal likelihood can be represented by the following integral~\cite{calderhead2009estimating}:
\begin{equation}
	\log  p(\boldsymbol{\xi}) = \int_{0}^{1}  \mathbb{E}_{\boldsymbol{\theta}|\boldsymbol{\xi},t} \left[\log  p(\boldsymbol{\xi}|\boldsymbol{\theta})\right] dt  ,
	\label{TI_int}
\end{equation}
with $\boldsymbol{\xi}$ being general data, $p(\boldsymbol{\theta})$ the parameter prior, $p(\boldsymbol{\xi}|\boldsymbol{\theta})$ the likelihood and $p(\boldsymbol{\xi})$ the marginal likelihood. 
The distribution used to evaluate the mean $E_{\boldsymbol{\theta}|\boldsymbol{\xi},t}[\cdot]$ is called power-posterior $p_{t}(\boldsymbol{\theta}|\boldsymbol{\xi})$ with temperature $t\in[0,1]$, defined as~\cite{friel2008marginal}:
\begin{equation}
	p_{t}(\boldsymbol{\theta}|\boldsymbol{\xi}) := \frac{p(\boldsymbol{\xi}|\boldsymbol{\theta})^{t} q(\boldsymbol{\theta})}{\int p(\boldsymbol{\xi}|\boldsymbol{\theta})^{t} q(\boldsymbol{\theta})  d\boldsymbol{\theta}}.
\end{equation}
The power-posterior can be interpreted as an intermediate distribution between prior and posterior, since $p_{t=0}(\boldsymbol{\theta}|\boldsymbol{\xi}) = p(\boldsymbol{\theta})$ and $p_{t=1}(\boldsymbol{\theta}|\boldsymbol{\xi}) = p(\boldsymbol{\theta}|\boldsymbol{\xi})$. For derivations and evaluation techniques of the integral (\ref{TI_int}), see~\cite{calderhead2009estimating}. 

\begin{figure}[ht]
	\centering
	\includegraphics[width=0.6 \textwidth]{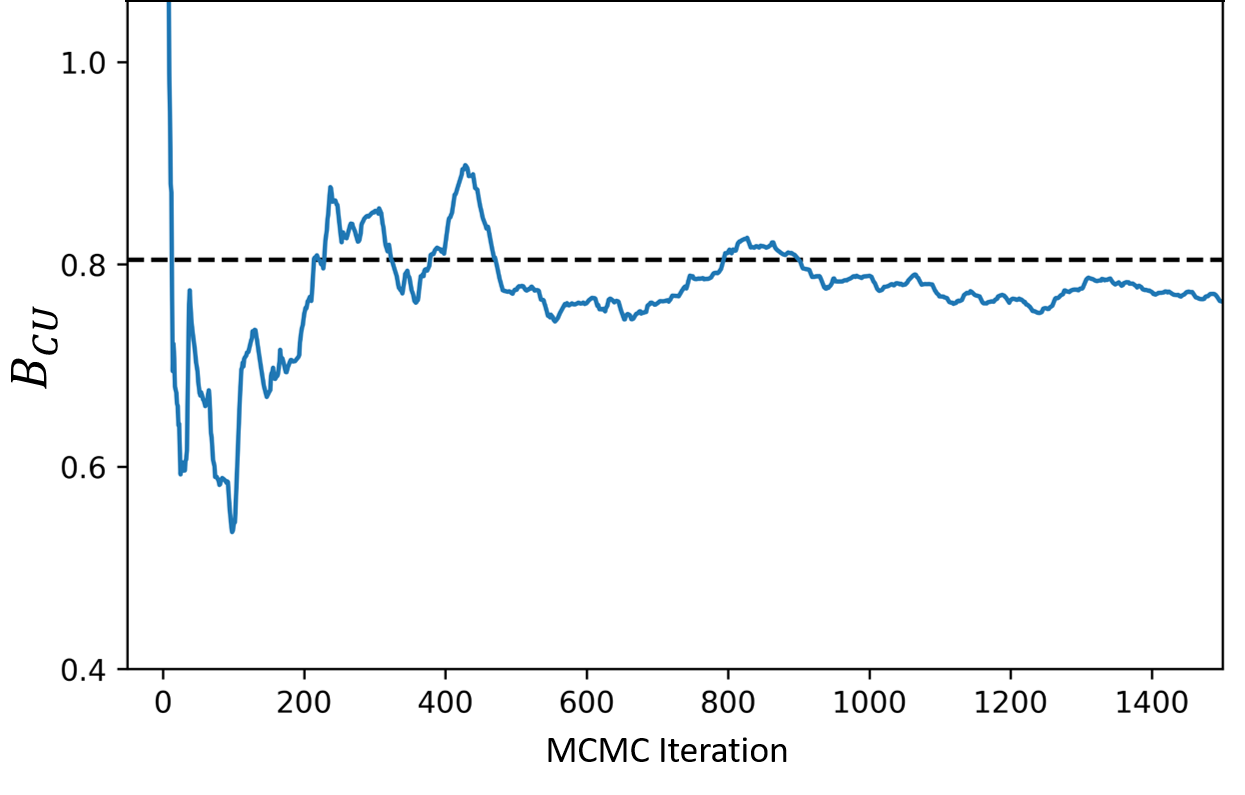}
	\caption[Calculation of Bayes factor $B_{CU}$ using the harmonic mean identity and thermodynamic integration]{Calculation of Bayes factor $B_{CU}$ using the harmonic mean identity (solid blue line) and thermodynamic integration (dashed black line) given two snapshots $\bi{N}(0)=(3,3)$, $\bi{N}(1)=(8,8)$ and flat priors $\theta_{\nu}\mathrel{\stackrel{\makebox[0pt]{\mbox{\normalfont\tiny iid}}}{\sim}}\mathcal{U}[0,3]$ for Models C and U. The harmonic mean estimate~(\ref{HM_est}) converges after only $B\approx 300$ MCMC iterations, although the convergence speed strongly depends on the snapshot data and on the sample size of the MC-step for estimating the snapshot-likelihood $L_{SS}(\{\bi{N}(t_{i})\}|\boldsymbol{\theta})$. We use thermodynamic integration only to estimate convergence speed, since this calculation is much more intensive than the harmonic mean identity.} 
	\label{HM_TI_compare}
\end{figure}

The Bayes factor calculation via thermodynamic integration is usually very accurate. On the other hand, it is computationally even more intensive than the harmonic mean identity and also more complicated to implement. We therefore try to avoid thermodynamic integration if other methods provide equally good estimates. As an example, Figure~\ref{HM_TI_compare} compares the Bayes factors obtained via the harmonic mean estimator (\ref{HM_est}) and via thermodynamic integration given two population snapshots. The harmonic mean estimator converges quite fast and provides equally good Bayes factor estimates than the method of thermodynamic integration.

\subsection{Product space search} 

\begin{figure}[ht]
	\centering
	\subcaptionbox{}{\raisebox{0.12\height}{\includegraphics[width=0.45 \textwidth]{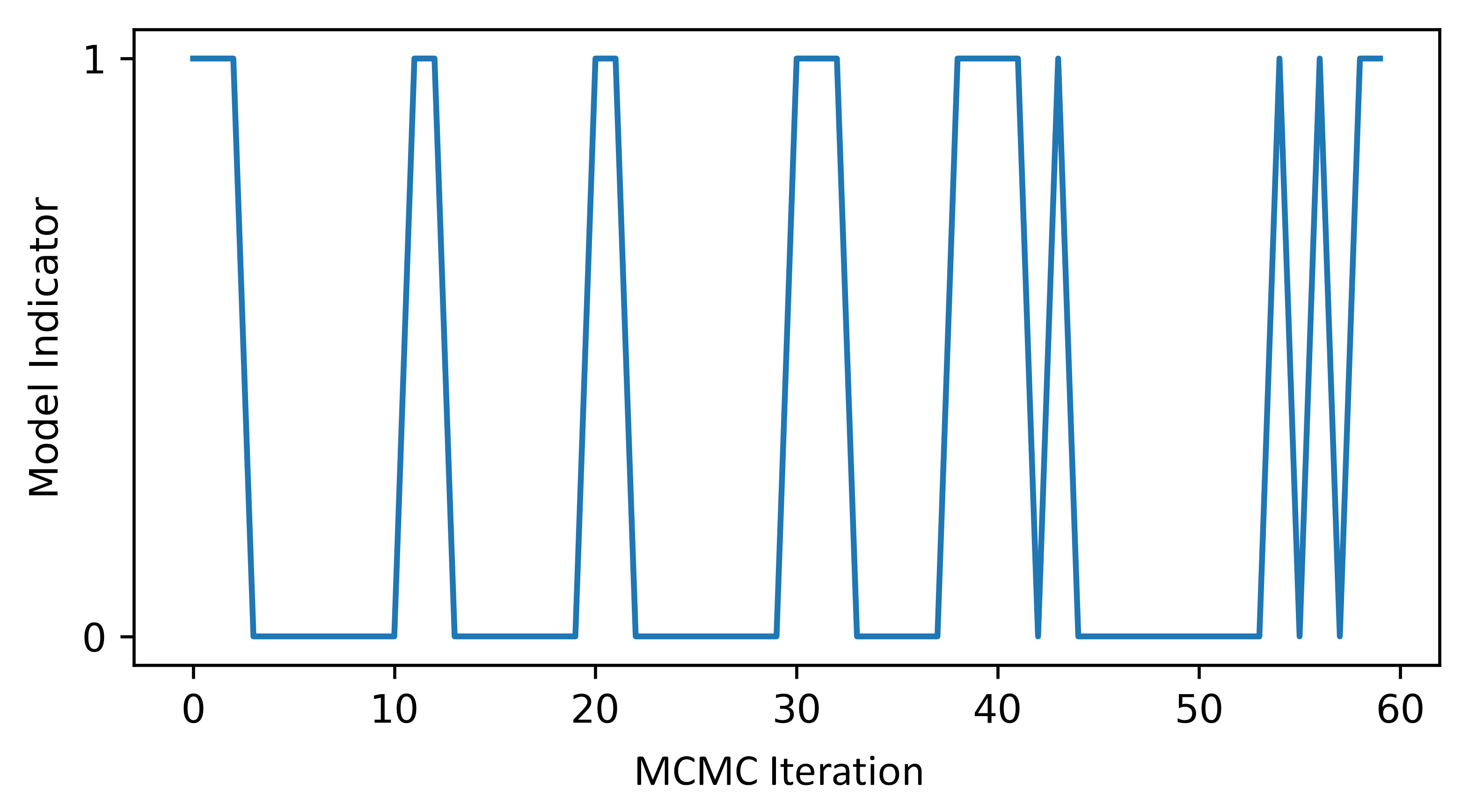}}}
	\quad
	\subcaptionbox{}{\includegraphics[width=0.45 \textwidth]{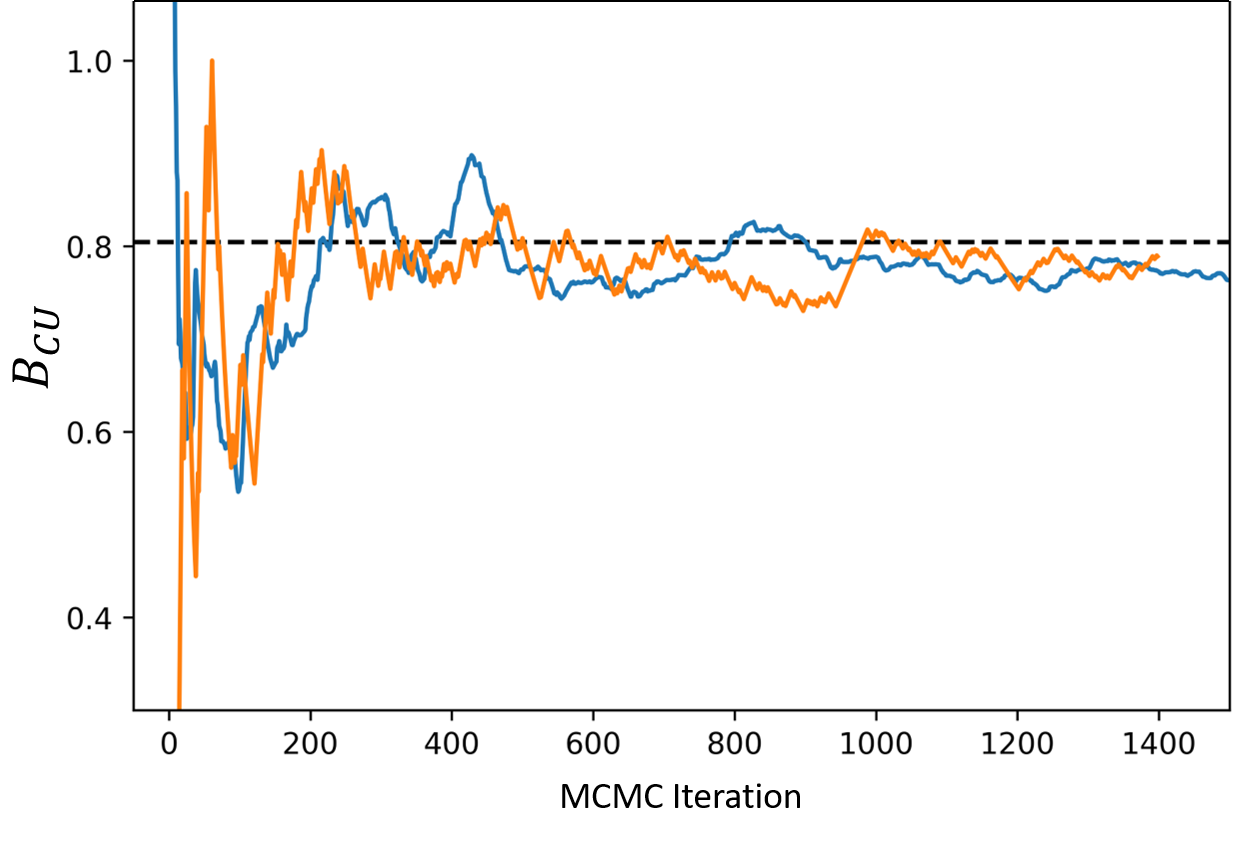}}
	\caption{Calculation of Bayes factor $B_{CU}$ using product space search given two snapshots $\bi{N}(0)=(3,3)$, $\bi{N}(1)=(8,8)$ and flat priors $\theta_{\nu}\mathrel{\stackrel{\makebox[0pt]{\mbox{\normalfont\tiny iid}}}{\sim}}\mathcal{U}[0,3]$ for Models C and U. (a) MCMC jumping between different Model C ($0$) and U ($1$) over a range of MCMC iterations. The Bayes factor can be estimated by the ratio of times the MCMC stayed in either Model C or Model U. (b) Estimation of $B_{XY}$ using product space search (solid orange line), the harmonic mean estimator (solid blue line) and thermodynamic integration (dashed black line).} 
	\label{prod_space}
\end{figure}

Another popular method to obtain Bayes factors is the \textit{product space search}~\cite{chib1995marginal}. 
Whereas the harmonic mean identity and thermodynamic integration are used to estimate the marginal likelihoods of Models C and U, the product space search estimates $B_{CU}$ directly. 
This method involves an MCMC algorithm not only exploring the parameter space of a single model, but jumping between competing models. In general, jumps between a whole set of competing models $\{M_{j}\}$ are allowed, but we will allow the algorithm to switch only between Models C and U. At the start of each iteration, we either stay in the current model or jump to the other model with a certain probability. In the end, the Bayes factor $B_{CU} \approx N(M=C)/N(M=U)$ can be estimated by the ratio of times the MCMC stayed in either Model C or Model U (see Figure~\ref{prod_space} (a)). 

We implement the product space search following~\cite[``Metropolised Carlin and Chib'']{dellaportas2002bayesian}. The performance of this method is compared to the harmonic mean estimator in Figure~\ref{prod_space}(b). Both methods have similar convergence speeds and variances. This is surprising, since the harmonic mean estimator has been criticised for being the ``Worst Monte Carlo Method Ever''~\cite{raftery2006estimating}. The reason being that although it is a consistent estimator, it may have infinite variance in some cases. This could not be observed in our system, where it has similar variance than more sophisticated methods such as the product space search.

\begin{figure}
	\centering
	\subcaptionbox{$\lambda_{AA}$-$\lambda_{AB}$-space}{\includegraphics[width=0.45 \textwidth]{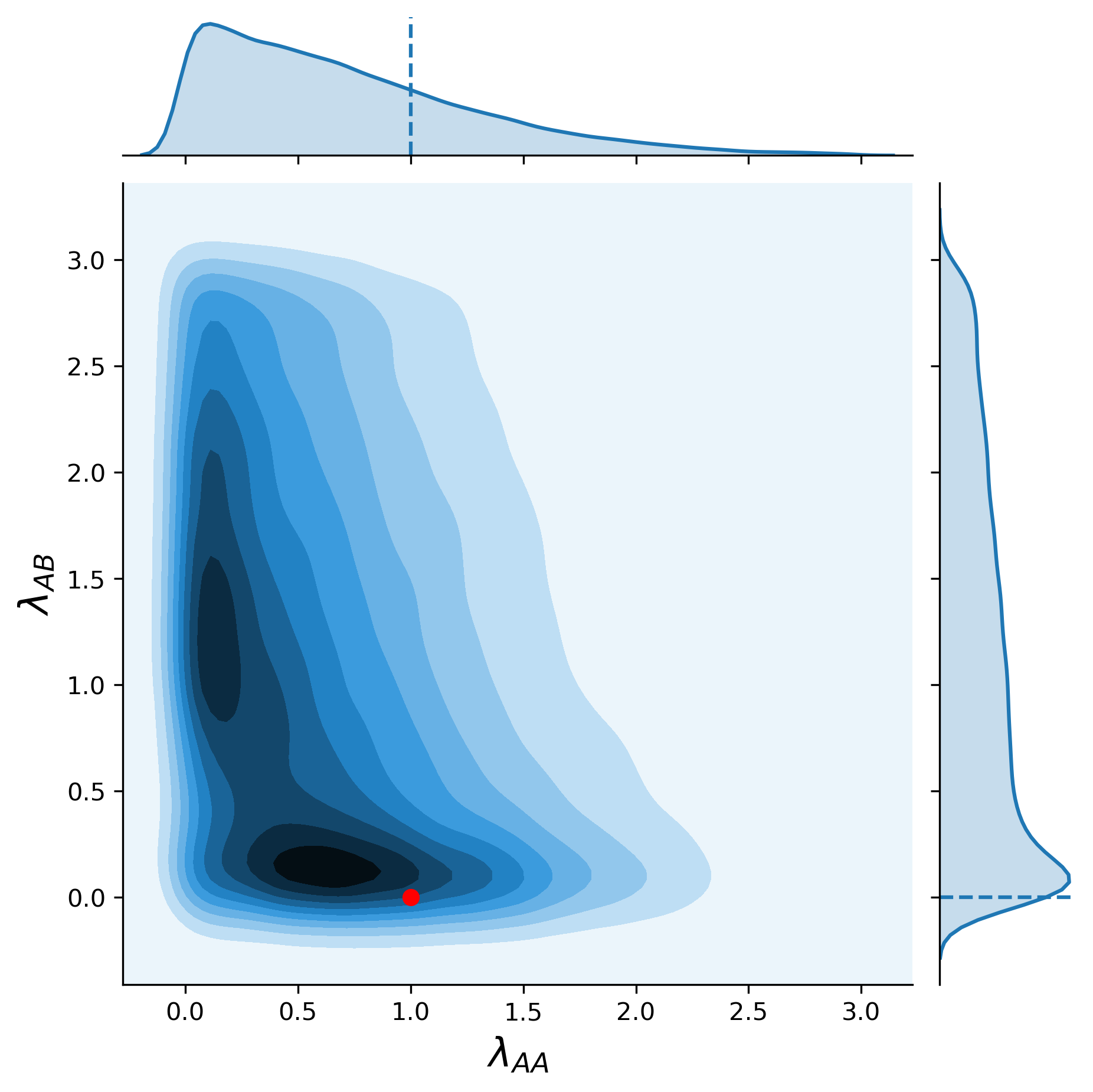}}
	\quad
	\subcaptionbox{$\lambda_{AB}$-$\lambda_{BB}$-space}{\includegraphics[width=0.45 \textwidth]{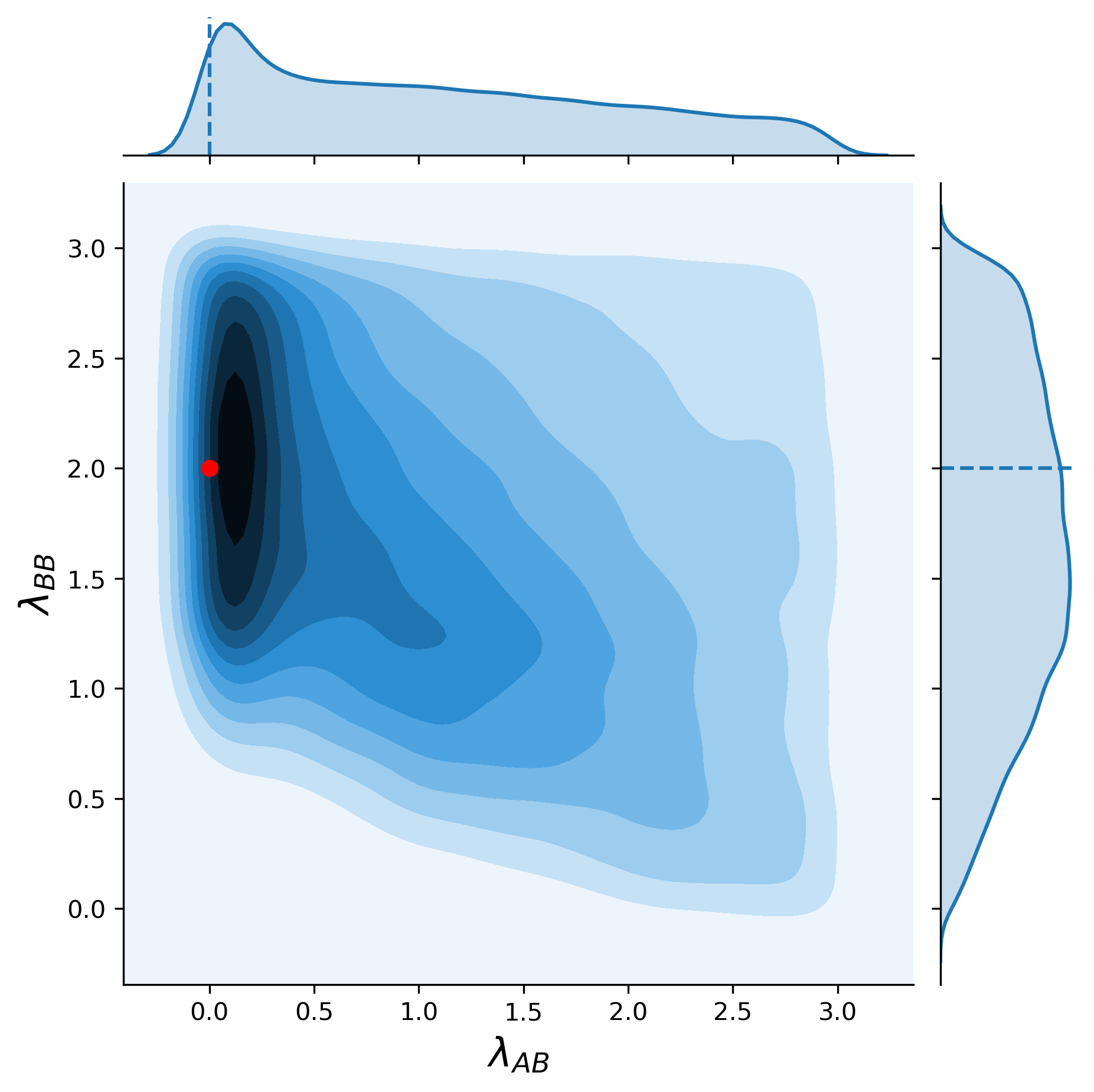}}
	\caption[Kernel density estimation of MCMC output]{\rev{Posterior distribution of Model C calculated based on five independent snapshots $\{\boldsymbol{N}(t=1.0)\}$ which were sampled from a system with parameters $\boldsymbol{\theta}=(1,0,2)$ with initial conditions $\boldsymbol{N}(t=0)=(5,5)$,  and uniform, independent priors $\theta_{\nu} \mathrel{\stackrel{\makebox[0pt]{\mbox{\normalfont\tiny iid}}}{\sim}} \mathcal{U}[0,3]$. The three-dimensional posterior for Model C is projected onto the two-dimensional planes with the corresponding marginal probability distributions at upper and right margins, shown as the KDE of MCMC samples. The true rate constants from which the two snapshot data were created is shown as a red dot and dashed, blue lines. The non-zero posterior density for negative rate constants is due to the KDE visualisation and does not represent actual points sampled by the MCMC. Because the population dynamics in model C are less sensitive to changes in the asymmetric division rate $\lambda_{AB}$ than to variations in $\lambda_{AA}$ and $\lambda_{BB}$, we observe a larger posterior variance along $\lambda_{AB}$.
}}
	\label{fig_KDE2}
\end{figure}

\clearpage

\section{Supplementary figures for ABC of epiblast stem cell data}

\begin{figure}[ht]
	\centering
	\includegraphics[width=0.7 \textwidth]{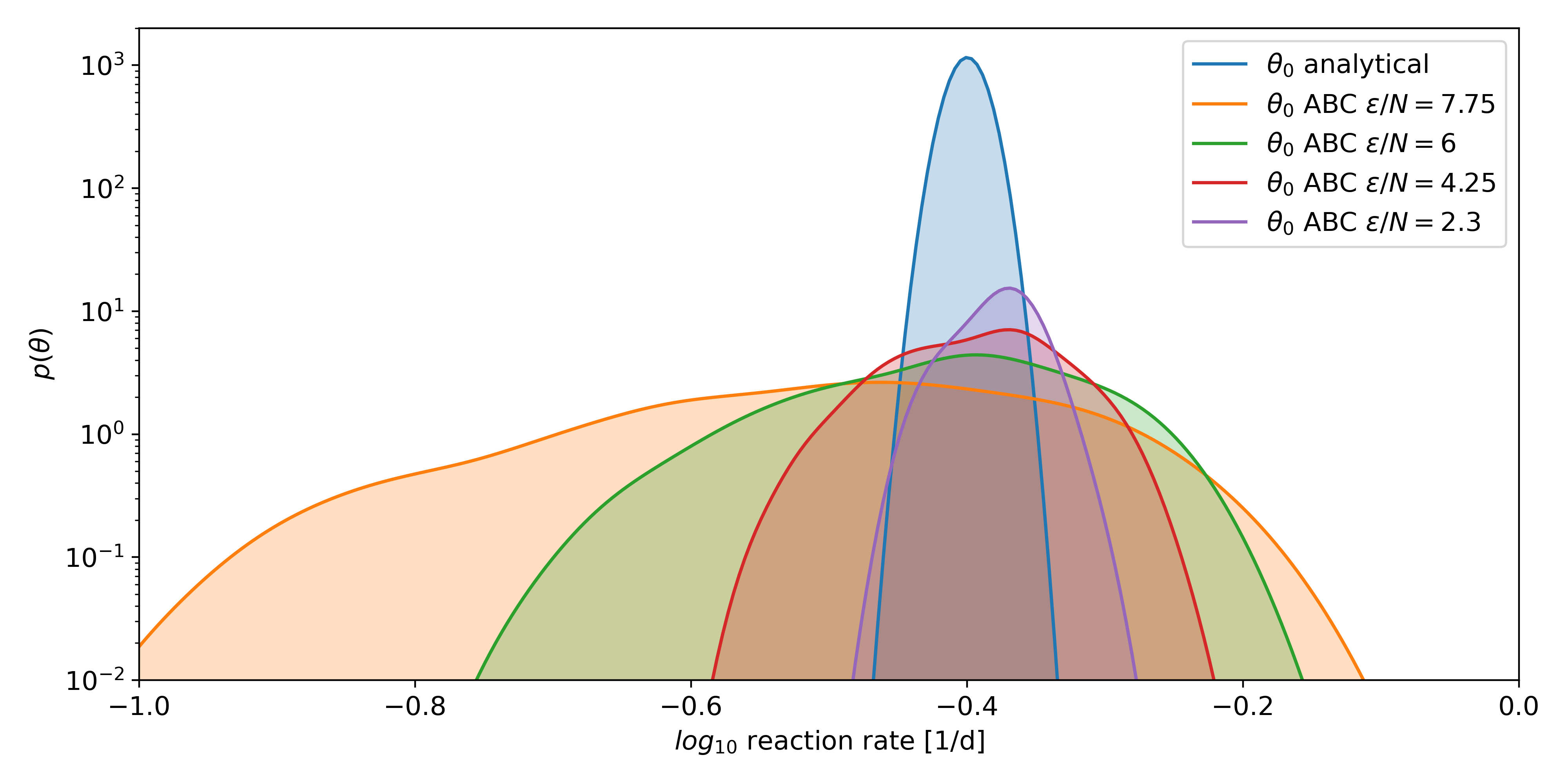}
	\caption{Comparison of SMC-ABC posterior distributions $p(\boldsymbol{\theta}|D(s(\mathcal{D}^{sim}),s(\mathcal{D}))< \epsilon)$ and the analytical posterior distribution $p(\theta_{0}|\mathcal{D})$ of model U obtained from dataset $\mathcal{D}_\mathrm{CHIR}$. For sufficiently small $\epsilon$, the SMC-ABC posterior deviates only slightly from the exact solution. For this specific case, the analytical posterior distribution can be calculated because the likelihood for $\theta_{0}$ in model U is known: $p(\theta_{0}|N) = e^{-\theta_{0} t}(1 - e^{-\theta_{0} t})^{N - 1}$, which is the probability that we observe a clone size of $N$ cells at time $t$.}
	\label{eps_convergence}
\end{figure}

\begin{figure}[ht]
	\centering
	\includegraphics[width=0.7 \textwidth]{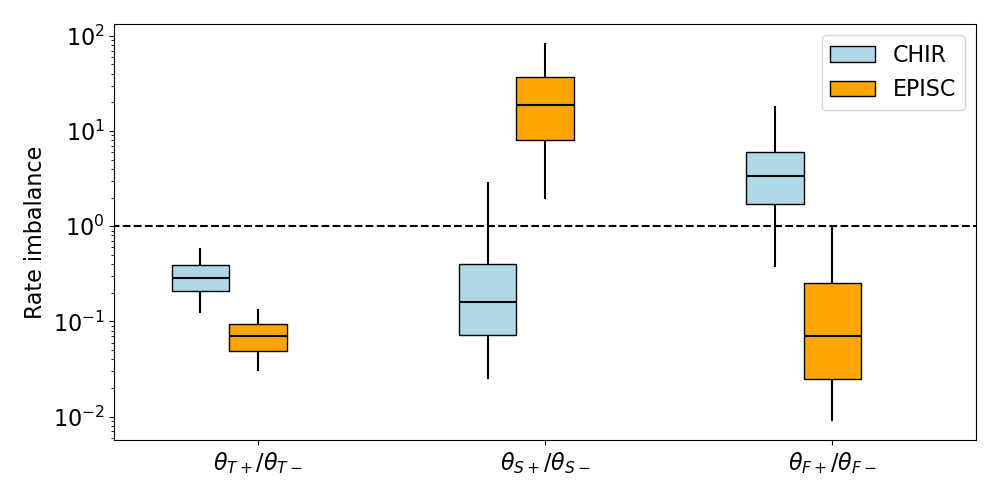}
	\caption{Median (horizontal line), 0.25/0.75 quantiles (box) and 0.05/0.95 quantiles (vertical lines) of the rate imbalance of Model U log-rates $\log_{10}(\theta_{k})$ obtained from $\mathcal{D}_\mathrm{CHIR}$ and $\mathcal{D}_\mathrm{EPISC}$.} 
	\label{CHIR_EPISC_compareY}
\end{figure}

\begin{figure}[ht]
	\centering
	\subcaptionbox{CHIR conditions}{\includegraphics[width=0.45 \textwidth]{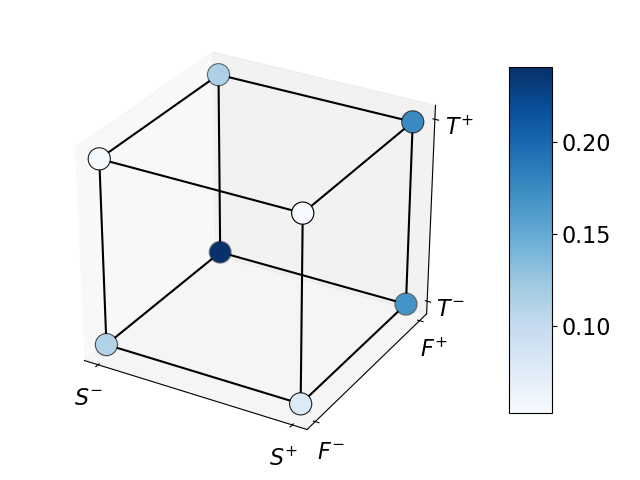}}
	\subcaptionbox{EPISC conditions}{\includegraphics[width=0.45 \textwidth]{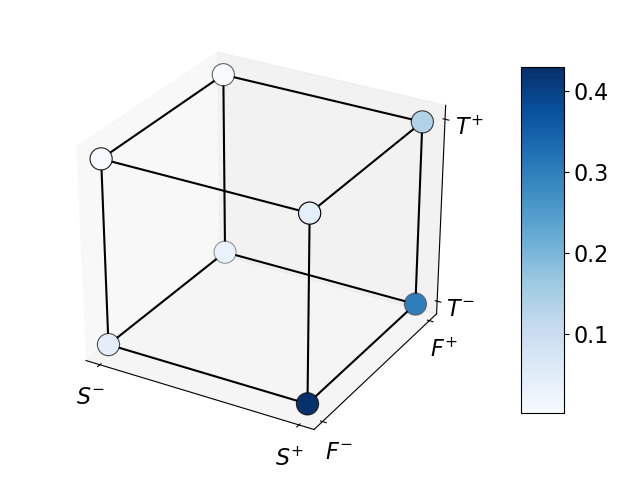}}
	\caption{Complete cell state distribution of initially unsorted stem cells after $t=3 d$ for different environmental conditions (CHIR/EPISC). The initial cell state distribution at $t=0 d$ is assumed to be uniform. The probability of the respective marker expression is indicated by the color map.} 
\end{figure}

\begin{figure}[ht]
	\centering
	\includegraphics[width=0.7 \textwidth]{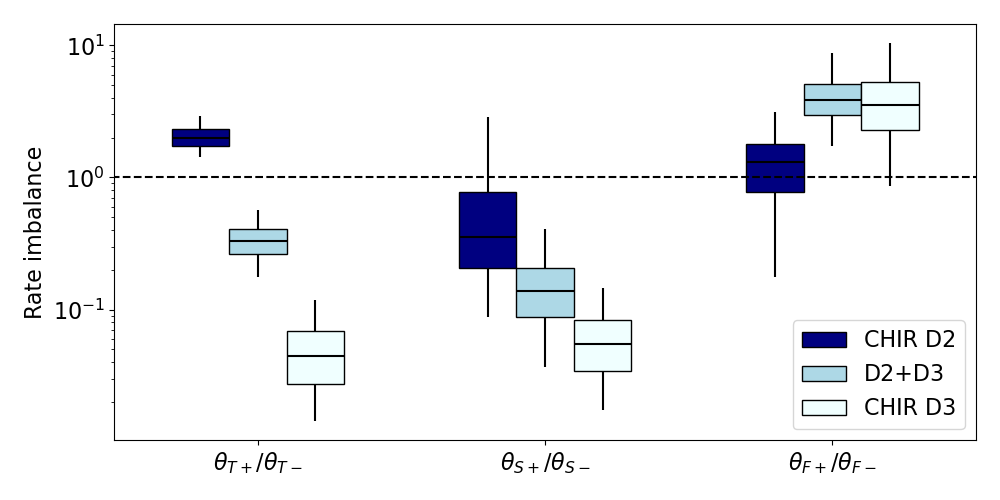}
	\caption{Median (horizontal line), 0.25/0.75 quantiles (box) and 0.05/0.95 quantiles (vertical lines) of the rate imbalance of Model U log-rates $\log_{10}(\theta_{k})$ obtained from $\mathcal{D}_\mathrm{CHIR}$. CHIR D2 and CHIR D3 shows inferred parameters using exclusively day 2 or day 3 data, respectively. CHIR D2+D3 was obtained by including all data points of $\mathcal{D}_\mathrm{CHIR}$.} 
	\label{CHIR_time_compareY}
\end{figure}

\begin{figure}
	\centering
	\includegraphics[width=0.8 \textwidth]{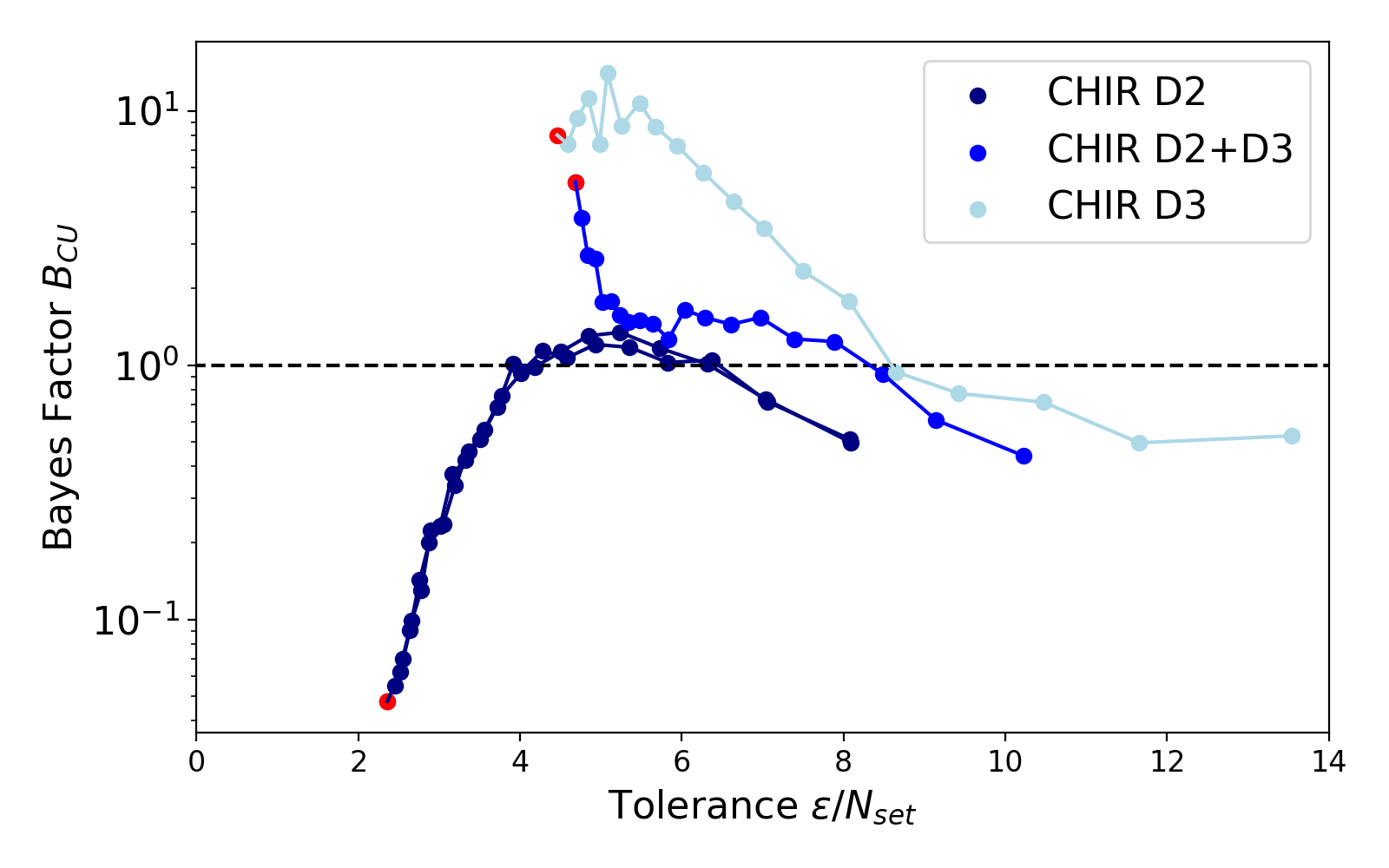}
	\caption{Bayes factors $B_{CU}$ of dataset $\mathcal{D}_\mathrm{CHIR}$ using either exclusively day 2 or day 3 data over a range of tolerance values $\epsilon$. Identical log-uniform priors were used for all rate constants in both models $\log_{10}(\theta_{k}) \sim \mathcal{U}[-2,0])$. The x-axis was re-scaled by the number of measurement series $N_{set}$ ($N_{set}=4$ for both D2 and D3). Red dots show the minimal epsilon value to which the SMC-ABC algorithm converged after $20$ iterations. The data up to day 2 favours model C dynamics ($B_{CU}>1$) while the population data at day 3 provides evidence in favour for model U dynamics ($B_{CU}<1$).}
	\label{chir_d2_d3}
\end{figure}

\clearpage

\section{Rejection-ABC given experimental stem-cell data}

We used the following Rejection-ABC algorithm to verify results for parameter inference and model selection (modified from~\cite{toni2008approximate}). The advantage of this algorithm is that it samples the entire parameter space uniformly and therefore does not risk getting stuck in local minima, unlike MCMC and SMC algorithms. Since Rejection-ABC is less efficient than SMC-ABC, calculations with this algorithm are only feasible at larger tolerance values $\epsilon$. 

\begin{algorithm}[H]
\caption{Rejection-ABC for model selection}
	
	\begin{algorithmic}[1]
		\STATE Set model $M^{\star}=M_{C}$ with probability $0.5$, otherwise set $M^{\star}=M_{U}$
		\STATE Sample parameter $\boldsymbol{\theta}^{\star}$ according to prior distribution $q(\boldsymbol{\theta}^{\star}|M^{\star})$
		\STATE Simulate a set of trajectories $\{\xi\}$ from $M^{\star}$ with parameters $\boldsymbol{\theta}^{\star}$ and obtain projected snapshot data $D^{sim}$
		\STATE If $D(s(\mathcal{D}^{sim}),s(\mathcal{D}))<\epsilon$, accept $(\boldsymbol{\theta},M^{\star})$, otherwise reject
		\STATE Go back to (1)
	\end{algorithmic}
\end{algorithm}

\begin{figure}
	\centering
	\subcaptionbox{Acceptance probability as a function of tolerance $\epsilon$ for Model C and dataset $\mathcal{D}_\mathrm{CHIR}$.}{\includegraphics[width=0.45 \textwidth]{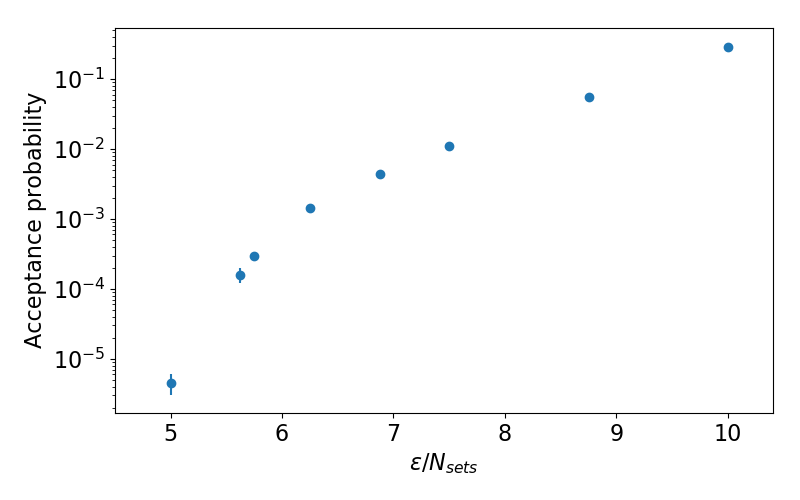}}
	\quad
	\subcaptionbox{Bayes factors $B_{CU}$ of datasets $\mathcal{D}_\mathrm{CHIR}$ and $\mathcal{D}_\mathrm{EPISC}$ for a range of tolerance values $\epsilon$.}{\includegraphics[width=0.45 \textwidth]{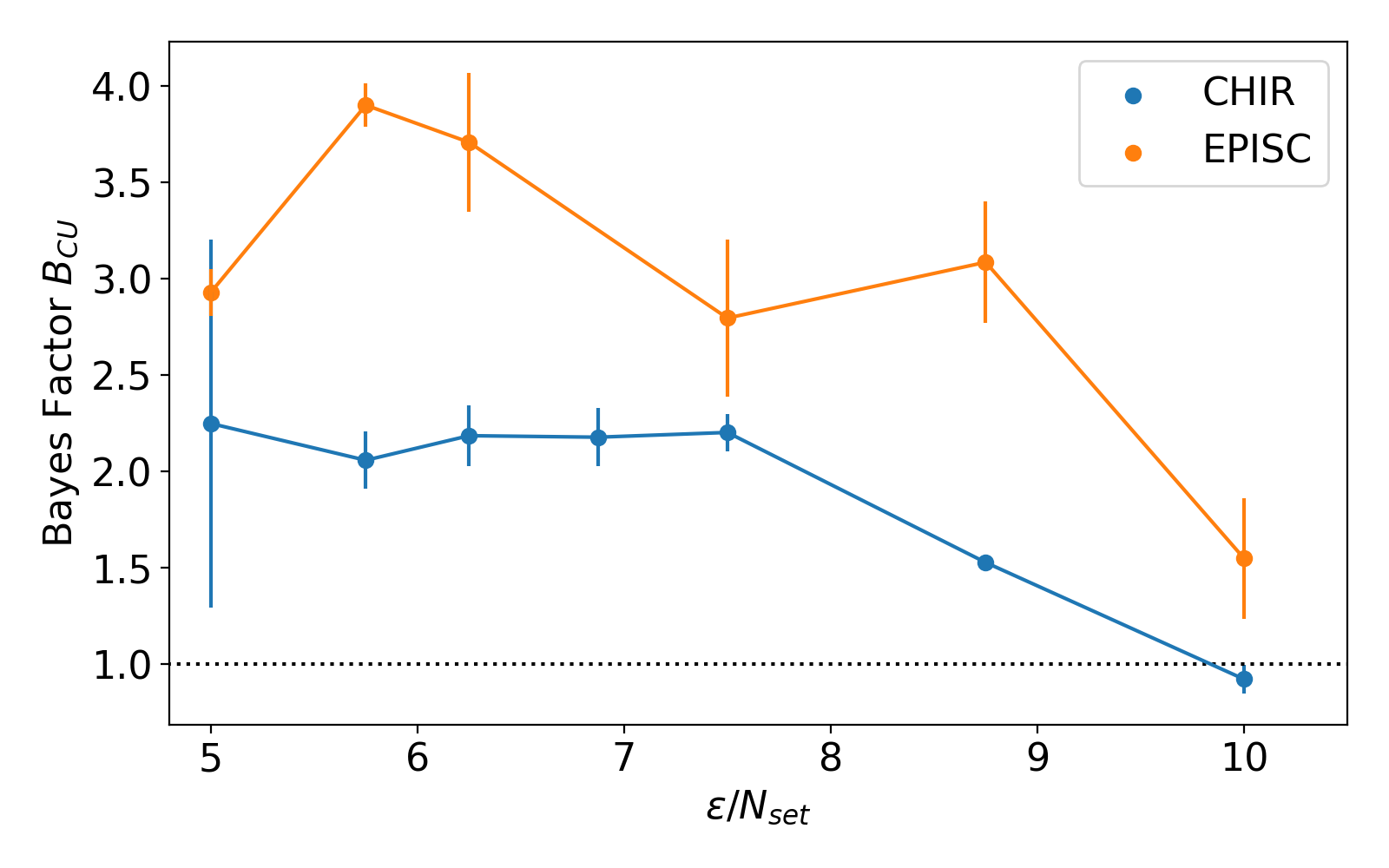}}
	\caption{Acceptance probability and Bayes factors for a range of tolerance values $\epsilon$ with identical log-uniform priors for all rate constants in both models $\log_{10}(\theta_{k}) \sim \mathcal{U}[-3,1])$. The x-axis was re-scaled by the number of measurement series $N_{set}$ in the respective dataset ($N_{set}=8$ in $\mathcal{D}_\mathrm{CHIR}$, $N_{set}=4$ in $\mathcal{D}_\mathrm{EPISC}$). Errorbars indicate the spread between five independent rejection-ABC runs.}
	\label{rej_epsilon}
\end{figure}

The rejection-ABC algorithm was run several times and $\epsilon$ was progressively lowered until the acceptance probability of samples reached a lower threshold of $10^{-5}$. While lowering $\epsilon$, the acceptance probability decreases faster than exponentially (see Figure~\ref{rej_epsilon} a). The Bayes factor $B_{CU}$ can be estimated by the ratio of the acceptance probabilities of models $M_{C}$ and $M_{U}$. This was done for both datasets $\mathcal{D}_\mathrm{CHIR}$ and $\mathcal{D}_\mathrm{EPISC}$ and the resulting Bayes factors are shown in Figure~\ref{rej_epsilon} b as a function of $\epsilon$. We chose identical log-uniform priors for all rate constants in both models $\log_{10}(\theta_{k}) \sim \mathcal{U}[-3,1])$ (in units of per day). Note that these priors have a slightly larger domain than the ones used for SMC-ABC.

For large $\epsilon$, the resulting posterior distributions differ only slightly from the uniform priors and Model C and Model U are equally likely ($B_{CU}\approx 1$). By decreasing $\epsilon$, the posterior distributions show distinct peaks and are very different for both models. Consequently by lowering $\epsilon$, $B_{CU}$ starts to deviate from one and eventually plateaus at a finite value $B_{CU}>1$. The large uncertainty for $\epsilon/N_{set}=5$ is due to a very low acceptance probability, leading to a small sample size of accepted parameters.

\begin{figure}[ht]
	\centering
	\includegraphics[width=0.9 \textwidth]{abb/EpiSC/EPISC_L12_abs_U[-2,0].png}
	\caption{Mean and standard deviation of Model C log-rates $\log_{10}(\theta_{k})$ calculated from posterior distributions given $\mathcal{D}_\mathrm{CHIR}$ using Rejection-ABC. Replacing the $L_{1}$-norm in eqn (\ref{diff_norm}) by the $L_{2}$-norm does not significantly change the shape of parameter posteriors given $\epsilon$ is small enough ($\epsilon/N_{set}=5$ used for $L_{1}$ and  $\epsilon/N_{set}=3$ used for $L_{2}$).} 
	\label{episc_L12}
\end{figure}

\clearpage

\section*{References}
\bibliographystyle{unsrt}
\bibliography{SMbib}

%% file: methods.tex

We begin this section by introducing two continuous-time Markov processes as minimal stochastic models for stem cell dynamics. These simple models are then generalized to describe the dynamics of epiblast stem cells (EpiSC) in more detail. We will close this section with a brief summary of Bayesian parameter inference and model selection.

\subsection{Minimal models}

We begin by defining minimal models of cell division and differentiation in a well-mixed population. We assume that the population dynamics can be adequately described by a continuous-time Markovian stochastic process where state transition rates depend only on the current state of the system. This in turn is specified by a set of population numbers $\{\textbf{N}\}$ of different cell states. In this section we will consider systems with two different cell states and assume that cells within the population divide and change state independently of one another.

\begin{figure}
\centering
\subcaptionbox{\label{test1}}{\includegraphics[width=0.9 \textwidth]{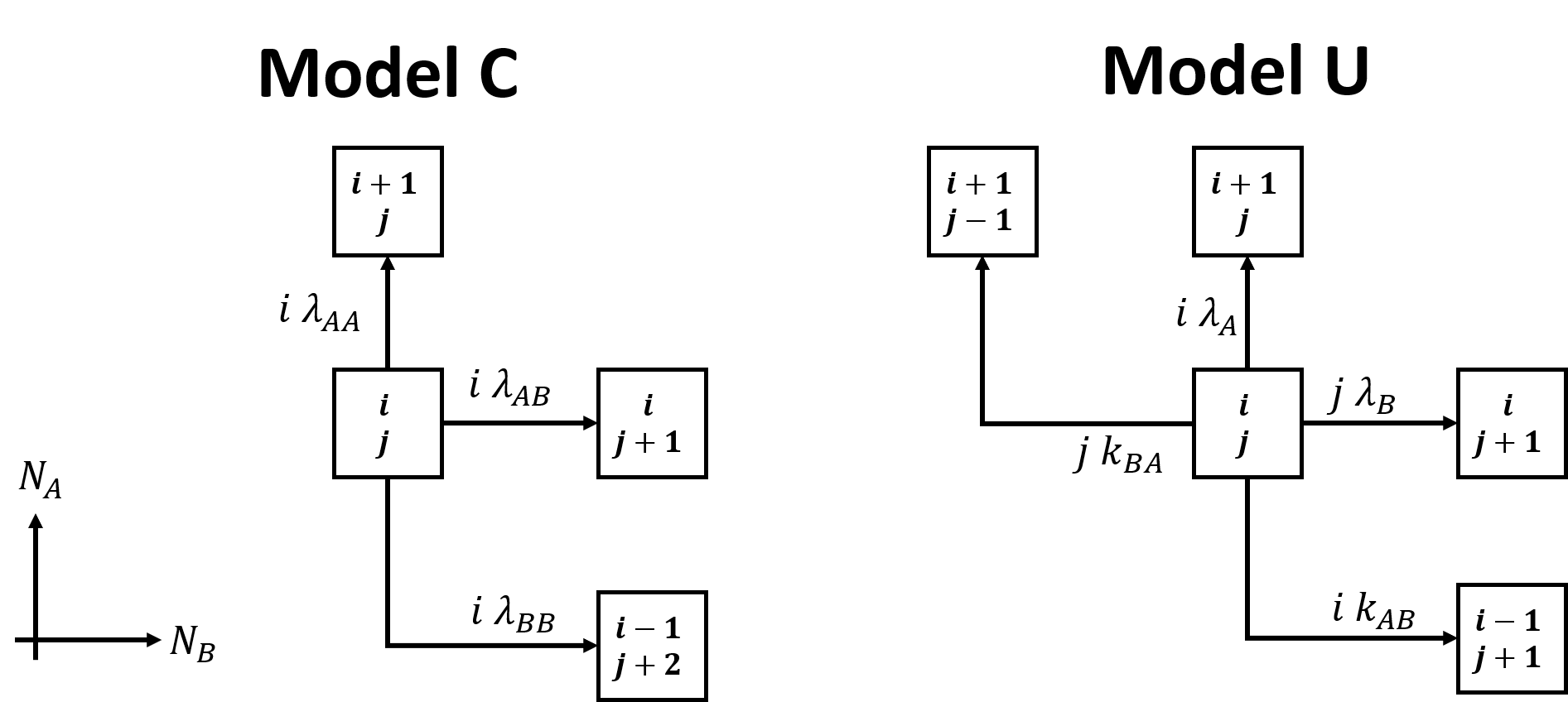}}
\par\bigskip\bigskip
\subcaptionbox{\label{test2}}{\includegraphics[width=0.35 \textwidth]{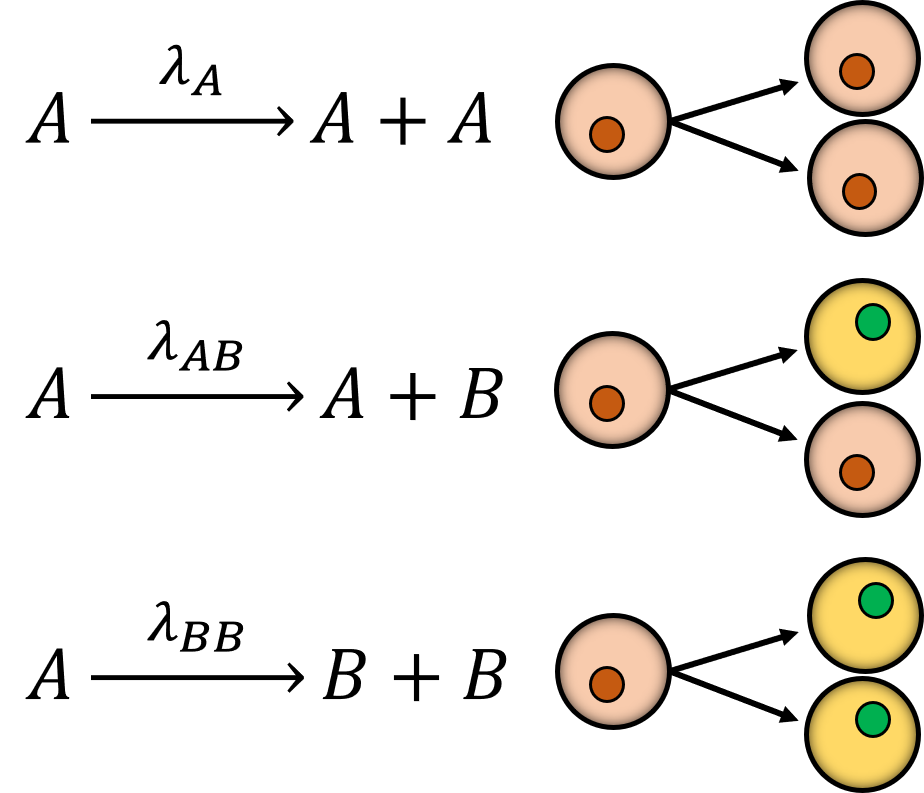}}
\qquad
\subcaptionbox{\label{test3}}{\includegraphics[width=0.35 \textwidth]{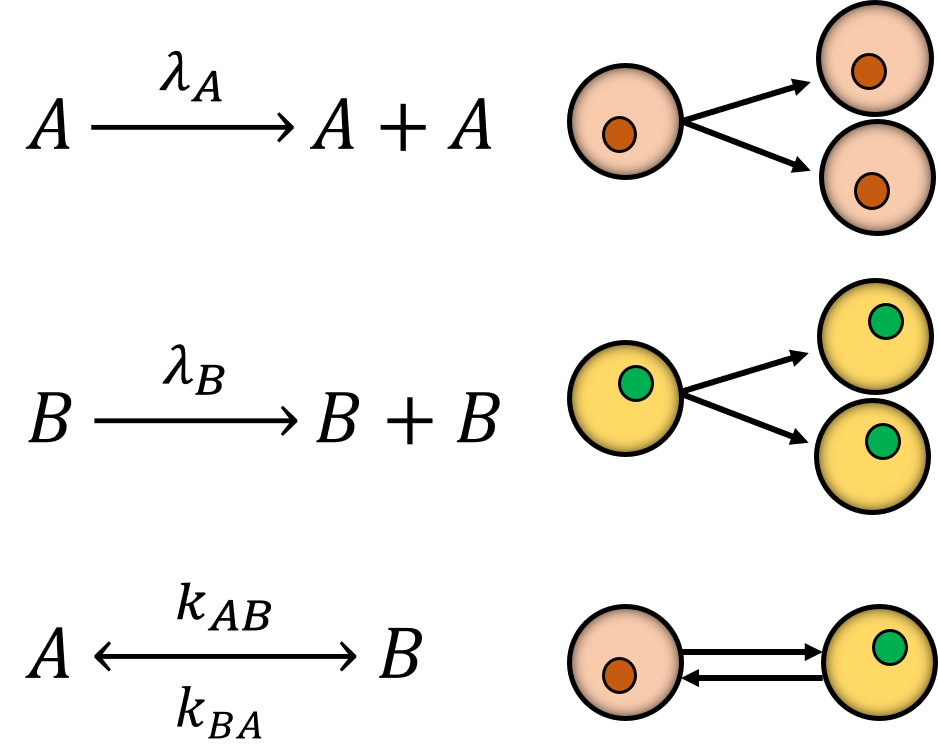}}
\caption[Cell transitions in Models C and U]{Cell transitions in Models C and U. (a) The state of the system is defined by the number of A-state cells ($i$) and B-state cells ($j$). Population dynamics can be interpreted as a 2D random walk with position-dependent jump rates. (b) Model C allows symmetric and asymmetric divisions. (c) Model U features only symmetric divisions, and state changes are made through reversible transitions that do not change cell number.}
\label{random_walk}
\end{figure}

We consider two distinct models of population dynamics: one in which cell differentiation is either coupled to cell division (Model C) one in which processes are uncoupled (Model U).
In both models there are two cell states, A and B. The essential difference is that in Model C, cells of a different state can be produced only through cell division, whereas in Model U cells can change state reversibly without dividing. 
The dynamics in the state space spanned by $N_A(t)$ and $N_B(t)$ are illustrated in Figure~\ref{random_walk}. Initial populations $N_{A}(0)$ and $N_{B}(0)$ will be denoted as $N_{A0}$ and $N_{B0}$, respectively.

\subsubsection{Model C}

Model C is defined by the following continuous-time transition rates:

\begin{equation}
[N_{A},N_{B}] \rightarrow \left\{ \begin{array}{ll}
{[N_{A}+1,N_{B}]}, & \quad \text{with rate $h_{1}^{C} := N_{A} \lambda_{AA}$,} \\
{[N_{A},N_{B}+1]}, & \quad \text{with rate $h_{2}^{C} := N_{A} \lambda_{AB}$}, \\
{[N_{A}-1,N_{B}+2]}, & \quad \text{with rate $h_{3}^{C} := N_{A} \lambda_{BB}$},
\end{array} \right.
\label{hazard_X}
\end{equation}
where the system's transition rates are denoted as $h_{k}^{C}$. In each of these three reactions, an A-state cell divides, producing two A-state cells, an A- and a B-state cell, or two B-state cells, at per-capita rates $\lambda_{AA}$, $\lambda_{AB}$, and $\lambda_{BB}$ respectively. In this model, B-state cells do not divide.

Let $p_{i,j}(t)$ be the probability that our cell population consists of $i$ A-state and $j$ B-state cells at time $t$, then the master equation for Model C is given by
\begin{eqnarray}
	\frac{dp_{i,j}}{dt} = &(i-1) \lambda_{AA} p_{i-1,j} + (i+1) \lambda_{BB} p_{i+1,j-2} + i \lambda_{AB} p_{i,j-1} \nonumber\\
	&- i (\lambda_{AA}+\lambda_{AB}+\lambda_{BB}) p_{i,j}.
\label{ME_X}
\end{eqnarray}
This master equation is a differential equation defining the time evolution of population probabilities $p_{i,j}(t)$ caused by stochastic transitions.

\subsubsection{Model U}

In Model U, the state transition rates are:

\begin{equation}
[N_{A},N_{B}] \rightarrow \left\{ \begin{array}{ll}
{[N_{A}+1,N_{B}]}, & \quad \text{with rate $h_{1}^{U} := N_{A} \lambda_{A}$,} \\
{[N_{A}-1,N_{B}+1]}, & \quad \text{with rate $h_{2}^{U} := N_{A} k_{AB}$}, \\
{[N_{A},N_{B}+1]}, & \quad \text{with rate $h_{3}^{U} := N_{B} \lambda_{B}$}, \\
{[N_{A}+1, N_{B}-1]}, & \quad \text{with rate $h_{4}^{U} := N_{B} k_{BA}$},
\end{array} \right.
\label{hazard_Y}
\end{equation}
with transition rates $h_{k}^{U}$. 
These reactions correspond to division of A-state and B-state cells into two daughter cells of the same state at per-capita rates $\lambda_A$ and $\lambda_B$, respectively, and to dynamic state changes of an A-state into a B-state at per-capita rate $k_{AB}$, and of a B-state into an A-state at per-capita rate $k_{BA}$.

The master equation of Model U is given by
\begin{eqnarray}
	\frac{dp_{i,j}}{dt} = &(i-1) \lambda_{A} p_{i-1,j} + (j-1) \lambda_{B} p_{i,j-1} + (j+1) k_{BA} p_{i-1,j+1} \nonumber\\
	&+ (i+1) k_{AB} p_{i+1,j-1}  - (i  \lambda_{A}+ j  \lambda_{B}+ i  k_{AB} + j  k_{BA}) p_{i,j}.
\label{ME_Y}
\end{eqnarray}

\subsubsection{Generalisations of Models C and U: Epiblast stem cell models}
\label{epiblast_models}

\begin{figure}
\centering
\subcaptionbox{\label{EpiSC_network}}{\includegraphics[width=0.6 \textwidth]{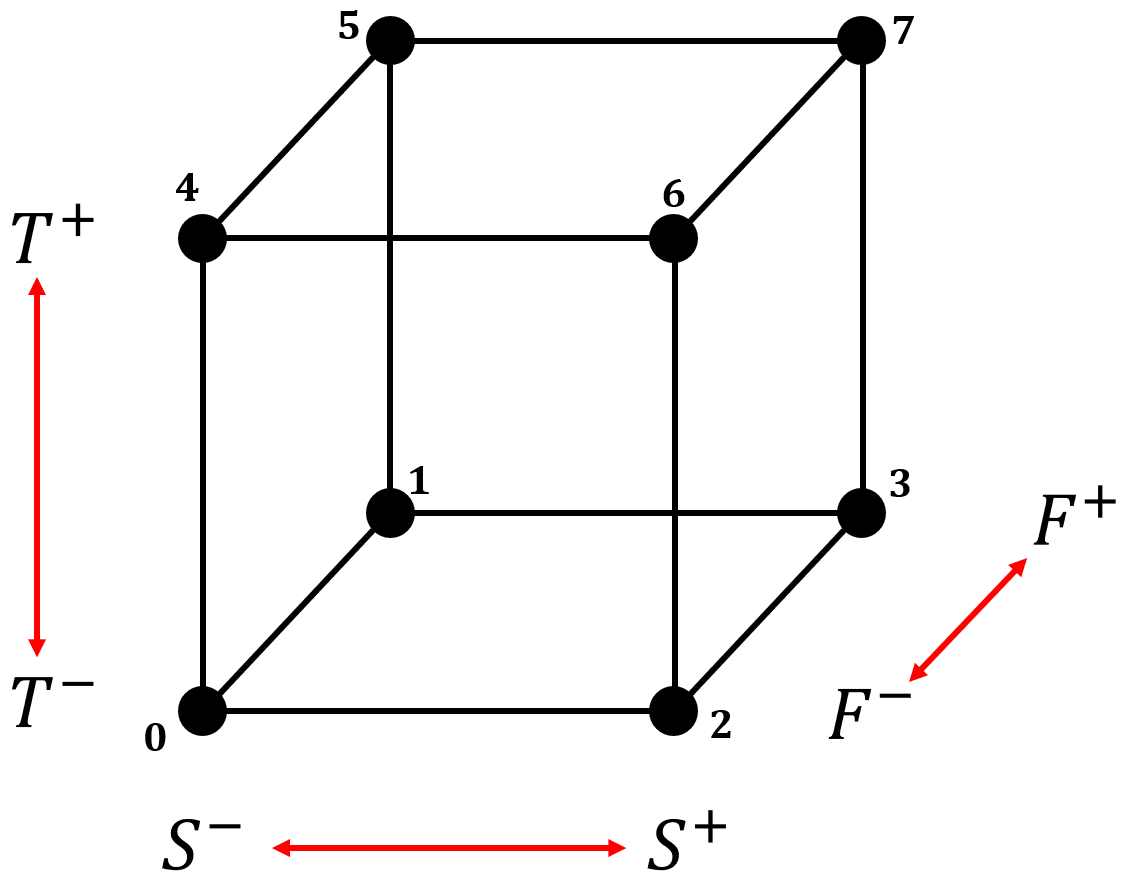}}
\par\bigskip\bigskip
\subcaptionbox{\label{extended_model_X}}{\includegraphics[width=0.4 \textwidth]{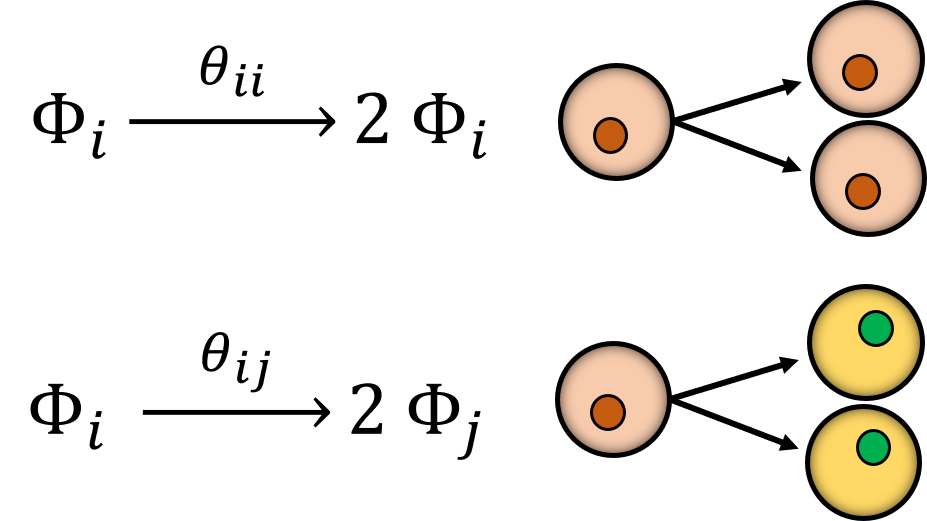}}
\qquad
\subcaptionbox{\label{extended_model_Y}}{\includegraphics[width=0.4 \textwidth]{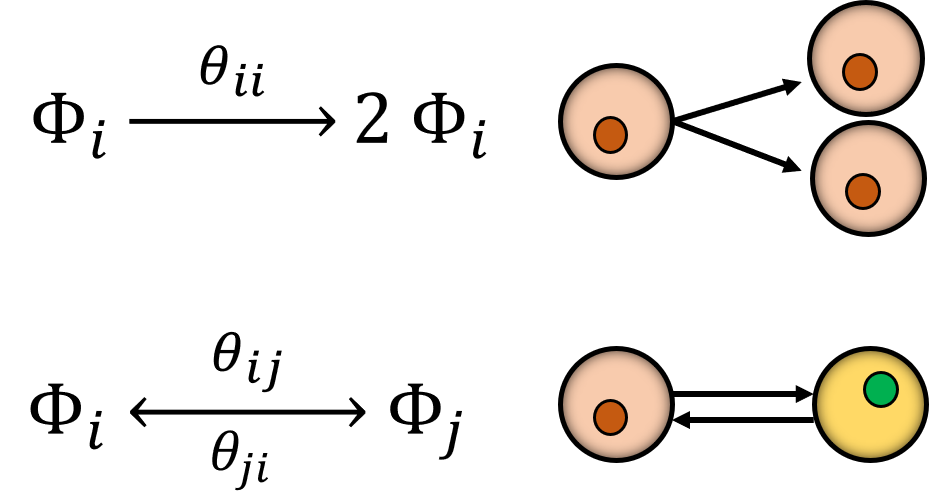}}
\caption{Dynamics in generalisations of Models C and U. (a) We characterise EpiSCs by the co-expression of three genes, T(Bra), Sox2 and Foxa2, corresponding to a space of eight possible cell states (cube vertices), $\mathcal{S} = \{ (T^{\pm}, S^{\pm}, F^{\pm}) \}$. A combination of these 8 states may be present in a heterogeneous EPiSC population. We assume that cell state transitions change the expression of only one gene at a time. This results in $12$ possible transitions between cell states (cube edges). (b) In the generalised Model C, cells in state $\Phi_i \in \mathcal{S}$ divide symmetrically with rate $\theta_{ii}$ into two daughter cells with state $\Phi_{i}$, and symmetrically with rate $\theta_{ij}$ into two daughter cells with a neighbouring state $\Phi_{j}$. (c) In the generalised Model U, cells in state $\Phi_i \in \mathcal{S}$ divide symmetrically with rate $\theta_{ii}$ into two daughter cells with state $\Phi_i$, and transition with rate $\theta_{ij}$ into a neighbouring state $\Phi_{j}$.} 
\end{figure}

While Models C and U are useful for illustrative purposes, we require more cell states to make use of experimental data on epiblast-derived stem cell (EpiSC) dynamics. Here, we define EpiSC states by the binary expression of the transcription-factor genes T(Bra), Sox2, and Foxa2, which is what the experimental dataset contains information about~\cite{tsakiridis2014distinct}. There are thus $8$ distinct cell states accessible for modelling, which we will denote by 
\begin{equation}
\mathbf{\Phi}_{i} := \left(
\begin{array}{c}
[T^{-},S^{-},F^{-}] \\\relax
[T^{-},S^{-},F^{+}] \\\relax
[T^{-},S^{+},F^{-}] \\\relax
[T^{-},S^{+},F^{+}] \\\relax
[T^{+},S^{-},F^{-}] \\\relax
[T^{+},S^{-},F^{+}] \\\relax
[T^{+},S^{+},F^{-}] \\\relax
[T^{+},S^{+},F^{+}] 
\end{array}
\right)_{i},
\label{EpiSC_states}
\end{equation}
where the index T,S,F represents gene T(Bra), Sox2 and Foxa2, respectively. For simplicity, we start with the assumption that cell state transitions change only one of the three genes $T,F,S$ at a time. This leads to $12$ allowed transitions between cell states, which can be conveniently visualised as a cube where vertices represent cell states and edges represent transition paths (see Figure~\ref{EpiSC_network}).

The generalisation of Model C retains the defining feature that new cell states can be created only at division of the parent. Specifically, we allow symmetric division of cells in each state, $\Phi_i \to 2\Phi_i$, at per-capita rate $\theta_{ii}$. Note that we can recover Model C by setting all but one of these rates to zero. 
To keep the number of parameters small, we accommodate only the other symmetric cell division processes, $\Phi_{i}\rightarrow 2\Phi_{j}$ for any $j$ sharing an edge with $i$, at per-capita rate $\theta_{ij}$. These dynamics are illustrated in Figure~\ref{extended_model_X}. 
The fact that we chose to include only symmetric cell divisions should not change the outcome of model comparison too much, considering the following argument: the asymmetric division process $\Phi_{i}\rightarrow \Phi_{i}+\Phi_{j}$ with  $i\neq j$ effectively increases the $\Phi_{j}$ population by one. The same outcome can also be achieved either by a symmetric division $\Phi_j \to 2\Phi_j$ or by the sequence of two symmetric cell divisions $\Phi_i \to 2\Phi_j$, $i\neq j$ and $\Phi_j \to 2\Phi_k$, $j\neq k$. 
  
Since cells in any state can divide, there are eight different reactions of the form $\Phi_{i} \rightarrow 2 \Phi_{i}$. 
State-changing divisions of the form $\Phi_{i} \rightarrow 2 \Phi_{j}$ can happen between all neighbouring cell states $(i,j)$ in the network shown in Figure~\ref{EpiSC_network}; there are $3 \times 8 = 24$ such reactions, since each cell state has three neighbouring states. 
The total number of reaction types in extended Model C is therefore $8+24=32$.

In principle therefore there are 32 parameters in this model. \rev{To reduce the size of the parameter space, we further assume that the self-renewal rates are the same for all cell states, i.e., $\theta_{ii}=\theta_{0}$ for all $i$. 
We further assume that switching each gene $M\in\{T,S,F\}$ on or off does not depend on the state of the other two genes which are not changed. This assumption implies the transition rates of genes are independent of each other, which is a first approximation only. We discuss relaxing this assumption in Section~\ref{futurework}.}
Each gene $M$ thus contributes two independent rate constants $\theta_{M+}$, $\theta_{M-}$ for switching on and off, respectively. The extended Model C therefore has $7$ free parameters:
\begin{equation}
	\boldsymbol{\theta} = (\theta_{0},\theta_{T+},\theta_{T-},\theta_{S+},\theta_{S-},\theta_{F+},\theta_{F-}) \, . 
\end{equation} 
We now turn to the generalisation of Model U, recalling that the original two-state model allowed symmetric division of both states, and dynamic differentiation in which one state could turn into the other. This is easily extended to $8$ cell states $\Phi_{i}$, wherein each cell state can perform symmetric cell division at rate $\theta_{ij}$, and dynamical state changes between all pairs of neighbouring cell states can occur at rates $\theta_{ij}$ and $\theta_{ji}$. See Figure~\ref{extended_model_Y}.

As in the extended Model C, there are eight symmetric division rates, which we again set to a universal value $\theta_{ii}=\theta_0$. Similarly the rates of dynamic cell state changes are defined by the gene that is switched on or off, which again leads to a total of $8+24=32$ reaction types and $7$ free parameters:
\begin{equation}
\boldsymbol{\theta} = (\theta_{0},\theta_{T+},\theta_{T-},\theta_{S+},\theta_{S-},\theta_{F+},\theta_{F-}).
\end{equation} 
An appealing feature of Models C and U from a classical model-selection perspective is that they have the same number of free parameters. A problem when comparing models with different numbers of free parameters is that the models which offer more free parameters will usually show better agreement with the data. However, this is because it can be more accurately fitted to the data, not because the underlying assumptions are more accurate in describing nature. Bayesian model selection generally eliminates this problem since it embodies Occam’s razor: simpler models have higher posterior probabilities than more complex models if both fit the data equally well \cite{jefferys1991sharpening}.

\subsection{Bayesian inference and model selection}
\label{method_model_comparison}

We use Bayesian methods to perform model selection and parameter inference. Given some experimental data $\boldsymbol{{\xi}}$, we seek the probability distribution $p(\boldsymbol{{\theta}}|\boldsymbol{{\xi}})$ over the parameters $\boldsymbol{{\theta}}$ of either Model C or Model U. This is achieved by appealing to Bayes' theorem,
\begin{equation}
p(\boldsymbol{{\theta}}|\boldsymbol{{\xi}}) = \frac{p(\boldsymbol{{\xi}}|\boldsymbol{{\theta}}) p(\boldsymbol{{\theta}})}{p(\boldsymbol{{\xi}})} \;.
\label{bayes}
\end{equation}
Here $p(\boldsymbol{{\theta}})$ encapsulates our \emph{prior} beliefs as to appropriate values for the parameters $\boldsymbol{{\theta}}$ (i.e., in the absence of any data), and $p(\boldsymbol{{\xi}}|\boldsymbol{{\theta}})$ is the \emph{likelihood} of the data $\boldsymbol{{\xi}}$ given a parameter choice. In this context, the likelihood is given by solving the master equation for the model under consideration. Lastly $p(\boldsymbol{{\xi}})$ is the probability of observing the data $\boldsymbol{{\xi}}$, and is usually obtained indirectly by ensuring that the \textit{posterior} distribution is correctly normalised. It is worth mentioning that data $\boldsymbol{{\xi}}$ can take very different forms, such as continuous trajectory data of cell population or population snapshots at a finite number of times. The form of $\boldsymbol{{\xi}}$ does not affect Bayes' theorem, but enters the calculation via the likelihood function $p(\boldsymbol{{\xi}}|\boldsymbol{{\theta}})$.

If not only the model parameters $\boldsymbol{{\theta}}$ are unknown, but also the model itself, we can use Bayesian methods for model comparison. 
Given data $\boldsymbol{{\xi}}$ and the two competing models $M_{C}$ and $M_{U}$, we wish to calculate the probability of each model $p(M_{j}|\boldsymbol{{\xi}})$ given the data.
Following Bayes' theorem, the \textit{posterior odds} of the models are given by
\begin{equation}
\frac{p(M_{C}|\boldsymbol{{\xi}})}{p(M_{U}|\boldsymbol{{\xi}})} = \frac{p(\boldsymbol{{\xi}}|M_{C}) q(M_{C})}{p(\boldsymbol{{\xi}}|M_{U}) q(M_{U})},
\end{equation}
where $q(M_i)$ is the prior belief that model $M_{i}$ is the appropriate description of the data. The selection of the ``best'' model is usually not based on posterior odds, but the ratio of posterior to prior odds: that is, how much more the data point to one model over the other, given our prior beliefs:
\begin{equation}
B_{CU} := \frac{p(M_{C}|\boldsymbol{{\xi}})/p(M_{U}|\boldsymbol{{\xi}})}{q(M_{C})/q(M_{U})} = \frac{p(\boldsymbol{{\xi}}|M_{C})}{p(\boldsymbol{{\xi}}|M_{U})}.
\label{BF}
\end{equation}
This ratio is called the \textit{Bayes factor}~\cite{kass1995bayes}. 
Following (\ref{BF}), the \textit{marginal likelihoods} $p(\boldsymbol{{\xi}}|M_{j})$ are the quantities of key interest. 
They can be obtained by integrating over the parameter space $\boldsymbol{\theta}^{j}$ of the corresponding model:
\begin{equation}
p(\boldsymbol{{\xi}}|M_{j}) = \int p(\boldsymbol{{\xi}}|\boldsymbol{\theta}^{j},M_{j})  p(\boldsymbol{\theta}^{j}|M_{j})  d\boldsymbol{{\theta}}^{j}  ,
\label{BF2}
\end{equation}
where $p(\boldsymbol{\theta}^{j}|M_{j})=q(\boldsymbol{\theta}^{j})$ is the parameter prior of model $M_{j}$.

All algorithms shown in this work were implemented using python3 with standard libraries (numpy, pandas, scipy, pyabc). All code can be accessed via GitHub (https://github.com/real-save/Stem-cell-inference).

%% file: results.tex

\section{Results}

\subsection{Coupling of differentiation and division leads to qualitatively different population dynamics}

Our first set of results pertain to how populations evolve over time in Models C and U in their original formulation with two cell states. In particular, we will see the utility of \emph{moment generating functions}. In the following we will derive equations of the probability generating function and mean expressions for $N_A$ and $N_B$ for each of Models C and U. In the last section we will also derive approximate expressions for mean populations $\langle \mathbf{\Phi} \rangle$ in the extended models.

\subsubsection{Population dynamics of Model C}\label{exact_X}

Applying the procedure of probability generating functions (see Appendix B) to the master equation \eqref{ME_X} for Model C yields
\begin{equation}
	\frac{\partial G}{\partial t} = \left[ \lambda_{AA}(z_{A}^{2}-z_{A}) + \lambda_{BB}(z_{B}^{2}-z_{A}) + \lambda_{AB}  (z_{A} z_{B}-z_{A}) \right] \frac{\partial G}{\partial z_{A}}.
	\label{PGF_X_PDF}
\end{equation}
This is a linear, first-order PDE for $G(\boldsymbol{z},t)$ and can be solved by the method of characteristics~\cite{courant1937methoden}. \rev{We assume the system initially consists of $N_{A0}$ A-state cells and $N_{B0}$ B-state cells, which translates to the initial condition $G(z_A,z_B,t=0)=z_A^{N_{A0}} z_B^{N_{B0}}$.} Using this initial condition, we find
\begin{equation}
\fl	G(z_{A},z_{B},t) = z_{B}^{N_{B0}} \left[ \frac{1}{\rho \lambda_{AA}} \tan\left(\frac{t}{\rho} + \arctan\left(\rho\lambda_{AA} z_{A} + \frac{\rho \xi}{2}\right) \right) - \frac{\xi}{2 \lambda_{AA}} \right]^{N_{A0}},
	\label{PGF_sol}
\end{equation}
where
\begin{eqnarray}
	\xi(z_{B}) &= \lambda_{AB}(z_{B}-1) - \lambda_{AA}-\lambda_{BB}, \nonumber\\
	\rho(z_{B}) &= \frac{2}{\sqrt{4 \lambda_{AA}\lambda_{BB} z_{B}^{2} - \xi^{2}}}.
\end{eqnarray}
The full time-dependence of mean population sizes then follow from \eqref{Genprop2}:
\begin{eqnarray}
	\langle N_{A}(t) \rangle &= N_{A0} e^{\Delta \lambda \: t}, \\
	\langle N_{B}(t) \rangle &= N_{B0} + N_{A0} \left( \frac{2 \lambda_{BB} + \lambda_{AB}}{\Delta \lambda}\right) \left( e^{\Delta \lambda \: t} - 1 \right),
	\label{mABX}
\end{eqnarray}
\rev{where $\Delta \lambda := \lambda_{AA}-\lambda_{BB}$ is the difference of symmetric division rates.} To understand these results, we note that cells in state A divide at per-capita rate $\lambda_{AA}$, and are lost at per-capita rate $\lambda_{BB}$. 
These competing processes lead to either an exponential growth ($\lambda_{AA}>\lambda_{BB}$) or exponential decay ($\lambda_{AA}<\lambda_{BB}$) in the mean number of cells in state A. 
Since there is no loss of cells from state B, the mean number of cells in state B either grows exponentially with time ($\lambda_{AA}>\lambda_{BB}$) or plateaus at a finite value ($\lambda_{AA}<\lambda_{BB}$). \rev{Mean population dynamics are therefore sensitive to changes in symmetric division rates since $\Delta \lambda$ controls the exponential growth/decay of mean populations. On the other hand, the rate of asymmetric cell divisions, $\lambda_{AB}$, weakly affects mean dynamics as it enters eqn \eqref{mABX} only as a constant prefactor.}

A particular feature of Model C is that there is an absorbing state. 
This arises because A-state cells can be lost via the process $A\rightarrow 2B$, and B-state cells do not divide. Therefore, this a finite probability that the number of cells in state A will eventually decrease to zero and the system dynamics will come to a halt. 
We can obtain $p_{\rm ext}(t)$, the \textit{extinction probability} that this has occurred by time $t$, from the moment generating function as follows: 
\begin{equation}
p_{\rm ext}(t) = {\rm Pr}\left[N_{A}(t)=0\right] = \sum_{j=0}^{\infty} p_{0,j}(t) = G\left(0, 1 ,t\right).
\end{equation}
From \eqref{PGF_sol} we find 
\begin{equation}
p_{\rm ext}(t) = \left( \frac{2 \lambda_{BB}}{ \lambda_{AA} + \lambda_{BB} + |\Delta \lambda| \coth \left( t |\Delta \lambda| / 2 \right) } \right)^{N_{A0}}.
	\label{pext1}
\end{equation}
In the special case $\lambda_{AA} = \lambda_{BB} = \lambda$ this simplifies to
\begin{equation}
p_{\rm ext}(t) = \left( \frac{\lambda t}{1+\lambda t} \right)^{N_{A0}}.
\label{pext2}
\end{equation}
The extinction probability is monotonically increasing with time and converges to a finite value $\lim_{t\to \infty} p_{\rm ext}(t) = p_{\rm ext}^{\rm max}$. For $\lambda_{BB} \geq \lambda_{AA}$, it follows $p_{\rm ext}^{\rm max}=1$ and A-state cells will always die out. For $\lambda_{BB} < \lambda_{AA}$ however, $p_{\rm ext}^{\rm max}=(\lambda_{BB}/\lambda_{AA})^{N_{A0}}<1$. 

\subsubsection{Population dynamics of Model U}\label{exact_Y}

We can solve Model U by following the same sequence of steps. 
We first arrive at the PDE

\begin{eqnarray}
\frac{\partial G}{\partial t} &=& \left[ \lambda_{A}(z_{A}^{2}-z_{A}) + k_{AB}(z_{B}-z_{A}) \right] \frac{\partial G}{\partial z_{A}} + {} \nonumber\\ && \quad \left[ \lambda_{B}(z_{B}^{2}-z_{B}) + k_{BA}(z_{A}-z_{B}) \right] \frac{\partial G}{\partial z_{B}}.
	\label{PGF_Y_PDF}
\end{eqnarray}
The solution is given by
\begin{eqnarray}
	G(z_{A},z_{B},t) = &\left[1+e^{\Gamma  t}  \left(c_{1}(t) (z_{A}-1)+c_{2}(t) (z_{B}-1)\right) \right]^{N_{A0}}\nonumber\\
	 &\quad \times \left[1+e^{\Gamma  t}  \left(c_{3}(t) (z_{A}-1)+c_{4}(t) (z_{B}-1)\right) \right]^{N_{B0}}.
	\label{PGF_Y}
\end{eqnarray}
Here, $c_{1}$, $c_{2}$, $c_{3}$, $c_{4}$ are functions of time $t$:
\begin{eqnarray}
	c_{1}(t) &= \cosh\left(\frac{\omega}{2} t\right) + \theta  \sinh\left(\frac{\omega}{2} t\right)  , \nonumber\\
	c_{2}(t) &= \frac{2 k_{AB}}{\omega} \sinh\left(\frac{\omega}{2} t\right)  , \nonumber\\
	c_{3}(t) &= \frac{2 k_{BA}}{\omega} \sinh\left(\frac{\omega}{2} t\right)  , \nonumber\\
	c_{4}(t) &= \cosh\left(\frac{\omega}{2} t\right) - \theta  \sinh\left(\frac{\omega}{2} t\right)  ,
\end{eqnarray}
and $\Gamma$, $\omega$, $\theta$ are constants depending on model parameters $(\lambda_{A}, \lambda_{B}, k_{AB}, k_{BA})$:
\begin{eqnarray}
	\Gamma &= \frac{1}{2} \left( (\lambda_{A}+\lambda_{B})-(k_{AB}+k_{BA})\right), \nonumber\\
	\omega &= \sqrt{(\lambda_{A}-\lambda_{B})^{2} + (k_{AB}+k_{BA})^{2} + 2 (\lambda_{A}-\lambda_{B}) (k_{BA}-k_{AB})}, \nonumber\\
	\theta &= \frac{1}{\omega} \left( (\lambda_{A}+k_{BA})-(\lambda_{B}+k_{AB})\right).
\end{eqnarray}
The mean numbers of cells in each state are given by
\begin{eqnarray}
\langle N_{A}(t) \rangle &= e^{\alpha t} \left[N_{A0}  \cosh(\beta  t) + \left( \frac{k_{BA}}{\beta} N_{B0} + \frac{\gamma}{2\beta} N_{A0}\right) \sinh(\beta  t)\right], \\
\langle N_{B}(t) \rangle &= e^{\alpha t} \left[N_{B0}  \cosh(\beta  t) + \left( \frac{k_{AB}}{\beta} N_{A0} - \frac{\gamma}{2\beta} N_{B0}\right) \sinh(\beta  t)\right],
\label{mABY}
\end{eqnarray}
where $\alpha$, $\beta$, $\gamma$ are constants depending on model parameters $(\lambda_{A}, \lambda_{B}, k_{AB}, k_{BA})$:
\begin{eqnarray}
	\alpha &= \frac{1}{2} \left( (\lambda_{A}+\lambda_{B}) - (k_{AB}+k_{BA}) \right), \\
	\beta &= \frac{1}{2}\sqrt{\gamma^{2}+4 k_{AB} k_{BA}}, \\
	\gamma &= (\lambda_{A}+k_{BA})-(\lambda_{B}+k_{AB})  .
	\label{mABY2}
\end{eqnarray}
Over long times, the population means grow approximately exponentially, as in Model C, but with a different rate $\alpha+\beta$. In contrast to Model C, however, the population means never decay since $\alpha + \beta \geq 0$ as can be easily shown. 
This is due to the absence of an absorbing state in Model U. 
It is worth mentioning that for any given parameter set $(\lambda_{A},\lambda_{B},k_{AB},k_{BA})$ in Model U, there exists a set of parameters $(\lambda_{AA},\lambda_{AB},\lambda_{BB})$ in Model C for which the asymptotic behaviour ($t\rightarrow \infty$) of the means $\langle N_{A} \rangle$, $\langle N_{B} \rangle$ is identical in both models (see Appendix A). 
Thus, we cannot use information about mean populations alone to distinguish Model U from Model C. 
The converse is not true, since $\langle N_{A} \rangle$ decays for some parameter regimes in Model C, which is impossible in Model U.
This is part of the rationale for appealing to Bayesian methods to infer model parameters and facilitate model comparison.

\subsubsection{Population dynamics of extended models C \& U}

As we showed in the last two sections, obtaining an exact solution of the probability generating function is already quite challenging for only two distinct cell states A and B. To obtain the probability generating function for the extended models including eight distinct cell states we would need to solve PDEs in 8 dimensions with 7 free parameters. Therefore we restrict the analysis of population dynamics in the extended models on obtaining approximate expressions of the mean populations $\langle \mathbf{\Phi} \rangle (t)$. Following the definition of allowed transitions in the extended models, on can derive a system of ODEs for the time evolution of moments $\langle \mathbf{\Phi} \rangle$ of the form \cite{schnoerr2017approximation}:
\begin{eqnarray}
	 \frac{d \langle \mathbf{\Phi} \rangle}{d t} = \mathbf{M} \cdot \langle \mathbf{\Phi} \rangle ,
	\label{mABYext}
\end{eqnarray}
where $\mathbf{M}$ is a $8 \times 8$ matrix which depends on the allowed transitions in the respective model (shown in Appendix B). The general solution of mean populations is a linear combination $\langle \mathbf{\Phi} \rangle (t) = \sum_{m} c_{m} \mathbf{v}_{m} e^{\lambda_{m} t}$ where $\lambda_{m}, \mathbf{v}_{m}$ are the m-th eigenvalue and eigenvector of matrix $\mathbf{M}$, respectively, and $c_{m}$ are constants depending on initial conditions.
\rev{For the extended models U and C, the matrix $\mathbf{M}$ is listed in the Supplementary Material, eqn \eqref{M_C} and \eqref{M_U}, and only has real eigenvalues.}
At late times $\lambda_{m} t \gg 1$, the term with the largest eigenvalue $\max(\{\lambda_{m}\})$ dominates the dynamics. For extended model U, the late time evolution according to the largest eigenvalue is given by
\begin{eqnarray}
	 \langle \mathbf{\Phi} \rangle(t) &\propto \mathbf{v}_{0} \cdot e^{\theta_{0} t}, \quad
	 \mathbf{v}_{0} &= \left( \begin{array}{c}
        \theta_{T-} \theta_{S-} \theta_{F-}\\\relax
        \theta_{T-} \theta_{S-} \theta_{F+}\\\relax
        \theta_{T-} \theta_{S+} \theta_{F-}\\\relax
        \theta_{T-} \theta_{S+} \theta_{F+}\\\relax
        \theta_{T+} \theta_{S-} \theta_{F-}\\\relax
        \theta_{T+} \theta_{S-} \theta_{F+}\\\relax
        \theta_{T+} \theta_{S+} \theta_{F-}\\\relax
        \theta_{T+} \theta_{S+} \theta_{F+}
\end{array}
\right).
	\label{mUlate1}
\end{eqnarray}
\rev{This result can be interpreted as follows: all cell states grow with the same rate $\theta_{0}$ and therefore the total population size also grows with rate $\theta_{0}$. The shape of $\mathbf{v}_{0}$ reflects that after sufficiently many division events, pairs of cell states are distributed according to their equilibrium distribution, $ \langle\Phi_{M+}\rangle/\langle\Phi_{M-} \rangle=\theta_{M+}/\theta_{M-}$, where $ \langle \Phi_{M\pm} \rangle$ is the mean population size of cells with gene $M\in\{T,S,F\}$ switched on and off, respectively.}
For extended Model C, we can find two kinds of approximations for the late time evolution of mean populations. If rates for switching genes on and off are very unbalanced, $\theta_{M+}\gg \theta_{M-}$ or $\theta_{M+}\ll \theta_{M-}$, it follows
\begin{eqnarray}
	 \langle \mathbf{\Phi} \rangle(t) &\propto \mathbf{e}_{i} \cdot e^{\theta_{0} t},
	\label{mUlate2}
\end{eqnarray}
where $\mathbf{e}_{i}$ equals one for those cell state favoured by the combination of imbalanced rates $\theta_{M+}\gg \theta_{M-}$ and zero for all other cell states. The effective growth rate of the total population equals $\theta_{0}$ and is therefore the same as in Model U. On the other hand, if rates for switching genes on and off are approximately equal, $\theta_{M+}\approx \theta_{M-}=\theta_{M}$, late time dynamics are given by
\begin{eqnarray}
	 \langle \mathbf{\Phi} \rangle(t) &\propto \mathbf{1} \cdot e^{(\theta_{0}+\theta_{T}+\theta_{S}+\theta_{F}) t} ,
	\label{mUlate3}
\end{eqnarray}
where $\mathbf{1}$ is a $7 \times 1$ all-ones vector. Cell states are therefore uniformly distributed and the overall population grows with effective growth rate $(\theta_{0}+\theta_{T}+\theta_{S}+\theta_{F})$. \rev{Differentiation events can therefore trigger increased population growth in extended model C. Populations grow with an increased rate $\Tilde{\theta_0}$, $\theta_0 \leq \Tilde{\theta_0} \leq (\theta_0 + \theta_T + \theta_F + \theta_S)$, which exact value  depends on the rate imbalance $\theta_{M+}/\theta_{M-}$ of the marker switching rates.}

\subsubsection{Summary}

In this section we derived exact solutions of the probability generating function for both Model C and Model U. This allows an efficient and systematic way of calculating moments of the population distribution. We explicitly derived expressions for the time evolution of mean populations in both models. \rev{Neglecting cases for which A-state populations decrease in Model C,} mean populations grow exponentially over long time scales in both models with rates depending on the underlying model parameters. These expressions will be used in the next section for deriving an efficient algorithm for reaction rate inference (see~\ref{Appendix_reactiontimes}) and for proposing experimental measurement procedures which optimise parameter inference (see Section~\ref{qual_parinf}). We showed that the extended stem cell models follow similar growth dynamics and presented approximate expressions for mean population at late times.

\subsection{Bayesian inference on minimal models}

In the previous section we analysed the time-evolution of the distribution over cell states $\boldsymbol{N}(t)$ as a function of model parameters $\mathbf{\theta}$. Now we will look at the inverse problem: obtaining the distribution over model parameters $\mathbf{\theta}$ given population data $\boldsymbol{N}$. In the first section we consider parameter inference given complete knowledge about population sizes $\boldsymbol{N}(t)$ at all times. In the second section we generalise parameter inference to cases where we have incomplete data and population sizes $\boldsymbol{N}(t_{i})$ are only known at certain time points $\{t_{i}\}$.

\subsubsection{Longer observations decrease uncertainty exponentially on complete data\label{completeData}}

We first suppose that we have complete knowledge of how the population sizes $\boldsymbol{N}(t)$ have changed over time. 
In other words, each cell division and differentiation event is evident in the trajectory $\boldsymbol{N}(t)$. 
Such data could, for example, come from live-imaging experiments in which each cell state can be identified and the full lineage be tracked. Since each reaction as defined by Eq.~\eqref{hazard_X}, \eqref{hazard_Y} leads to a unique change in the number of cells in each state, we can infer the type of an occurred reaction from changes in $\boldsymbol{N}(t)$. 
\rev{We define $\tau_i$, $\nu_i$ and $h_{\nu_i}(\tau_i)$ as the reaction times, the corresponding reaction types and the transition rates, respectively.}
We define the combined transition rate $h_0(t)$ as the sum over transition rates for all possible reactions that can take place at time $t$ (which will depend on the state of the system at that time). 
Since each reaction event is modelled as a Poisson process with corresponding rate $h_{\nu_i}$, the likelihood of a sequence of $n$ reactions is given by~\cite{wilkinson2011stochastic}
\rev{
\begin{equation}
L_{\rm comp}(\xi|\boldsymbol{\theta}) =  \left[\prod_{i=1}^{n} h_{\nu_i}(\tau_{i-1})\right] \cdot e^{-\sum_{i=0}^{n} h_{0}(\tau_{i})\cdot(\tau_{i+1}-\tau_{i})} ,
\end{equation}
}
where the subscript ${\rm comp}$ denotes complete data and $\tau_{0}=0$, $\tau_{n+1}=T$. Now, in the models we consider here, each  transition rate $h_{\nu_i}$ is proportional to a per-capita rate $\theta_{\nu_i}$ multiplied by the number of cells in state $\Phi_{\nu}$, which is the cell state associated with a reaction of type $\nu$. 
If we define now $N_{\nu}(t)$ as the total number of cells in state $\Phi_\nu$ at time $t$, we have that
\begin{equation}
\label{LCD}
L_{\rm comp}(\xi|\boldsymbol{\theta}) \propto \prod_\nu \left[ \theta_{\nu}^{r_\nu} \exp\left( - \theta_\nu \int_0^T N_{\nu}(t) dt \right) \right], 
\end{equation}
where the product is over each type of reaction $\nu$ (i.e. division or differentiation of each cell state) and $r_{\nu}$ is the number of times that reaction $\nu$ occurs during the observation window $0 \leq t \leq T$.

Note that what matters for Bayesian inference is the dependence of the likelihood on the parameters $\boldsymbol{\theta}$. Therefore factors that do not depend on these parameters (such as the population sizes $N_{\nu}$ that multiply each of the transition rates $\theta_k$) can be subsumed into the constant of proportionality in \eqref{LCD}.

A key feature of \eqref{LCD} is that it factorises into terms that each depend on a different reaction rate $\theta_\nu$. 
This means that each parameter value can be inferred independently, assuming that parameter priors are independent: $p(\boldsymbol{\theta})=\prod_{\nu} p(\theta_\nu)$. 
That is, the posterior distribution of $\theta_\nu$ is
\begin{equation}
\label{ptncd}
p(\theta_\nu|\xi) \propto \theta_{\nu}^{r_\nu} \exp\left( - \theta_\nu \int_0^T N_{\nu}(t) dt \right) p(\theta_\nu),
\end{equation}
where $p(\theta_\nu)$ is the prior distribution over the reaction rate $\theta_\nu$. As before, the constant of proportionality can be obtained by normalising the posterior distribution $p(\theta_\nu|\xi)$. 
It is convenient for the prior distribution to be taken conjugate to the likelihood, which means that the product of prior and likelihood is of the same functional form as the prior. 
Here, the likelihood is a gamma distribution, and therefore the conjugate prior is also a gamma distribution~\cite{moscuoroums1985distrlbution}. 
If the prior is a gamma distribution with shape $a_\nu$ and rate $b_\nu$,
then the posterior is a gamma distribution with shape $a'_{\nu} = a_{\nu} + r_{\nu}$ and rate $b'_{\nu} = b_\nu + \int_0^T N_{\Phi_\nu}(t) dt$, with mean $\langle \theta_\nu |\xi\rangle = a'_{\nu} / b'_{\nu}$ and variance ${\rm Var}(\theta_\nu|\xi) = a'_{\nu} / b'^{\,2}_{\nu}$~\cite{kotz2004continuous}. 

The confidence of parameter estimation can be quantified by the dispersion of the posterior distribution $p(\theta_{\nu}|\xi)$, measured by the \emph{coefficient of variation} $c_{v}$ (COV)
\begin{equation}
	c_{v}(\theta_{\nu}|\xi) = \frac{\sqrt{{\rm Var}(\theta_{\nu}|\xi)}}{\langle \theta_{\nu}|\xi \rangle} = \frac{1}{\sqrt{a_{\nu}+r_{\nu}}}  .
\end{equation}
For experimental design it is essential to estimate how much data is needed to achieve parameter inference of a certain precision. 
More precisely, if we observe $n$ identical systems over a time period $T$, we would like to know how the COV scales on average with sample size $n$ and observation time $T$. 
As shown in Appendix C, the asymptotic scaling of the mean COV with observation time $T$ in Model C is given by
\begin{eqnarray}
		\langle c_{v} \rangle(\theta_{\nu}|\xi_{1},\ldots,\xi_{n}) \propto
		\left\{
		\begin{array}{ll}
		\frac{\exp\left(-\frac{\Delta \lambda}{2}  T\right)}{\sqrt{N_{A0}  n}} \quad & \Delta \lambda >0, \\[10pt]
		\frac{1}{\sqrt{N_{A0}  n}} \quad & \Delta \lambda < 0,
		\end{array} 
		\right.
		\label{X_cv_T}
\end{eqnarray}
where $\Delta\lambda = \lambda_{AA}-\lambda_{BB}$. For Model U, we obtain
\begin{eqnarray}
\langle c_{v} \rangle(\theta_{\nu}|\xi_{1},\ldots,\xi_{n}) \propto
\frac{\exp\left(-\frac{(\alpha+\beta)}{2}  T\right)}{n}, 
\label{Y_cv_T}
\end{eqnarray}
where $\alpha$ and $\beta$ are functions of rate constants as given by  \eqref{mABY2}. Note that the mean COV of $\theta_{\nu}$ depends on the rate constant $\theta_{\nu}$ itself. This result can be useful for experimental design if a rough estimate for the magnitude of $\theta_{\nu}$ is available. If not, one has to rely only on the scaling with sample size $n$. 

To summarise, if we have a complete history of all the population sizes $N_\nu(t)$ over time, we can infer the reaction rate $\theta_{\nu}$ by counting how many times the reaction $\nu$ occurs, and by integrating the size of the population that drives each reaction, $N_{\nu}(t)$, over time $t$. Scaling laws for the confidence of parameter estimation can be obtained as a function of sample size $n$ and observation time $T$. The uncertainty of the parameter estimation decreases exponentially with observation time, but only decays as a power law with sample size. Besides parameter inference, the aim of experiments is often to compare competing models based on data $\xi$. In the complete-data scenario, reaction types can be directly inferred from changes in population. Hence, if reactions are observed which are impossible in either Model C or Model U, we can rule out the respective model and model selection is trivial. This will be very different in scenarios where we only have access to snapshot data, as we now show.

\subsubsection{Efficient sampling enables inference for analytically intractable snapshot data}

The results of the previous section depend on having complete knowledge of population sizes over time. It is unlikely that this level of detail will be available in practice. More likely are a set of \emph{snapshots}, that is, measurements of $\boldsymbol{N}$ at a finite set of times $t_0=0, t_1, t_2, \ldots, t_n$.

In the previous section we obtained an expression for the posterior distribution $p(\boldsymbol{\theta}|\xi)$ given a completely-specified trajectory $\xi$. To obtain the corresponding quantity from snapshot data, we integrate over all trajectories $\xi$ that are consistent with the observations $\{\boldsymbol{N}(t_i)\}$, weighted by the probability $p(\xi|\boldsymbol{N}(t_i))$ which is conditioned on passing through the observation points. That is,
\begin{equation}
p(\boldsymbol{\theta}|\{\boldsymbol{N}(t_i)\}) = \int D\xi\, p(\boldsymbol{\theta}|\xi) p(\xi|\{\boldsymbol{N}(t_i)\})  \;.
\end{equation}
Since the underlying model is Markovian, we have that
\begin{equation}
p(\xi|\boldsymbol{N}(t_i)) = \prod_{i=1}^{n} p(\xi_i | \boldsymbol{N}(t_{i-1}), \boldsymbol{N}(t_i)),
\end{equation}
where $\xi_i$ is the portion of the trajectory corresponding to the time interval $[t_{i-1},t_i]$. 

\begin{figure}
	\centering
	\includegraphics[width=0.9\textwidth]{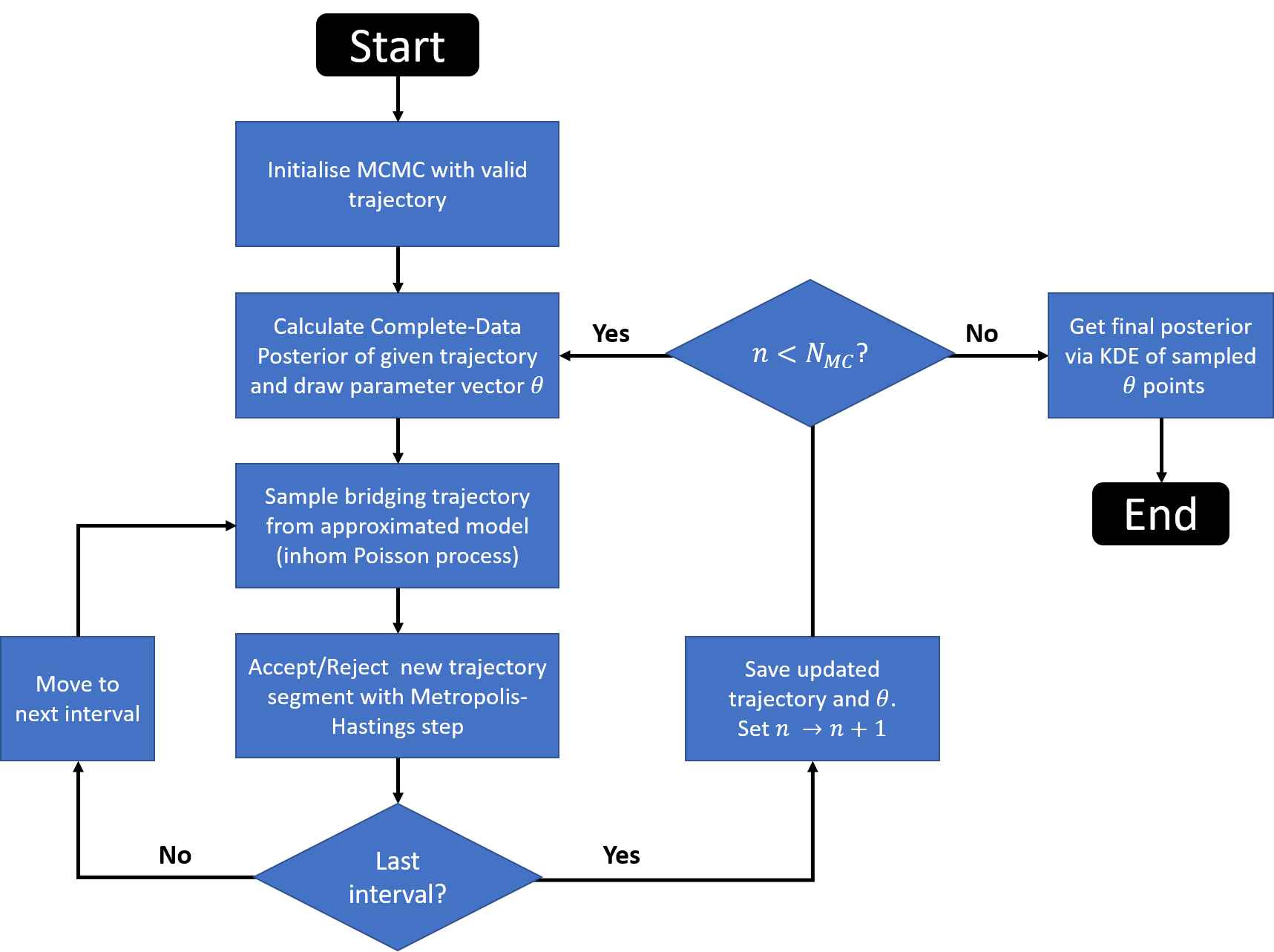}
	\caption[MCMC algorithm for posterior sampling]{Sample path Markov chain Monte Carlo (MCMC) algorithm for sampling $N_{MC}$ reaction rate values $\boldsymbol{\theta}$ from the posterior $p(\boldsymbol{\theta}|\{\boldsymbol{N}(t_{i})\})$, given snapshot-data $\{\boldsymbol{N}(t_{i})\}$. Algorithm motivated by~\cite{wilkinson2011stochastic}.} 
	\label{fig_algo}
\end{figure}

This observation implies that the posterior distribution can be sampled using the following algorithm (see also Figure~\ref{fig_algo}).

\begin{algorithm}[H]
	\caption{Sample path MCMC for parameter inference}
	\begin{algorithmic}[1]
		\STATE Initialise algorithm with valid trajectory segments $\{ \xi_i \}$, a parameter set $\boldsymbol{\theta}$ and set $i=1$.
		\STATE Propose trajectory segment $\xi_i^{\star}$ between time points $t_{i-1}$ and $t_i$ with fixed endpoints $\boldsymbol{N}(t_{i-1})$ and $\boldsymbol{N}(t_i)$.
		\STATE Metropolis-Hastings step: replace original segment $\xi_i$ by $\xi_i^{\star}$ with acceptance probability $\min(1,A_i)$, with acceptance ratio $A_i$ from eqn \eqref{acc-prob}.
		Otherwise keep original segment $\xi_i$.
		\STATE If $i<n$, set $i=i+1$ and go to (2), otherwise go to (5)
		\STATE Form an updated complete trajectory $\xi$ by joining the updated segments $\xi_i$.
		\STATE Sample a new parameter set $\boldsymbol{\theta}$ from the complete-data posterior $p\left(\boldsymbol{\theta}|\xi\right)$.
		\STATE Output the current $\boldsymbol{\theta}$, set $i=1$ and go back to (2).
	\end{algorithmic}
\end{algorithm}

Previously we used the Gillespie algorithm to sample random trajectories with a fixed starting point. However, to propose new trajectory segments $\xi_i^\star$ in step 2 of the algorithm, we need to sample a random trajectory where both end points are fixed. This is a much harder problem and we will use an approach based on the Metropolis-Hastings-Green algorithm~\cite{wilkinson2011stochastic, brooks2011handbook} to achieve this. The basic idea is to draw $\xi_i^\star$ from a distribution that approximates the true distribution, and adjust the acceptance ratio $A$ accordingly.

To construct the approximate distribution, we make use of the exact results for the mean population sizes that were derived in Sections~\ref{exact_X} and \ref{exact_Y}. We found that typically these grow or decline exponentially. Therefore we consider a set of reactions $\nu$ whose transition rates are \emph{deterministic} functions of time,
\begin{equation}
	h_{\nu}^{*}(t) =  N_{\nu}(0) \theta_{\nu} \left(\frac{N_{\nu}(T)}{N_{\nu}(0)}\right)^{t/T}, 
	\label{exp_hazard}
\end{equation} 
where $t$ is the time from the start of the trajectory segment, and $T$ is the length of the trajectory segment. The simplification here is that the probability a reaction occurs at time $0\le t\le T$ is not affected by the sequence of reactions that has occurred up to time $t$, as would be the case for the true distribution. This amounts to sampling $r_{\nu}$ reactions of type $\nu$ from an inhomogeneous Poisson process with the time-dependent transition rates $h_\nu^*$. The procedure is to first draw the numbers of reactions $r_\nu$, and then sample the reaction times from an inhomogeneous Poisson process. To account for the approximation in sampling the trajectory, the acceptance ratio in the Sample Path MCMC algorithm must be chosen accordingly, which is shown in detail in Appendix D.

\subsubsection{Optimal snapshot timing can improve parameter inference} \label{qual_parinf}

We now apply the Sample Path MCMC algorithm to understand how well it performs given specific snapshots. 
Recall that this algorithm generates a sequence of rate-constant vectors, $\boldsymbol{\theta}$, which (when converged) are distributed according to the correct posterior distribution $p(\boldsymbol{\theta}|\{\boldsymbol{N}(t_{i})\})$.

\begin{figure}
	\centering
	\subcaptionbox{$\lambda_{AA}$-$\lambda_{AB}$-space}{\includegraphics[width=0.45 \textwidth]{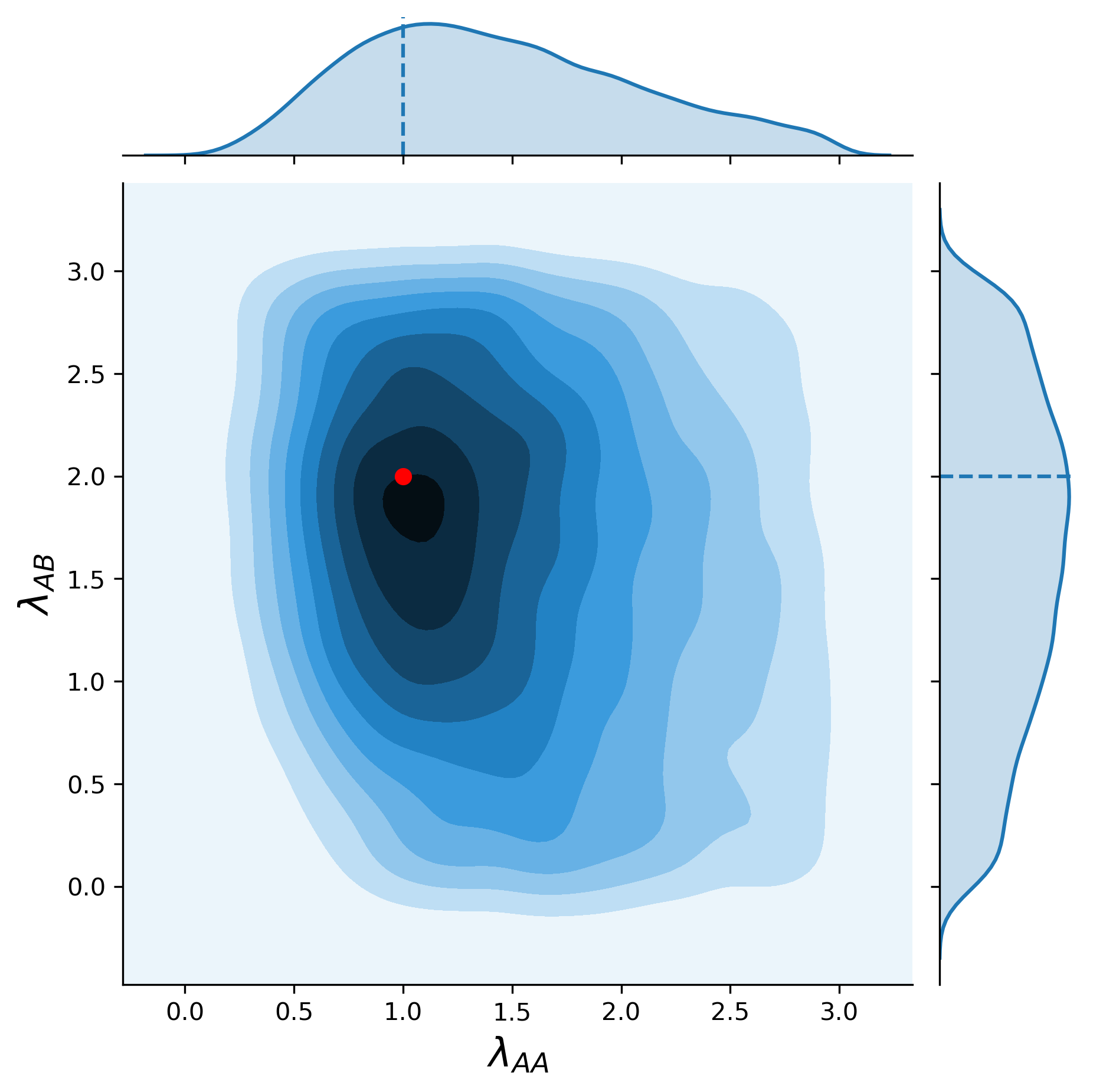}}
	\quad
	\subcaptionbox{$\lambda_{AB}$-$\lambda_{BB}$-space}{\includegraphics[width=0.45 \textwidth]{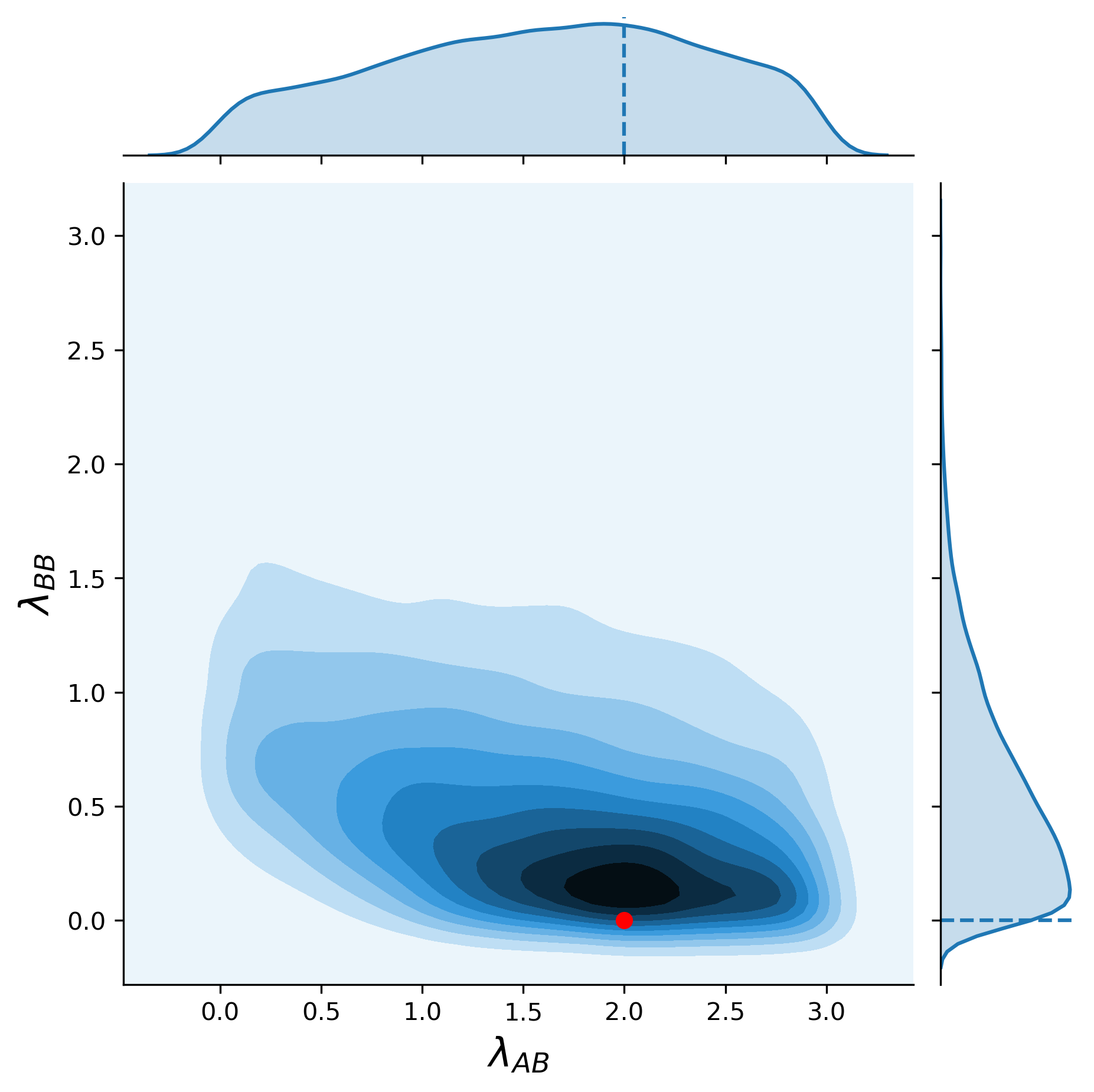}}
	\caption[Kernel density estimation of MCMC output]{\rev{Posterior distribution of Model C calculated based on five independent snapshots $\{\boldsymbol{N}(t=2.0)\}$ which were sampled from a system with parameters $\boldsymbol{\theta}=(1,2,0)$ with initial conditions $\boldsymbol{N}(t=0)=(1,1)$,  and uniform, independent priors $\theta_{\nu} \mathrel{\stackrel{\makebox[0pt]{\mbox{\normalfont\tiny iid}}}{\sim}} \mathcal{U}[0,3]$. The three-dimensional posterior for Model C is projected onto the two-dimensional planes with the corresponding marginal probability distributions at upper and right margins, shown as the KDE of MCMC samples. The true rate constants from which the two snapshot data were created is shown as a red dot and dashed, blue lines. The non-zero posterior density for negative rate constants is due to the KDE visualisation and does not represent actual points sampled by the MCMC. Because the population dynamics in model C are less sensitive to changes in the asymmetric division rate $\lambda_{AB}$ than to variations in $\lambda_{AA}$ and $\lambda_{BB}$, we observe a larger posterior variance along $\lambda_{AB}$.
}}
	\label{fig_KDE}
\end{figure}

Our first test is a single run of Model C, in which the initial condition is $\boldsymbol{N}(0) = (1,1)$ and after time $t=1.5$ has evolved to $\boldsymbol{N}(1.5)=(3,7)$ with reaction rates $\boldsymbol{\theta} = (\lambda_{AA},\lambda_{AB},\lambda_{BB}) = (1,0,1)$. 
With a large amount of data, one would expect the posterior distribution to be peaked around this point. 
\rev{Two-dimensional projections of the three-dimensional posterior distributions obtained via kernel density estimation (KDE) are shown in Figure~\ref{fig_KDE} and Figure~\ref{fig_KDE2}, and shows that the parameter estimates are reasonably well constrained despite the small amount of data (2 snapshots) used for the inference.} 
Throughout, we use Scott's Rule~\cite{scott2015multivariate} for KDE bandwidth estimation. 
Unlike the complete-data scenario (Section \ref{completeData}), the likelihood (and so the posterior) \textit{does not} factorise into independent component likelihoods (cf.~Eq.~\eqref{ptncd}). 
This means given snapshot-data, rate constants are correlated and rate inference \textit{cannot} be carried out separately. 
The alignment of the high-density regions in Figure~\ref{fig_KDE} along the diagonal clearly indicate correlations between $\lambda_{AA}$ and $\lambda_{BB}$.

We now investigate how the parameter estimates are affected by the length of time, $\Delta t$, between the two snapshots. 
We anticipate that when $\Delta t$ is small, only a small number of reactions could reasonably have occurred, and so one would not expect to lose much from having only the trajectory endpoints (as opposed to the full trajectory). 
On the other hand, when $\Delta t$ is large, we would expect estimates based on snapshots to be less tightly constrained than those based on full trajectories.

\begin{figure}
	\centering
	\subcaptionbox{$\Delta t = 0.5$}{\includegraphics[width=0.4 \textwidth]{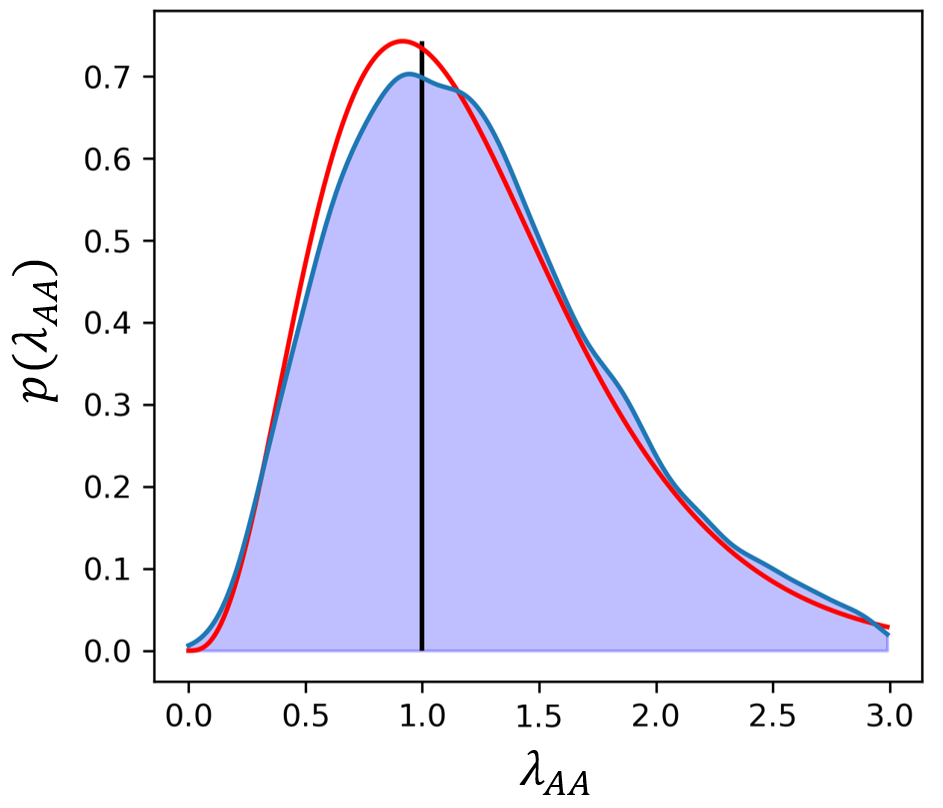}}
	\quad
	\subcaptionbox{$\Delta t = 1.5$}{\includegraphics[width=0.4 \textwidth]{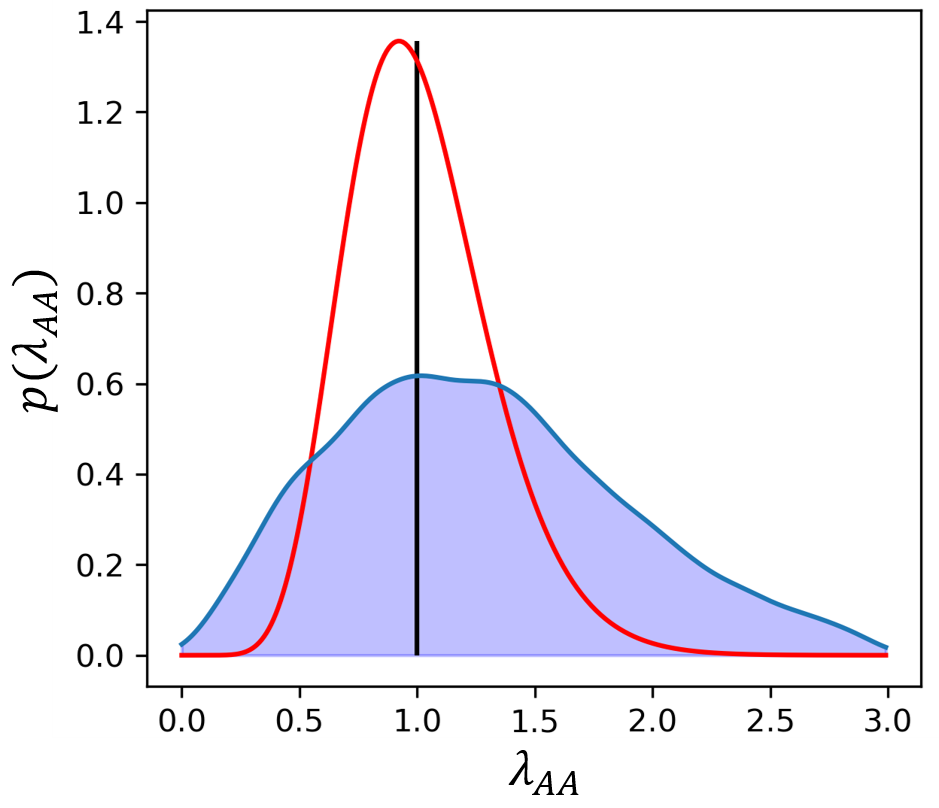}}
	\caption[Complete-data posterior (red curve) given a trajectory sampled from Model C with parameters $\boldsymbol{\theta}=(1,1,1)$]{Complete-data posterior (red curve) given a trajectory sampled from Model C with parameters $\boldsymbol{\theta}=(1,1,1)$ (black, vertical lines), $\boldsymbol{N}_{0}=(1,1)$ on time interval $[0,\Delta t]$. The snapshot posterior (blue, filled curve) is calculated based on population snapshots of the underlying trajectory at times $t_{0}=0$ and $t_{1}=\Delta t$. Snapshot posteriors are less accurate than their complete-data counterpart, especially for large snapshot separation $\Delta t$. In the simulation we used uniform priors for rate constants $\theta_{k}\stackrel{iid}{\sim}\mathcal{U}[0,3]$.} 
	\label{fig_var_dt}
\end{figure}

Our procedure is to generate a trajectory from Model C of length $\Delta t$, starting from $\boldsymbol{N}(0) = (1,1)$ and with $\boldsymbol{\theta} = (\lambda_{AA},\lambda_{AB},\lambda_{BB}) = (1,1,1)$. 
We then perform the parameter inference with both the full trajectory, and with the two endpoints. 
The resulting posterior distributions for the parameter $\lambda_{AA}$ for different $\Delta t$ are shown in Figure~\ref{fig_var_dt}.

\begin{figure}
	\centering
	\includegraphics[width=0.6 \textwidth]{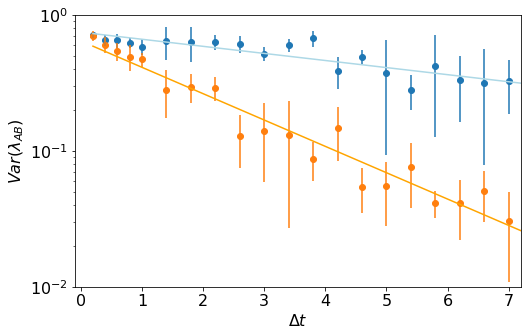}
	\caption{Variance of complete-data posterior (orange) and snapshot posterior (blue) as a function of snapshot separation $\Delta t$. Trajectories were sampled from Model C with parameters $\boldsymbol{\theta}=(1,1,1)$, $\boldsymbol{N}_{0}=(1,1)$ on time interval $[0,\Delta t]$. Error bars show the spread of the posterior variance calculated from 8 independently sampled trajectories, while $\Delta t$ was kept constant. The variance of the snapshot posterior decays at a much smaller rate with observation time than the complete-data posterior (solid lines show exponential fits).} 
	\label{fig_var_scaling_dt}
\end{figure}

The snapshot posteriors (blue, filled curves) are always less accurate with respect to the true parameters than their complete-data counterpart (red lines), as one would expect. 
As more reaction events are contained in the full trajectory $\xi$ for longer times, it provides more information about the system and the variance of the complete-data posterior decreases. 
The \textit{marginal information} stored in snapshot-data increases too with reaction numbers, however at a much smaller rate than the full trajectory $\xi$ (see Figure~\ref{fig_var_scaling_dt}).

Suppose now we have the ability to take more than two snapshots, and have freedom to choose when they are taken.
Is it more informative to space them out evenly, or to cluster them together? 
Given that the total rate at which reactions occur is proportional to the number of cells in the population, we expect to learn more from closely-spaced observations when the population size is large than when it is small.

\begin{figure}
	\centering
	\includegraphics[width=0.6 \textwidth]{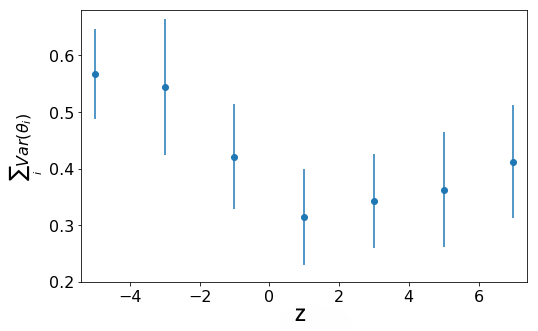}
	\caption{Trace of covariance matrix of snapshot posterior as a function of snapshot spacing parameter $z$. Trajectories were sampled from Model C with parameters $\boldsymbol{\theta}=(1,1,0)$, $\boldsymbol{N}_{0}=(1,1)$ on time interval $[0,3]$ and $n=7$ snapshots where taken according to equation~\eqref{SS_space_X} with spacing parameter $z$. Error bars show the spread of the posterior variance calculated from eight independently sampled trajectories. As predicted, the posterior variance is on average minimised for $z\approx \Delta \lambda =1$.} 
	\label{fig_var_scaling_multi}
\end{figure}

\begin{figure}
	\centering
	\subcaptionbox{Uniform snapshot spacing ($z=0$)}{\includegraphics[width=0.8 \textwidth]{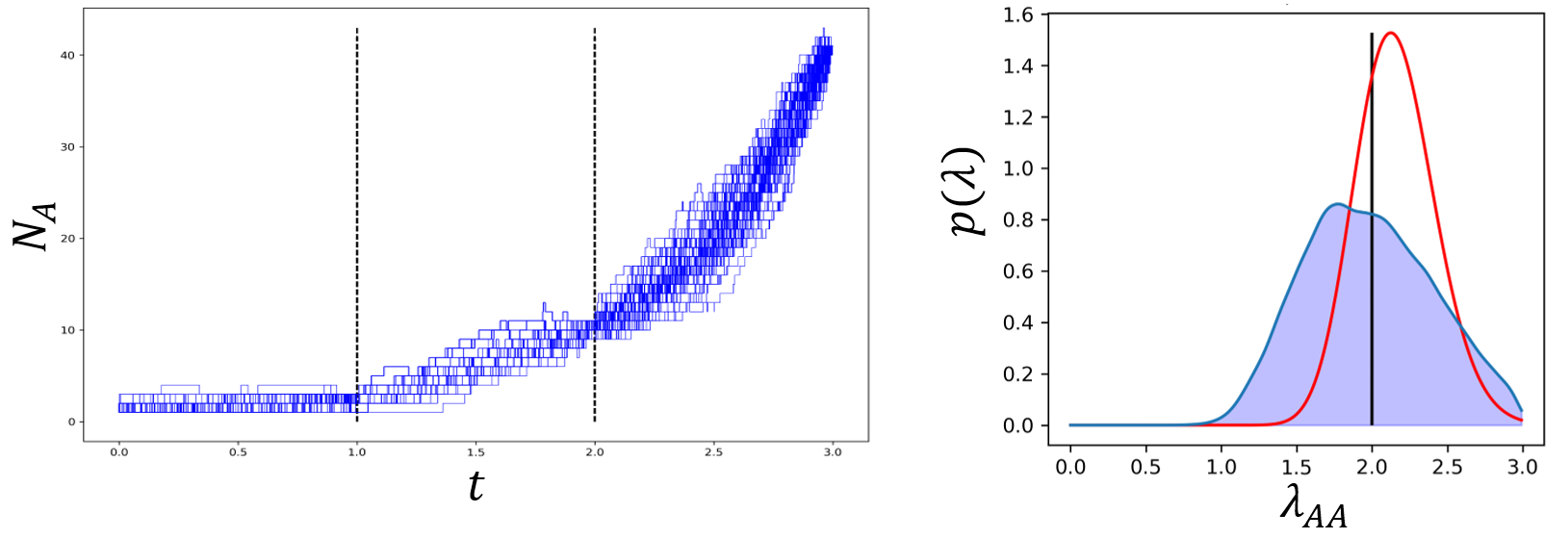}}
	\quad
	\subcaptionbox{Optimal snapshot spacing ($z=\Delta \lambda = 1$)}{\includegraphics[width=0.8 \textwidth]{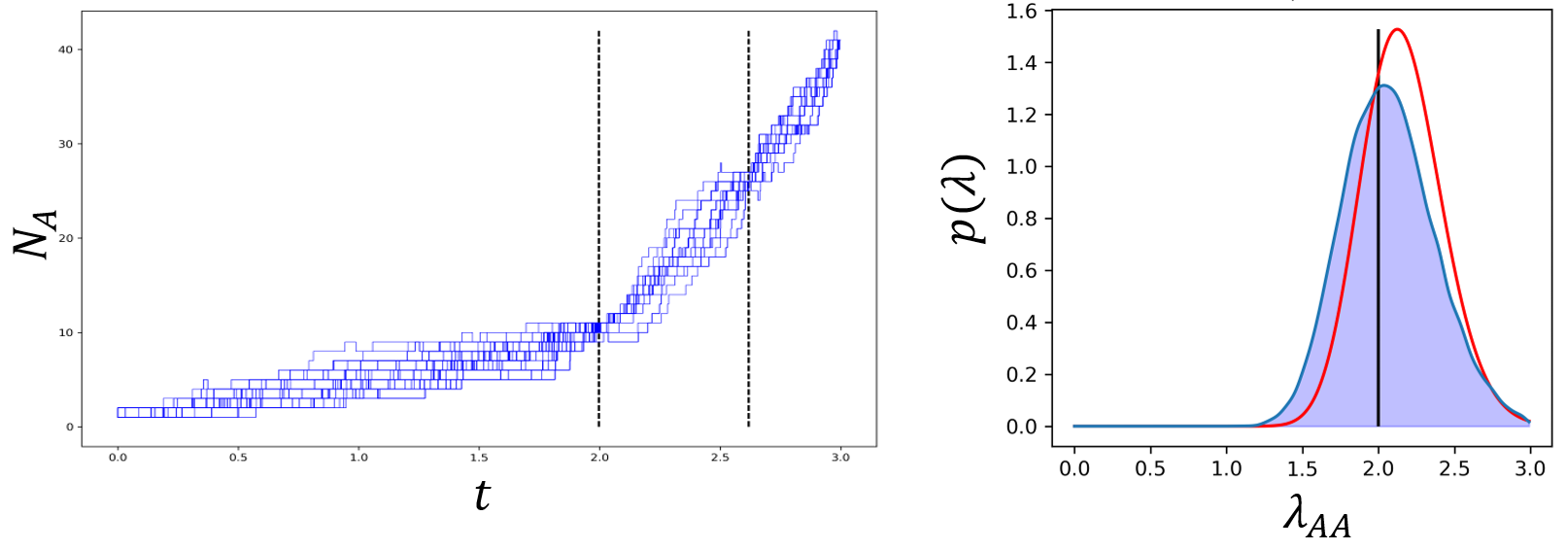}}
	\caption[Uniform vs. optimal snapshot sampling]{Left panels: sampled bridging trajectories between population snapshots (vertical, dashed line) obtained via the sample path algorithm (see Figure~\ref{fig_algo}). Right panels: complete-data posterior (red curve) given a trajectory sampled from Model C with parameters $\boldsymbol{\theta}=(2,1,1)$ (black, vertical line), $\boldsymbol{N}_{0}=(1,1)$ on time interval $[0,3]$. The snapshot posterior (blue, filled curve) is calculated based on $n=5$ population snapshots which where taken according to equation~\eqref{SS_space_X} with spacing parameter $z$.} 
	\label{fig_var_spacing}
\end{figure}

More precisely, we can attempt to maximise the information content of multiple snapshots by keeping the mean number of reactions in between them constant. 
For example, in Model C we know that the total reaction rate is $N_A(t) ( \lambda_{AA} + \lambda_{AB} + \lambda_{BB} )$ and from equation~\eqref{mABX} that the mean number of A-state cells grows exponentially as $N_{A0} e^{\Delta\lambda  t}$, where $\Delta \lambda = \lambda_{AA} - \lambda_{BB}$.  We can therefore deduce that the mean number of reactions between times $t_{i-1}$ and $t_i$ is
\begin{equation}
	\langle r_{0} \rangle_{i} = N_{A0} \frac{\lambda_{AA}+\lambda_{AB}+\lambda_{BB}}{\Delta \lambda} \left( e^{\Delta \lambda t_{i}} - e^{\Delta \lambda t_{i-1}}\right).
\end{equation}
Setting this number equal for all $i$, with the constraint that $t_0=0$ and $t_n=T$, we find that the $i^{\rm th}$ snapshot should be taken at time
\begin{equation}
	t_{i} = \frac{1}{\Delta\lambda} \ln\left[1+\frac{i}{n} \left(e^{\Delta\lambda  T}-1\right)\right]  \; ,
	\label{SS_space_X}
\end{equation}
where $n$ is the total number of snapshots taken in the time interval. 
\rev{We can compare the variance of posterior distributions obtained from differently spaced snapshots by replacing $\Delta \lambda$ with a variation parameter $z$ in equation~\eqref{SS_space_X}. For $z<0$ snapshots are concentrated towards the start of the observation window, $z=0$ represents uniformly spaced snapshots and $z>0$ leads to a higher snapshot density towards the end of the observation window (see Figure~\ref{fig_var_spacing} for example).} In Figure~\ref{fig_var_scaling_multi} we compare the variance of posterior distributions obtained from differently spaced snapshot times over a range $-5 \leq z \leq 7$. We see that parameter estimates are most precise if the average number of reactions between snapshots are kept constant ($z=\Delta \lambda$), as proposed. 
This confirms our intuition that one is likely to be best served by sampling larger populations more frequently. 
A similar analysis is possible for Model U, which yields the same optimal snapshot times as in Eq.~\eqref{SS_space_X}, but $\Delta \lambda$ is replaced by $\alpha+\beta$ from Eq.~\eqref{mABY2}.

\subsection{Application to epiblast stem cell data}

Having established that parameter inference and model selection is viable with relatively limited amounts of data in the context of Models C and U with two cell states, we turn now to the more challenging case of epiblast stem cell data with eight states as described in Section~\ref{epiblast_models}. This may elucidate whether cell division and state changes are independent or coupled in this particular biological system.

\subsubsection{Data structure}

The structure of our dataset is largely influenced by measurement restrictions of the experimental setup (Schumacher et al., in preparation). The experiment starts with single cells of only partially known state: prior to the experiments cells have been sorted for the expression of marker $T^{\pm}$. From previous experiments measuring steady-state populations of $T^{\pm}$-sorted stem cells~\cite{tsakiridis2014distinct} we can calculate the distribution of marker expressions for $F,S$ as $p(F^{+}|T^{+}) \approx 0.49$, $p(F^{+}|T^{-}) \approx 0.26$, $p(S^{+}|T^{+}) \approx 0.75$, $p(S^{+}|T^{-}) \approx 0.46$, with complementary probabilities $p(F^{-}|T)=1-p(F^{+}|T)$ and $p(S^{-}|T)=1-p(S^{+}|T)$, and marker expressions are assumed to be independent, so that $p(F,S|T) = p(F|T) p(S|T)$.

After time $\Delta t$, individuals cells will have grown into small colonies (clones). These are fixed, the number of cells in each clone counted and the expression of either $T,F$ or $T,S$ are measured. We can therefore only obtain information about the sum of underlying populations, where the sum is taken over the unknown marker (either $S$ or $F$) in a given measurement. As an example: the number of cells possessing marker expression $[T^{+},F^{+}]$ is given by the sum of cell populations with marker expression $[T^{+},F^{+},S^{-}]$ and $[T^{+},F^{+},S^{+}]$: $N_{[T+,F+]}=N_{[T+,F+,S-]}+N_{[T+,F+,S+]}$. In general,
\begin{equation}
N_{[M1,M2]} = \sum_{M3} N_{[M1,M2,M3]},
\end{equation}
where $M1$, $M2$ are measured marker expressions and $M3$ is the unknown marker. To compare snapshot data obtained from simulations to experimental data, we first have to calculate the \textit{projected populations} $\{N_{[M1,M2]}\}:=\mathcal{D}^{sim}$ before comparing it to measured populations from experiments. Population snapshot data was experimentally recorded from cell colonies subject to two different environmental conditions: CHIR ($\mathcal{D}_\mathrm{CHIR}$), which refers to the presence of the differentiation factor Chrion, and EPISC ($\mathcal{D}_\mathrm{EPISC}$), which refers to the absence of this factor~\cite{tsakiridis2014distinct}. The first data set $\mathcal{D}_\mathrm{CHIR}$ consists of eight measurement series recorded at different times and for marker expressions:
\begin{enumerate}
	\item Initial marker $T^{+}$, marker measured after $\Delta t=2 d$ and $\Delta t=3d$: $T,F$; 
	\item Initial marker $T^{+}$, marker measured after $\Delta t=2 d$ and $\Delta t=3 d$: $T,S$;
	\item Initial marker $T^{-}$, marker measured after $\Delta t=2 d$ and $\Delta t=3 d$: $T,F$; 
	\item Initial marker $T^{-}$, marker measured after $\Delta t=2 d$ and $\Delta t=3 d$: $T,S$.
\end{enumerate}
The second data set $\mathcal{D}_\mathrm{EPISC}$ consists of four measurement series recorded at different times and for marker expressions:
\begin{enumerate}
	\item Initial marker $T^{+}$, marker measured after $\Delta t=3 d$: $T,F$; 
	\item Initial marker $T^{+}$, marker measured after $\Delta t=3 d$: $T,S$;
	\item Initial marker $T^{-}$, marker measured after $\Delta t=3 d$: $T,F$; 
	\item Initial marker $T^{-}$, marker measured after $\Delta t=3 d$: $T,S$.
\end{enumerate}
Each measurement series in both sets contains between $98$ and $142$ observations of projected population snapshots.

\subsubsection{Inference using ABC}

The main challenge in applying the MCMC algorithm developed above to the EpiSC cell state network is the high dimensionality of the state space. The state space of our system is given by population numbers $\{N_{\Phi}\}$ of all $8$ cell states $\Phi_{i}$. We need to evaluate snapshot likelihoods of the form $L_{SS}(\{N_{\Phi}(\Delta t)\}|\boldsymbol{\theta})$ by an MC estimate, so a sufficiently large ensemble of trajectories is necessary to populate most parts of state space $\{N_{\Phi}\}$. The experimental data $\mathcal{D}_\mathrm{CHIR}$ and $\mathcal{D}_\mathrm{EPISC}$ contain up to $8$ cells per cell state and therefore the MC estimate would need to sample up to $8^{8} \approx 2\cdot10^{7}$ possible system states. Unfortunately, sampling a sufficiently large ensemble from $\sim 10^{7}$ system states turned out to be computationally not feasible. We therefore turn to \textit{Approximate Bayesian Computation} (ABC)~\cite{toni2008approximate}. In our case, ABC speeds up calculations by several order of magnitudes by comparing data based on lower-dimensional \textit{summary statistics} only, at the expense of introducing an additional error by approximating posterior distributions. 

We choose the following summary statistics: the mean $\langle \boldsymbol{N}_{\Phi} \rangle$, median ${\rm med}\{\boldsymbol{N}_{\Phi}\}$, and standard deviation $\boldsymbol{\sigma}_{\Phi}$. We use a Sequential Monte Carlo algorithm (SMC-ABC) for parameter inference and model selection \cite{sisson2007sequential, klinger2018pyabc}. We verify our results also using a Rejection-ABC algorithm (modified from~\cite{toni2008approximate}) for parameter inference and model selection, which is presented in Appendix E.

In ABC, a distance function $D(s(\mathcal{D}^{sim}),s(\mathcal{D}))$ classifies how ``close'' the summary statistic of observed and simulated data are. A tolerance $\epsilon\geq 0$ controls the level of agreement between both summary statistics and $\epsilon>0$ introduces the eponymous approximation in ABC. 
We choose the following summary statistic for a measurement series $d$: $\boldsymbol{s}_{i}(d)=[\langle N_{i} \rangle, {\rm Var}(N_{i}), {\rm Med}(N_{i})]^{T}$, where $i \in \{1,2,3,4\}$ is indexing the projected populations (expression of either T and F or T and S, starting from either T+ or T- cells).
This summary statistic captures the centre, spread, and skewness of the respective distribution $p(N_{i})$. \rev{We use the sample median instead of the skewness as it can be calculated more easily than the third central moments needed for skewness estimation.}
We use the following norm $D(s(\mathcal{D}^{sim}),s(\mathcal{D}))$ to compare summary statistics of experimental and simulated data:
\begin{equation}
    D(s(\mathcal{D}^{sim}),s(\mathcal{D})) = \sum_{k} \sum_{i=1}^{4} ||\boldsymbol{s}_{i}(d_{k}^{sim})-\boldsymbol{s}_{i}(d_{k})||_{1},
    \label{diff_norm}
\end{equation}
where $d_{k}^{sim}$ and $d_{k}$ is the summary statistic of the $k$-th measurement series of simulated and experimental data, respectively, and $||.||_{1}$ is the $L_{1}$-norm. The choice of norm does not significantly affect the shape of parameter posteriors if the acceptance threshold $\epsilon$ is sufficiently small (see Figure~\ref{episc_L12}). Note that $k \in \{1,\ldots,8\}$ for data set ${D}_\mathrm{CHIR}$ and $k \in \{1,\ldots,4\}$ for data set ${D}_\mathrm{EPISC}$. The output of this ABC algorithm is a set of parameters $\{\boldsymbol{\theta}\}$ distributed according to $p(\boldsymbol{\theta}|D(s(\mathcal{D}^{sim}),s(\mathcal{D}))< \epsilon)$. For sufficiently small $\epsilon$, this distribution should be a good approximation of the true posterior $p(\boldsymbol{\theta}|s(\mathcal{D}))$ which we confirmed by comparing the ABC posterior of cell division rate $\theta_{0}$ in model U with the analytical solution (see Figure~\ref{eps_convergence}).

For parameter inference and model selection we choose identical log-uniform priors for all rate constants in both models $\log_{10}(\theta_{k}) \sim \mathcal{U}[-2,0])$ (in units of per day). The upper cut-off can be justified by inspection of posterior distributions, which are close to zero above this threshold, $p(\theta_{k}>1)\approx 0$. Cell transitions of a certain type should occur sufficiently often over the observation period of the experiment to be considered in our model. By constraining the prior distribution $p(\theta_{k}<0.01)=0$, we only discard very rare cell transitions which would occur less than once every 100 days in a single cell. Considering that total population sizes in the experiment are around 10 cells and the observation period is three days, this lower cut-off seems reasonable.

\begin{figure}
	\centering
	\includegraphics[width=0.8 \textwidth]{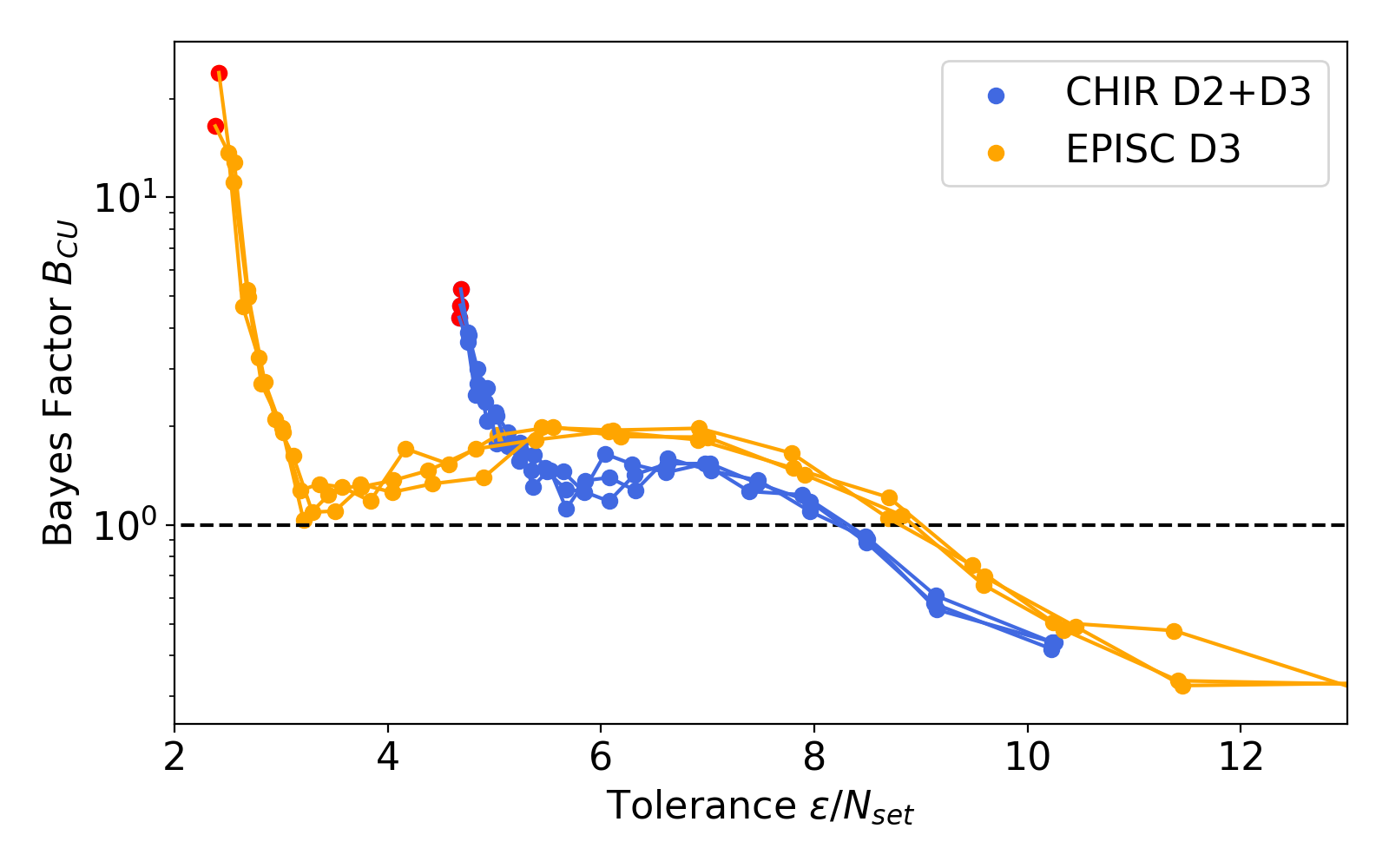}
	\caption{Bayes factors $B_{CU}$ of datasets $\mathcal{D}_\mathrm{CHIR}$ and $\mathcal{D}_\mathrm{EPISC}$ over a range of tolerance values $\epsilon$ with identical log-uniform priors for all rate constants in both models $\log_{10}(\theta_{k}) \sim \mathcal{U}[-2,0])$. The x-axis was re-scaled by the number of measurement series $N_{set}$ in the respective dataset ($N_{set}=8$ in $\mathcal{D}_\mathrm{CHIR}$, $N_{set}=4$ in $\mathcal{D}_\mathrm{EPISC}$). Red dots show the minimal epsilon value to which the SMC-ABC algorithm converged after $20$ iterations, resulting in $B_{CU}= 4.8 \pm 0.5$ for $\mathcal{D}_\mathrm{CHIR}$ and $B_{CU}= 20 \pm 4$ for $\mathcal{D}_\mathrm{EPISC}$. Three independent SMC-ABC runs were performed (shown as multiple lines), and the error of $B_{CU}$ was estimated by the spread between runs.}
	\label{episc_epsilon}
\end{figure}

We run the SMC-ABC algorithm, which progressively lowers tolerance $\epsilon$ until the acceptance probability becomes too small and $\epsilon$ cannot be decreased much further without impractical computational costs. The Bayes factor $B_{CU}$ can be calculated for every SMC-iteration from the marginal posterior distributions of models $M_{C}$ and $M_{U}$. This was done for both datasets $\mathcal{D}_\mathrm{CHIR}$ and $\mathcal{D}_\mathrm{EPISC}$ and the resulting Bayes factors are shown in Figure~\ref{episc_epsilon} as a function of $\epsilon/N_{set}$. We normalised the tolerance value $\epsilon$ by the number of measurement series $N_{set}$ in a dataset since the our distance function Eq.~\eqref{diff_norm} sums over the deviation of every measurement series $1\leq k \leq N_{set}$. For the lowest achievable tolerance (see red dots in Figure~\ref{episc_epsilon}) we obtain Bayes factors of $B_{CU}= 4.8 \pm 0.5$ for $\mathcal{D}_\mathrm{CHIR}$ and $B_{CU}= 20 \pm 4$ for $\mathcal{D}_\mathrm{EPISC}$). This constitutes moderate to strong evidence in favour of Model C dynamics, i.e. a coupling of cell division and state transitions.

\begin{figure}
	\centering
	\subcaptionbox{
	}{\includegraphics[width=0.7 \textwidth]{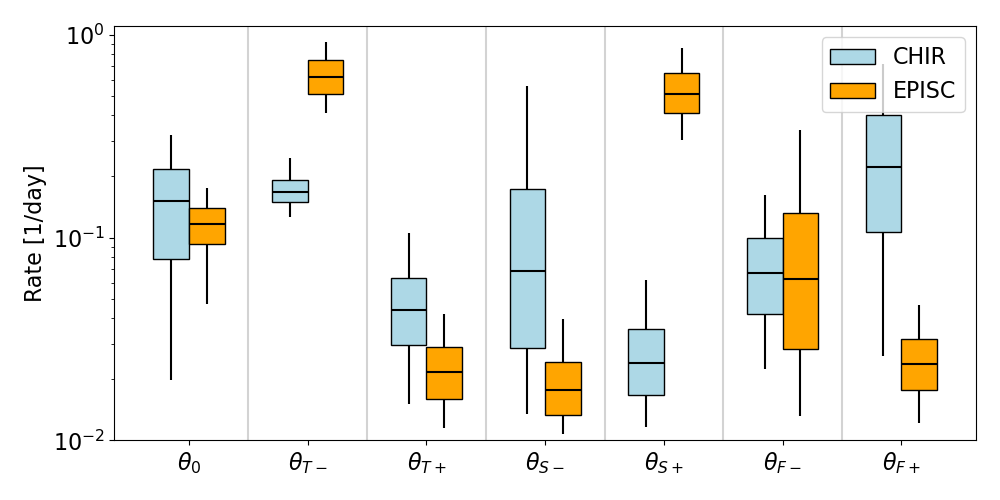}}
	\quad
	\subcaptionbox{
	}{\includegraphics[width=0.7 \textwidth]{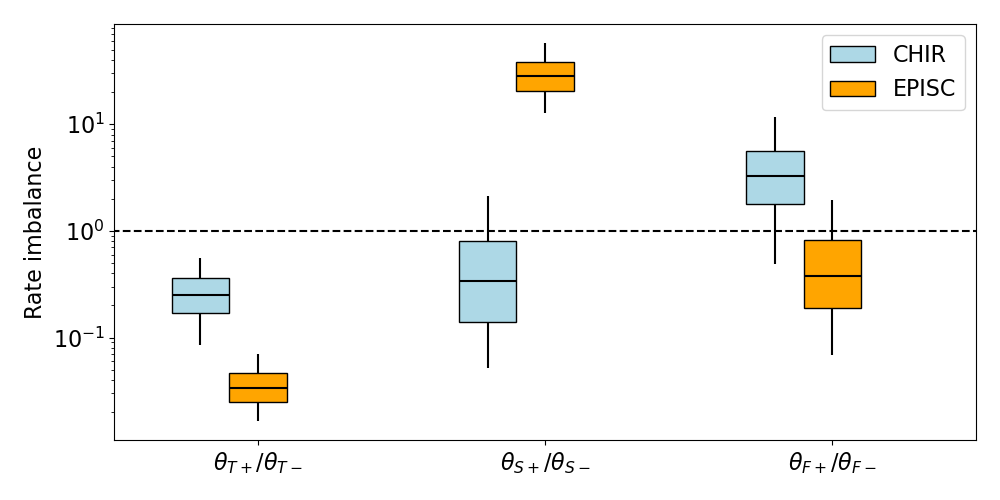}}
	\caption{Median (horizontal line), 0.25/0.75 quantiles (box) and 0.05/0.95 quantiles (vertical lines) of (a) posterior distributions and (b) rate imbalance of Model C posterior rates $\log_{10}(\theta_{k+}/\theta_{k-})$ given dataset $\mathcal{D}_\mathrm{CHIR}$ and $\mathcal{D}_\mathrm{EPISC}$.} 
	\label{episc_dat_compare}
\end{figure}

The posterior distribution of Model C log-rates $\log_{10}(\theta_{\nu})$ are shown in Figure~\ref{episc_dat_compare}(a) for both data sets. While the cell division rate $\theta_{0}$ depends only slightly on the experimental conditions ($\log_{10}(\theta_{0})\approx-1.1$), the rates for switching markers on and off differ considerably in different experimental conditions. The rate imbalance $\theta_{M+}/\theta_{M-}$ of marker switching rates is shown in Figure~\ref{episc_dat_compare} b, which implies:

\begin{itemize}
	\item Cells in CHIR conditions are mostly in state $[T^{-},S^{-},F^{+}]$. Although only a small sub-population of cells show marker expression $T^{+}$ in CHIR conditions, there are still more $T^{+}$ cells than under EPISC conditions which is consistent with the raw data from \cite{tsakiridis2014distinct}. The rate-imbalance of markers $F$ and $S$ are quite small, so in steady-state there may be significant sub-populations with marker expression $S^{+}$ or $F^{-}$.
	\item Cells in EPISC conditions are biased towards the state $[T^{-},S^{+},F^{-}]$. The rate-imbalance of marker $F$ is quite small, so in steady-state there may be sub-populations with marker expression $F^{+}$.
\end{itemize}

\begin{figure}
	\centering
	\subcaptionbox{$T^{-}$-sorted, CHIR conditions}{\includegraphics[width=0.45 \textwidth]{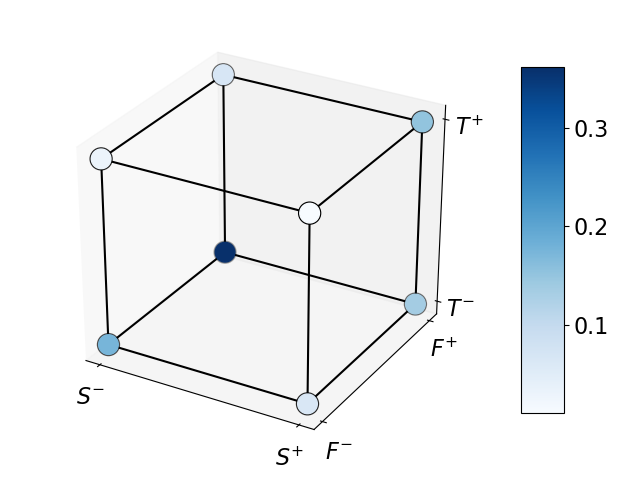}}
	\subcaptionbox{$T^{+}$-sorted, CHIR conditions}{\includegraphics[width=0.45 \textwidth]{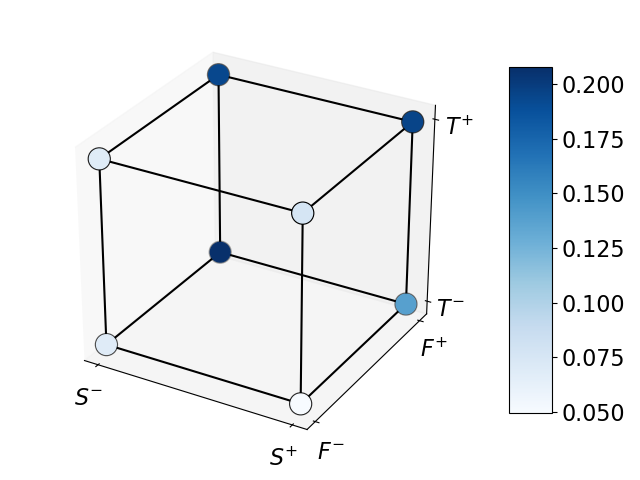}}
	\subcaptionbox{$T^{-}$-sorted, EPISC conditions}{\includegraphics[width=0.45 \textwidth]{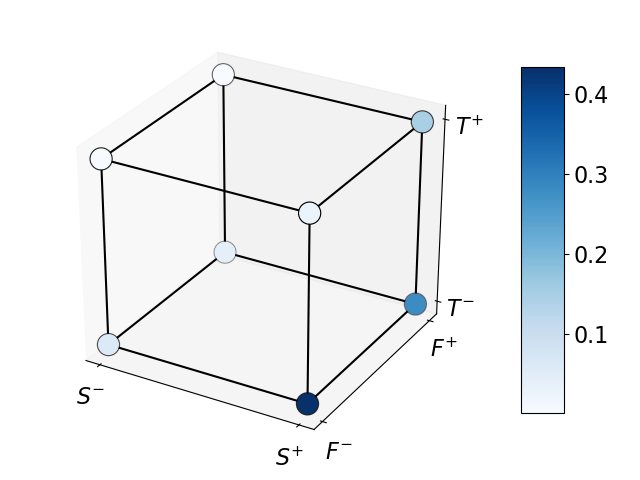}}
	\subcaptionbox{$T^{+}$-sorted, EPISC conditions}{\includegraphics[width=0.45 \textwidth]{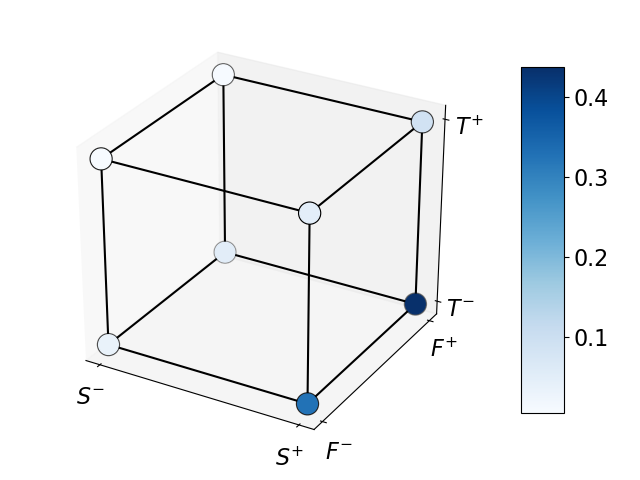}}
	\caption{Complete cell state distribution after $t=3 d$ for different environmental conditions (CHIR/EPISC) and initial cell conditions ($T^{-}$-sorted/$T^{+}$-sorted stem cells at $t=0 d$). The initial cell state distribution of $T^{\pm}$-sorted stem cells was obtained from steady-state populations from previous experiments ~\cite{tsakiridis2014distinct} and model C dynamics with inferred log-rates (see Figure~\ref{episc_dat_compare}) were used to obtain cell-state distribution at $t=3 d$. The probability of the respective marker expression is indicated by the colour map.} 
	\label{cellstat_distr}
\end{figure}

It is worth mentioning that although Model U was shown to be less likely, the posterior distribution of Model U shows a very similar rate imbalance (see Figure~\ref{CHIR_EPISC_compareY}). Furthermore, the inferred log-rates of model C allow us to obtain the complete distribution over cell states, which is not directly accessible by the experiments. As shown in Figure~\ref{cellstat_distr}, the cell state distribution depends on the initial cell conditions ($T^{-}$-sorted/$T^{+}$-sorted stem cells) and the environmental conditions (CHIR/EPISC). The cell state distribution for initially unsorted stem cells (assuming every cell state is equally likely) is shown in Appendix F.

\begin{figure}
	\centering
	\subcaptionbox{Model C posterior distribution of log-rates $\log_{10}(\theta_{k})$.}{\includegraphics[width=0.7 \textwidth]{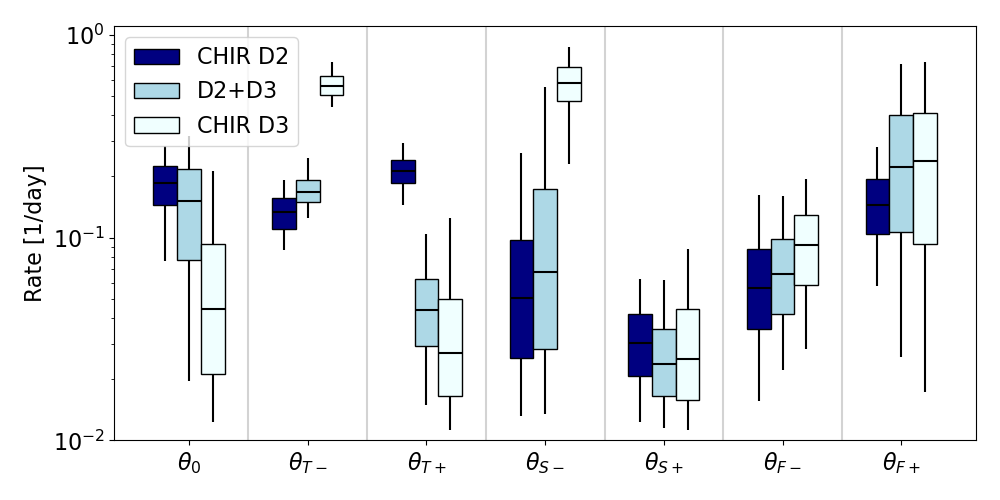}}
	\quad
	\subcaptionbox{Rate imbalance $\theta_{M+}/\theta_{M-}$ of posterior distributions for Model C.}{\includegraphics[width=0.7 \textwidth]{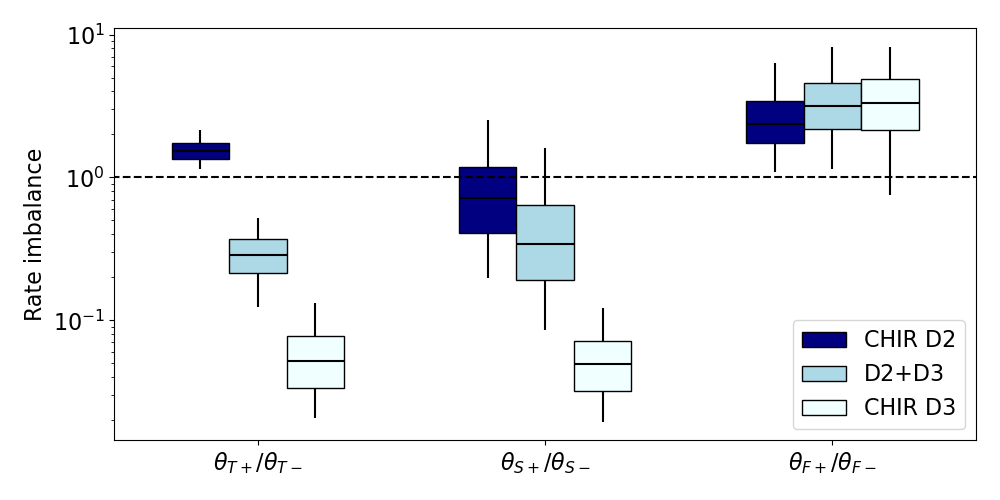}}
	\caption{Median (horizontal line), 0.25/0.75 quantiles (box) and 0.05/0.95 quantiles (vertical lines) of posterior distributions of Model C log-rates $\log_{10}(\theta_{k})$ obtained from $\mathcal{D}_\mathrm{CHIR}$. CHIR D2 and CHIR D3 shows inferred parameters using exclusively day 2 or day 3 data, respectively. CHIR D2+D3 was obtained by including all data points of $\mathcal{D}_\mathrm{CHIR}$ and coincides with the results shown in Figure~\ref{episc_dat_compare}.} 
	\label{CHIR_time_compareX}
\end{figure}

So far we assumed cell transitions happen at a fixed rate $\theta_{k}$ which does not vary in time. We can test this assumption by using experimental data $\mathcal{D}_\mathrm{CHIR}$ which recorded population snapshots at two different times $\Delta t = 2d$ and $\Delta t = 3d$. We used exclusively day 2 and day 3 data to obtain parameter posteriors separately for population dynamics at early times $0<t<2d$ and late times $2d<t<3d$, respectively. The inferred reaction rates are compared in Figure~\ref{CHIR_time_compareX}, which indicates:
\begin{itemize}
	\item The universal reproduction rate $\theta_{0}$ and marker rates $\theta_{S+}$, $\theta_{S-}$ do not show a substantial time dependence and stay constant over the course of three days.
	\item The rate imbalance of marker $T^{+}$, $\theta_{T+}/\theta_{T-}$, significantly decreases at day 3 compared to day 1 and day 2 (difference $>2\sigma$). The rate imbalance of marker $F^{+}$, $\theta_{F+}/\theta_{F-}$, seems to slightly increase with time, although the difference between day 2 and day 3 is not very substantial (difference $<\sigma$).
\end{itemize}

Again, the posterior distribution of Model U shows a very similar time dependence of reaction rates (see Figure~\ref{CHIR_time_compareY}).

%% file: discussion.tex

In this paper, we studied the population dynamics of two stochastic models of cell division and differentiation, or more generally, cell state changes: one in which cell state transitions are coupled to cell division, model C, defined by Eq.~\eqref{ME_X}, and one in which these two processes are independent, model U, defined by Eq.~\eqref{ME_Y}.
We solved the corresponding master equations analytically by using probability generating functions, which allow a systematic calculation of moments of the population distribution.
The resulting solutions for both Models C and U allowed us to analyse the dynamics of the population mean, based on which we showed that one cannot always differentiate between the two models.
In Model C, there is a finite probability that the proliferating sub-population will eventually die out and the dynamics will come to a halt.
The probability of extinction~\eqref{pext1} can be used to approximately infer the underlying reaction rate of model C solely based on the fraction of systems in which extinction was observed.
Contrary, in Model U both populations grow without bounds and extinction is impossible for finite reaction constants.
This characteristic feature of extinction in Model C might be useful to classify experimental observations or to infer cell division rates when there is negligible cell death.

For Bayesian parameter inference we were able to obtain analytical expressions for the likelihood if continuous-time observations $[0,T]$ of all populations are available, e.g. as would be obtained from live imaging. Using conjugate priors, we obtained the posterior distributions analytically. When confronted with incomplete data in form of population snapshots, Bayesian inference is more challenging, and we have to rely on numerical results. As estimating the snapshot likelihood directly is computationally costly, we used a more elaborate, but efficient sample path method for obtaining posterior distributions of rate constants given snapshot-data. We further showed that the information content of snapshot-data depends on the times these snapshots were taken, and proposed an optimal spacing strategy which maximises the confidence of parameter inference.

For parameter and model inference based on experimental EpiSC data, we had to extend Models C and U from two distinct cell populations to in total $8$ cell states. Since our sample path method turned out to be computationally not feasible, we used Approximate Bayesian Computation (ABC) which speeds up calculation by several orders of magnitude. We used ABC rejection sampling and ABC-SMC to obtain posterior parameter distributions and Bayes factors for selecting between the two competing Models C and U. The Bayes factors provide moderate to strong evidence in favour of model C in which cell division and differentiation are coupled (see Figure~\ref{episc_epsilon}). The posterior distribution of rate constants $\mathbf{\theta}$ quantify how cell fate changes depending on the in vitro culture conditions (CHIR/EPISC, see Figure~\ref{episc_dat_compare}). For cells in CHIR condition, we quantified how the rates of cell state transitions might change over time (see Figure~\ref{CHIR_time_compareX}). As we only have data from multiple time-points for the CHIR conditions, we cannot say whether this apparent time dependence is caused by a cell-intrinsic timing mechanism or whether the environment the cells experience changes over time because the cell colony is growing. Since cells kept under EPISC conditions don't shift their expression of transcription factors over comparable timescales~\cite{tsakiridis2014distinct}, it would seem more likely that these rates depend on the local environment and that cell-cell interactions are important, so that reaction rates change depending on the state of neighbouring cells. Interestingly, the data for cells under CHIR conditions indicates that cell division and differentiation are initially independent of each other (Model U) up to day 2, after which cell division and differentiation appear to be coupled (Model U, see Figure~\ref{chir_d2_d3}). To account more rigorously for this time-dependent dynamics in systems which are not well-mixed or with strong cell-cell interactions, one would have to consider more complex models such as non-homogeneous Markov models
or explicit representations of dynamic gene expression~\cite{Manning2019}.

\subsection{Limitations and alternative approaches}
\subsubsection{Assumption of Markovian dynamics}
We began our investigation with minimal models of dynamics with two cell states. Although Models C and U were motivated by biological arguments, the stochastic, two-dimensional jumping processes we defined could be applied to a variety of systems. The assumption of Markovian dynamics is often used due to a range of useful properties which makes the models easier to analyse. Often, especially in the context of cell dynamics, the Markov assumption is at best a rough approximation~\cite{Stumpf2017,gavagnin2018invasion}. Waiting times $\tau$ between jumps are distributed according to an exponential distribution, which is biologically unrealistic: On a molecular level, many transport processes are required for a cell to replicate. Therefore, waiting times between two division events cannot be arbitrarily small~\cite{gavagnin2018invasion}. 

Indeed, Recent work has suggested that experimentally observed cell cycle time distributions differ substantially from a markovian exponential distribution and are better described by the Erlang distribution~\cite{gavagnin2018invasion}. How do population dynamics change for a non-Markovian process? In this scenario cell populations cannot grow arbitrarily fast, there is a time-dependent upper bound on population sizes $N<N_{lim}(t)$. State probabilities $p_{i,j}(t)=p(N_{A}(t)=i,\: N_{B}(t)=j)$ are therefore zero for population numbers $(i,j)$ exceeding limit $N_{lim}(t)$. Cell population dynamics with Erlang distributed waiting times are non-Markovian stochastic processes with discrete states in continuous time. Non-Markovian processes like this can in general be analysed by converting them into Markov processes by inclusion of supplementary variables~\cite{cox1955analysis}.

\subsubsection{Insight from minimal models} For Bayesian inference given snapshot data, our main focus was on building a Bayesian framework which can be applied to experimental data. However, the algorithms for parameter inference and model selection originally proposed for Model C and U were computationally too intensive for the more complicated models with larger cell state networks. Nonetheless, the presented optimal snapshot spacing and scaling behaviour of Bayes factors with snapshot separation should generally be useful for experimental design. Measurement should be taken at times distributed non-uniformly according to equation~(\ref{SS_space_X}). \rev{It is worth mentioning that equation~\eqref{SS_space_X} is only valid if all $n$ snapshots are taken from the same system. If $n$ snapshots were taken from independent systems, equation~\eqref{SS_space_X} cannot be applied and the posterior variance scales with snapshot separation $\Delta t$ as shown in  Figure~\ref{fig_var_scaling_dt}. Since the posterior of a set of independent trajectories is the product over posteriors obtained from individual trajectories, one should try to maximise  $\Delta t$ for every trajectory individually.} 

\rev{Here we have presented models for growing in vitro colonies of stem cells, for which neglecting cell death is a reasonable approximation. There are many biological systems in which cell death may be relevant, e.g. under homeostatic self-renewal of tissues, in which total cell numbers have to kept constant, or even some applications in development where cell death is used for tissue size control \cite{kursawe2015capabilities,ambrosini2017apoptotic}. For these systems, the proposed models can be easily extended by including either a cell death process of the form $A \rightarrow \emptyset$. Similarly, constraints on the population size can be implemented by introducing a cell division or death rate which depends on the total population size $N$, which acts as a crowding feedback mechanism \cite{greulich2016dynamic}. This could represent contact inhibition or competition for niche access, for example. In spatial extensions of the models presented here, one could consider locally coupling division to differentiation and death events, as is relevant for example in epidermal homeostasis \cite{rompolas2016spatiotemporal,mesa2018homeostatic}. Any such mechanisms can be easily added to the models and the same methods for inference and model selection can be applied. A disadvantage of these additional mechanisms is that the corresponding master equations would likely be intractable and one would have to rely solely on numerical results.}

\subsubsection{Sufficiency of summary statistics} We carried out parameter inference and model selection based on experimental EpiSC data using an ABC algorithm which works with summary statistics of the data. We obtained Bayes factors of $B_{CU}\approx 4.8$ for $\mathcal{D}_\mathrm{CHIR}$ and $B_{CU}\approx 20$ for $\mathcal{D}_\mathrm{EPISC}$ (see Fig.~\ref{episc_epsilon}), which indicates moderate to strong evidence in favour of Model C. This result, however, has to be interpreted with care: 

First, model selection based on Bayes factors depends on the prior distribution accurately reflecting the state of knowledge about the underlying model parameters $\boldsymbol{\theta}$. Here we chose log-uniform priors $\theta_{k} \sim log(\mathcal{U}[-2,0])$, which reflects not knowing the magnitude of the rate $\boldsymbol{\theta}$. The exact value of Bayes factors might therefore change slightly depending on the choice of prior.

Secondly, because we used summary statistics of the data instead of the full set of observations, we did not obtain the ``true'' Bayes factor based on data $\mathcal{D}$, but one based on the summary statistic $s(\mathcal{D})$ of data. Considering the full data would mean to look at each experimental replicate individually by calculating the parameter posterior given a single experimental realisation and iterate over all data points, using the posterior distribution as the prior distribution for the next data point.
The ``true'' Bayes factor and the one obtained using summary statistics only coincide if the summary statistic $s(\mathcal{D})$ is \textit{sufficient} for comparing Models C and U: $p(\mathcal{D}|s(\mathcal{D}),M_{X})=p(\mathcal{D}|s(\mathcal{D}),M_{Y})$. There are only few cases in which summary statistics are known to be sufficient , e.g.~\cite{grelaud2009abc}. The more usual case is that we have to work with arbitrary summary statistics, which involve an unknown loss of information if they are insufficient. This ``curse of insufficiency'' is the reason why some researchers caution against a naive use of ABC for model selection~\cite{marin2015likelihood}. There are indeed three levels of approximation errors: one due to MC estimation, one due to the non-zero ABC tolerance $\epsilon>0$, and one due to insufficient summary statistics. Alternatives to model selection with insufficient statistics were explored by~\cite{pudlo2015reliable}, which selected models via a machine learning approach (random forests).  \rev{An exhaustive investigation how the choice of summary statistics affects the shape of the posterior distributions is beyond the scope of this present work, but an interesting prospect for follow-on work.}

Despite concerns about sufficiency of our summary statistics, we verified that the summary statistic used here is able to discriminate between extended Models C and U. Using Rejection-ABC, we obtained Bayes factors of $B_{CU}\approx 3.5$ on synthetic CHIR data $\mathcal{D}_\mathrm{CHIR}^\mathrm{sim}$ ($\epsilon/N_{set}=3.75$) and $\langle B_{CU} \rangle \approx 3.7$ on synthetic EPISC data $\mathcal{D}_\mathrm{EPISC}^\mathrm{sim}$ ($\epsilon/N_{set}=6$, note that one would likely achieve higher Bayes Factors at lower $\epsilon$ with more efficient sampling schemes). Synthetic data was generated from model C with reaction rates set to posterior means obtained from experimental data (see Figure~\ref{episc_dat_compare}). It is worth noting that for simulated data, the acceptance probability for a given $\epsilon$ value is much higher than for experimental data. This suggests that the Models C and U do not perfectly describe the experimental data. More realistic models for cell state changes are therefore discussed in the following section.

\subsubsection{Building models for single cell biology}
Several approaches that aim to go beyond pseudotime ordering and connect single cell gene expression data to dynamic models have recently emerged. Examples include fitting drift-diffusion equations on a $k$-nearest neighbour graph~\cite{Weinreb2017} or along one-dimensional pseudotime trajectories~\cite{Fischer2019}, simulating cell trajectories in dimensionally reduced gene expression space based on RNA velocity estimates~\cite{Kimmel2020}, and continuous state Markov process from single-cell time series~\cite{Taylor-King2020}. Here, we have presented an alternative approach to these which allows for rigorous quantification of uncertainty. Working in a Bayesian framework further allows us to integrate multiple sources of data relatively easy, e.g. to combine bulk gene expression and single-cell data, which we discuss further below.

\subsection{Future work\label{futurework}}

The models proposed in this work assumed that rates for switching genes on and off are independent of each other, thereby neglecting regulatory interactions between genes. A natural next step is to consider models with interdependencies between transcription factors, i.e., where the rate of transitioning into the on/off state for one transcription factor can depend on the expression state of another transcription factor. The number of possible pairwise dependencies alone leads to a model comparison problem that is computationally too intensive typical ABC inference approaches. However, systematic Bayesian model comparison can be performed on bulk population data, where likelihood evaluation is tractable, before parameterising the models on clonal data (Schumacher et al., in preparation). \rev{For Bayesian inference based on experimental EpiSC data, we had to extend Models C and U from two distinct cell populations to in total $8$ cell states. In order to reduce the number of free parameters, we only considered symmetric cell divisions, arguing that this should not significantly change population dynamics. Analysing Model C we found that mean population dynamics are less sensitive to changes in the asymmetric cell division rate than to changes in the rate of symmetric cell division, which further supports our heuristic argument. A detailed investigation about how symmetric vs asymmetric cell divisions would affect population dynamics in the extended model, and how well the relative contribution of these two modes of division can be inferred from experimental data would be an interesting prospect for follow-on work.}

Here we have have applied Bayesian inference and model comparison on in vitro data. It would be interesting to compare estimates of transition rates with in vivo case, where, in certain contexts, lineage tracing data can be used to identify clones. A potential application of our model comparison approach in this context would be to compare the number of stages in stem/progenitor cell lineages to best explain observed clone size distributions~\cite{Picco2019}.

Contemporary experiments increasingly measure the expression of many more genes than considered here, for which one should consider models with a higher number of genes. As we showed in the last section, however, exact Bayesian inference is already computationally challenging for models considering only three genes. Since the number of cell states grows exponentially with the number of considered genes, one would run into computational scaling issues for methods such as ours. Thus, dimensionality reduction or careful consideration of the state space will be crucial to extend the inference of cell state transition rates to single-cell transcriptomic data.

The presented work focused on total population sizes of cell states where individual cells are independent of each other, which implies a spatially homogeneous or well-mixed system. In many biological systems cell fate however depends on the local environment and signalling of cells. Here, there is potential to draw on and add to recent work inferring motility as well as proliferation in cell populations and tissue patterning~\cite{Lambert2017a,Guerrerodev176297,simpson2020practical}. The extension of these population models to spatially heterogeneous models including interactions between neighbouring cells, similar to \cite{greulich2016dynamic}, could thus be promising for inferring rates of differentiation in embryonic development.

\section*{Acknowledgements}
AGF was supported by a Vice-Chancellor’s Fellowship from the University of Sheffield, LJS was supported by a Chancellor's Fellowship from the University of Edinburgh.